\documentclass[prd,
superscriptaddress,preprintnumbers,tightenlines,nofootinbib, eqsecnum]{revtex4-2}

\usepackage{amsmath}
\usepackage{amsfonts}
\usepackage{amssymb}
\usepackage{bm}
\usepackage{hyperref}
\usepackage{mathrsfs}
\usepackage{graphicx}
\usepackage{subcaption} 
\usepackage{makecell, booktabs} 
\usepackage{empheq}
\usepackage{ulem}
\usepackage[usenames]{color}
\usepackage{ytableau}
        
\allowdisplaybreaks

\DeclareSymbolFontAlphabet{\mathrsfs}{rsfs}
\DeclareMathAlphabet{\mathcal}{OMS}{cmsy}{m}{n}

\newcommand{\nn}{\nonumber}
\newcommand\calO{\mathcal{O}}
\newcommand{\dd}{\mathrm{d}}
\newcommand{\di}{\mathrm{i}} 
\newcommand{\de}{\mathrm{e}} 

\newcommand{\da}{\mathrm{a}}
\newcommand{\db}{\mathrm{b}}
\newcommand{\dc}{\mathrm{c}}

\newcommand{\calX}{\mathcal{X}}
\newcommand{\calY}{\mathcal{Y}}
\newcommand{\calC}{\mathcal{C}}
\newcommand{\calD}{\mathcal{D}}

\newcommand{\be}{\begin{equation}}
\newcommand{\ee}{\end{equation}}
\newcommand{\bse}{\begin{subequations}}
\newcommand{\ese}{\end{subequations}}

\definecolor{darkgreen}{rgb}{0,0.5,0}

\hypersetup{
 unicode=false,          
 pdftoolbar=true,        
 pdfmenubar=true,        
 pdffitwindow=false,     
 pdfstartview={FitH},    
 pdftitle={My title},    
 pdfauthor={Author},     
 pdfsubject={Subject},   
 pdfcreator={Creator},   
 pdfproducer={Producer}, 
 pdfkeywords={keyword1} {key2} {key3}, 
 pdfnewwindow=true,      
 colorlinks=true,       
 linkcolor=red,          
 citecolor=cyan,        
 filecolor=magenta,      
 urlcolor=darkgreen,           
 linktocpage=true
}

\makeatletter
\g@addto@macro\bfseries{\boldmath}
\makeatother

\begin{document}
	
\title{Innermost stable circular orbit of  arbitrary-mass compact binaries \\at 
fourth post-Newtonian order}

\author{Luc \textsc{Blanchet}}\email{luc.blanchet@iap.fr}
\affiliation{$\mathcal{G}\mathbb{R}\varepsilon{\mathbb{C}}\calO$, 
	Institut d'Astrophysique de Paris, \\UMR 7095, CNRS, Sorbonne Universit{\'e},
	98\textsuperscript{bis} boulevard Arago, 75014 Paris, France}

\author{David \textsc{Langlois}}\email{david.langlois@apc.in2p3.fr}
\affiliation{Université Paris Cité, CNRS, Astroparticule et Cosmologie, 75013 Paris, France}

\author{Etienne \textsc{Ligout}}\email{etienne.ligout@apc.in2p3.fr}
\affiliation{$\mathcal{G}\mathbb{R}\varepsilon{\mathbb{C}}\calO$, 
	Institut d'Astrophysique de Paris, \\UMR 7095, CNRS, Sorbonne Universit{\'e},
	98\textsuperscript{bis} boulevard Arago, 75014 Paris, France}
\affiliation{Université Paris Cité, CNRS, Astroparticule et Cosmologie, 75013 Paris, France}

\date{\today}

\begin{abstract}
	We compute by means of post-Newtonian (PN) methods the innermost stable circular orbit (ISCO) of arbitrary-mass (in particular, comparable-mass) compact binaries. Two methods are used with equivalent results: dynamical perturbation of the conservative equations of motion in harmonic coordinates, and dynamical perturbation of the conservative Hamiltonian in ADM coordinates. The perturbation of the non-local tail term at  4PN order in both approaches is carefully investigated. Our final gauge invariant result for the location of the ISCO at  4PN order is close to the numerical value of the ISCO shift computed by the gravitational self-force (GSF) approach in the small mass-ratio limit, and is also in good agreement with the full numerical-relativity calculation in the case of equal masses. The PN method followed here is considered in standard Taylor-expanded form, without any resummation techniques applied. As a complement, we also compute explicitly the gauge transformation from harmonic coordinates to ADM coordinates up to 4PN order, including the tail contribution therein.
\end{abstract}


\maketitle

\section{Introduction}\label{section:introduction}

The conservative equations of motion of compact binary systems (with arbitrary masses and without spins) have been developed up to  fourth post-Newtonian (4PN) order $\sim c^{-8}$ beyond the Newtonian acceleration. The methods in use are: the Hamiltonian formalism in Arnowitt-Deser-Misner (ADM) coordinates~\cite{JaraS15,BiniD13,DJS14,DJS15eob}; the Fokker Lagrangian in harmonic coordinates~\cite{BBBFMa,BBBFMb,BBBFMc,MBBF17,BBFM17}; and the effective field theory (EFT) method~\cite{GLPR16,FS19,FPRS19,BlumMMS20a}. See also~\cite{BlumMMS21,BiniDG20b} for partial results at 5PN and 6PN orders. The equations of motion at 4PN order (and the ten invariants associated with the Poincar\'e group) are valid in a general frame and for general orbits (bound or unbound). The radiation reaction dissipative terms are also known up to 4.5PN order~\cite{IW93,IW95,GII97,PW02,NB05,BFT24} and can be added separately.

After reduction to the frame of the center of mass (CM) and further restriction to circular orbits, a natural question to ask is whether there is an analogue of the Innermost Stable Circular Orbit (ISCO) for the self-gravitating system of two arbitrary, in particular comparable, masses, in analogy with the ISCO of the Schwarzschild spacetime, \textit{i.e.} 
\begin{align}\label{eq: ISCO_Schw}
r^\text{Schw}_\text{ISCO}=\frac{6GM}{c^2}\,.
\end{align}
This question of the dynamical stability of circular orbits in the PN approximation, was
initiated by Kidder, Will and Wiseman~\cite{KWWisco} who considered a general perturbation of the equations of motion in harmonic coordinates at the 2PN order. The problem of stability has also been studied using the ADM Hamiltonian formalism~\cite{Schaferorleans}. Note that in the PN context, the notion of ISCO differs from an alternative notion called MECO (minimum energy circular orbit) or sometimes ICO (innermost circular orbit), which is defined by the minimum of the (negative) invariant binding energy for circular orbits~\cite{B02ico,BCV03a,LBB12}. The ISCO and ICO are formally equivalent if we control the full PN series, however they become distinct notions when computed numerically, with uncontrolled higher PN terms neglected, as discussed in Ref.~\cite{BCV03a} in the Hamiltonian formalism.

Up to 3PN order the problem of the dynamical perturbation was solved by Blanchet and Iyer~\cite{BI03CM}, using both the equations of motion in harmonic coordinates and the Hamiltonian in ADM coordinates. This led to a gauge invariant criterion for the stability of circular orbits, valid for arbitrary binary's masses, using the standard PN Taylor expansion without resummation. The interesting feature of this criterion is that it reduces to the Schwarzschild ISCO in the test mass limit,
Eq.~\eqref{eq: ISCO_Schw}, at any known PN order, with  corrections depending on the mass ratio starting only from the 2PN order.

The 3PN gauge-invariant criterion~\cite{BI03CM} has been confronted by Favata~\cite{F11a} to various other (analytical and numerical) calculations of the ISCO and shown to perform well. In particular it was found that the 3PN ISCO is in good agreement with the shift of the ISCO due to finite mass ratio effects between the two bodies, as computed from the conservative first order gravitational self-force (GSF) approach by Barack and Sago~\cite{BarackS09} and Le~Tiec \textit{et al.}~\cite{LBB12} (see also~\cite{BarackS07,BarackS11,Akcay12}). Favata~\cite{F11a} also explored various resummed PN-based approaches like the effective-one-body (EOB) method~\cite{BuonD99,BuonD00,DNorleans}, but concluded that such methods do not perform better than the standard non-resummed PN Taylor expansion, unless they are calibrated by hand to the results of numerical relativity. The stability criterion has also been generalized to include spin effects at 3PN order, and compared with conservative finite mass corrections to the ISCO of a Kerr black hole~\cite{F11b}.

In this paper, motivated by the good performance of the standard (non-resummed) PN expansion for predicting the ISCO of arbitrary-mass compact binaries, we extend the gauge-invariant criterion to the next 4PN order, using both the 4PN equations of motion in harmonic coordinates~\cite{BBBFMa,BBBFMb,BBBFMc,MBBF17,BBFM17} and the 4PN Hamiltonian in ADM coordinates~\cite{JaraS15,BiniD13,DJS14,DJS15eob}. We only consider the conservative part of the PN dynamics, \textit{i.e.} we ignore the dissipative radiation-reaction piece.
Note that the GSF calculation, with which  we want to  compare our result, is also restricted to conservative effects~\cite{BarackS09}. An important property of the equations of motion and Hamiltonian at 4PN order is the presence of a non-local-in-time contribution associated with gravitational wave tails propagating in the far zone~\cite{BD88,BD92,FStail,DJS14,GLPR16}. Thus, in this paper we address the delicate issue of the dynamical perturbation of the circular orbit due to the non-local 4PN tail term.

Since our calculation uses both the equations of motion and the Hamiltonian at 4PN order, it is also instructive to revisit the contact transformation or shift of the particle's trajectories linking the two descriptions of the motion in harmonic coordinates and in ADM coordinates. In particular this transformation entails removing the accelerations from the harmonic Lagrangian (that are known to appear from the 2PN order) to transform it into an ordinary Lagrangian from which we obtain an ordinary Hamiltonian. The analysis we perform in this paper closely follows and extends to 4PN order the method of~\cite{ABF01}, and can be seen as a particular case of a general algorithm for removing accelerations in a PN Lagrangian as described in~\cite{DS85,DS91}. We also point out that there is a contribution of the final contact transformation which is necessary in order to relate together the non-local 4PN tail terms in the two approaches. 

Our main result is the following gauge-invariant criterion for the stability of circular orbits at the 4PN order, given by $C_\text{ISCO} > 0$ with 
\begin{align}\label{eq: criterion result in intro}
C_\text{ISCO} &= 1 -6 x+14\nu\, x^2 + \nu \left[ \frac{397}{2}-\frac{123 \pi^2}{16} -14\nu  \right] x^3\nn\cr
&  + \nu \Bigg[ -\frac{215729}{180} + \frac{58265 \pi^2}{1536} +\frac{5024}{15}\,\gamma_{\rm E} +\frac{2512}{15}\,\mathrm{ln}\left(x \right) +\frac{1184}{15}\,\mathrm{ln}(2)+\frac{2916}{5}\,\mathrm{ln}(3) \\
         & \qquad \quad + \left( -\frac{4223}{6}+\frac{451\pi^2}{16}\right) \nu  +\frac{196}{27}\,\nu^2 \Bigg]x^4 + \calO(x^5)\,,
\end{align}
where $\nu=\mu/m$ is the binary's symmetric mass ratio (with $\mu$ the reduced mass and $m=m_1+m_2$ the total mass) and the quantity
\begin{align}\label{eq: def_x}
	x \equiv \left(\frac{G m \Omega_0}{c^3}\right)^{2/3}\,,
\end{align}
denotes the dimensionless PN parameter associated with the orbital frequency $\Omega_0$ of the circular orbit. The frequency $\Omega_0$, and thus $x$, is a directly measurable quantity, which is gauge invariant with respect to the large class of coordinate systems that are asymptotically Minkowskian at infinity.

The plan of the paper is as follows. In the next section, Sec.~\ref{section:stability analysis}, we present the general strategy to compute the linear stability criterion, based on two different approaches: the linearisation of the equations of motion in harmonic coordinates or the linearisation of the Hamiltonian equations written in ADM coordinates. The following section, Sec.~\ref{sec tail part}, is devoted to the calculation of the tail contributions and of their linear variations, in both formalisms. In Sec.~\ref{section:results}, we put together the results of the two previous sections to compute, independently in the two formalisms, the explicit expressions of the stability criterion up to 4PN order, first in a coordinate-dependent form and finally in the gauge-independent version, using the parameter~\eqref{eq: def_x}. As a supplementary material, we present in Sec.~\ref{section:transformation} the explicit contact transformation between harmonic and ADM coordinates, including the tail term up to 4PN order, and confirm the overall consistency of our calculation. The next section, Sec.~\ref{section:comparison}, is devoted to a discussion of the criterion we have obtained, comparing it with the criterion at lower PN orders and with numerical calculations, in particular the GSF ISCO shift computed in~\cite{BarackS09, LBB12}, and also with a distinct EOB criterion. The paper ends with a short conclusion in Sec.~\ref{section:conclusion}. Three appendices provide some details of our computation.

\section{Stability analysis for circular orbits}\label{section:stability analysis}

We present the general strategy to study the linear stability of the  system, first using the equations of motion in harmonic coordinates and then the Hamiltonian formalism. 

\subsection{Analysis based on equations of motion}\label{subsection:analysis EOM}

The equations of motion of compact binary systems (without spins) at  4PN order, in the center of mass (CM) frame\footnote{In the CM frame the individual positions $\bm{y}_1$ and $\bm{y}_2$ of the particles are related to the relative position $\bm{x}\equiv \bm{y}_{1}-\bm{y}_{2}$ and velocity $\bm{v}\equiv \bm{v}_{1}-\bm{v}_{2} = \dd \bm{x}/\dd t$. We denote $r\equiv\vert\bm{x}\vert$, together with $\bm{n}=\bm{x}/r$ and $\dot{r}=(nv)=\bm{n}\cdot\bm{v}$. The total mass is $m=m_1+m_2$; the reduced mass $\mu=m_1m_2/m$; and the symmetric mass ratio $\nu\equiv \mu /m = m_1 m_2/(m_1+m_2)^2$.
}
are of the form
\begin{equation}
	\label{eq: EOM Lagrangian first form}
	\frac{\dd v^i}{\dd t}=-\frac{Gm}{r^2} \Bigl[ \bigl(1+\mathcal{A}\bigr) n^i + \mathcal{B} v^i \Bigr] + a_{\rm tail}^{i}\,,
\end{equation}
where the functions $\mathcal{A}$ and $\mathcal{B}$ encompass the contributions of the ``instantaneous'' (local-in-time) part of the 4PN dynamics, while $a_{\rm tail}^{i}$ denotes the non-local 4PN tail term. The long explicit expressions of $\mathcal{A}$ and $\mathcal{B}$ at 4PN order, given in harmonic coordinates, are relegated to Appendix~\ref{Appendix: explicit expressions of A and B}. 

The tail acceleration $a_{\rm tail}^{i}$ is derived  from the variation of the Lagrangian
\begin{equation}\label{L_tail}
	\mathcal{L}_{\rm tail} =\frac{G^2 M}{5 c^8 }Q_{ij}^{(3)}\, \mathcal{T}_{ij}^{(3)}\,,
\end{equation}
where  
$M$ is the constant ADM mass of the system (which reduces  to the total mass $m=m_1+m_2$ at leading order), $Q_{ij}^{(3)}$ denotes the third time derivative of the ordinary Newtonian symmetric trace-free quadrupole moment of the binary system,\footnote{Since all terms involving such integrals are multiplied by $1/c^8$,  it is sufficient, at 4PN order, to consider only the Newtonian version of the quadrupole moment.} and $\mathcal{T}_{ij}^{(3)}$ stands for the expression
\begin{equation}\label{Tij}
\mathcal{T}_{ij}^{(3)}(t) \equiv \mathop{\mathrm{Pf}}\limits_{2r/c}\int_{-\infty}^{+\infty} \frac{ Q_{ij}^{(3)}(t')}{\lvert t-t' \rvert}\, \dd t'\,,
\end{equation}
where Pf is the Hadamard Partie finie.\footnote{\label{footnote: partie finie def} In general, for any well-behaved function $f(t)$ going to zero sufficiently fast when $t \rightarrow \pm \infty$, this Partie finie is defined by
\begin{equation*}
	\mathop{\mathrm{Pf}}\limits_{\tau_0}\int_{-\infty}^{+\infty} \frac{f(t')}{\lvert t-t' \rvert}\, \mathrm{d}t' \equiv \int_{0}^{\infty} \mathrm{d}\tau \, \mathrm{ln}\left( \frac{\tau}{\tau_0}\right) \left[ f^{(1)}(t-\tau)-f^{(1)}(t+\tau) \right]\,.
\end{equation*}
In the present formalism, we assume that the quadrupole moment is constant in the remote past (for $t\leqslant -\mathcal{T}$ where $-\mathcal{T}$ is the instant of formation of the compact binary), and we extend this assumption to the remote future as well (for $t\geqslant +\mathcal{T}$), so the integral~\eqref{Tij} is well-defined at infinity.
} 
In the rest of this section, the  term $a_{\rm tail}^{i}$ will be kept generic, 
its explicit form being required only  later,  in Sec.~\ref{sec tail part}. 

Let $\Omega$ be the orbital frequency associated with the motion of the binary system, $n^{i}$ the unit radial vector and $\lambda^{i}$ the unit tangential vector in the orbital plane, oriented in the sense of the motion, so that $\dot{n}^i = \Omega \lambda^i$ and $\dot{\lambda}^i = - \Omega n^i$. Setting $u \equiv \dot{r}$, the velocity vector can be decomposed as
$v^i=u \,n^i + r\,\Omega\, \lambda^i$. Then Eq.~\eqref{eq: EOM Lagrangian first form} amounts to the following couple of equations:
\begin{subequations}\label{eq: coupled system EOM Lagrangian}
   \begin{align}
          &\Dot{u} =- \frac{Gm}{r^2} \bigl(1+\mathcal{A}+\mathcal{B}u \bigr) + r \Omega^2 +  a^{\rm tail}_{n}\,,\\
          &\Dot{\Omega} =-\Omega\left( \frac{2u}{r}+\frac{Gm}{r^2} \mathcal{B} \right) +  \frac{1}{r} \,  a^{\rm tail}_{\lambda}\,,
   \end{align}
\end{subequations}
where we have defined $a^{\rm tail}_{n} \equiv \boldsymbol{n} \cdot \boldsymbol{a}_{\rm tail}$ and $a^{\rm tail}_{\lambda} \equiv \boldsymbol{\lambda} \cdot \boldsymbol{a}_{\rm tail}$. 
  On a circular orbit — here and thereafter indicated by the subscript $0$ — the orbital frequency is given by the generalized Kepler's law
\begin{equation}
	\label{eq: frequency lagrangian formalism generic expression}
	\Omega_0^2 = \frac{Gm}{r_0^3}\left(1+\mathcal{A}_0 \right)- \frac{1}{r_0}\,a^{\rm tail}_{n,0}\,.
\end{equation}

To investigate the problem of the stability of circular orbits against dynamical perturbations, we start from a circular orbit defined by the constants $r_0$, $\Omega_0$ (and, implicitly, $u_0=0$), and consider a small perturbation to a new orbit characterized  by $r=r_0+\delta r$, $\Omega=\Omega_0+\delta \Omega$ and $u=\delta u$. Following~\cite{KWWisco, BI03CM}, let us vary Eqs.~\eqref{eq: coupled system EOM Lagrangian} in order to find the evolution equations governing the perturbations $\delta r$, $\delta \Omega$ and $\delta u$. Neglecting the dissipative contributions at 2.5PN and 3.5PN orders, $\mathcal{A}$ is a function of $u^2$ (we have $v^2 = u^2+r^2 \Omega^2$), so that the partial derivative of $\mathcal{A}$ with respect to $u$ vanishes at $u=0$. Thus, 
\begin{equation}\label{eq:delta A}
        \delta \mathcal{A}= \left(\frac{\partial \mathcal{A}}{\partial r} \right)_{0} \delta r + \left(\frac{\partial \mathcal{A}}{\partial \Omega} \right)_{0}  \delta \Omega\,.
\end{equation}
Moreover, because $\mathcal{B} \propto u$, we simply have 
\begin{equation}\label{eq:delta B}
     \delta \mathcal{B} =\left(\frac{\partial \mathcal{B}}{\partial u} \right)_{0} \,  \delta u\,.
\end{equation}
Using~\eqref{eq:delta A}--\eqref{eq:delta B} together with Eq.~\eqref{eq: frequency lagrangian formalism generic expression}, the perturbations of $\dot{r}=u$ and of the equations of motion~\eqref{eq: coupled system EOM Lagrangian} lead to a linear system of differential equations of the form
\begin{subequations}\label{eq: differential system perturbation Lagrangian formalism}
	\begin{align}
		\Dot{\delta r} &= \delta u\,, \\
		\Dot{\delta u} &= \alpha_0\,  \delta r  + \beta_0\,  \delta \Omega\,,  \\
		\Dot{\delta \Omega} &= \gamma_0\,  \delta u \,,
	\end{align}
\end{subequations}
where we have introduced the coefficients
\begin{subequations}\label{eq: stability coefficients}
    \begin{align}
        & \alpha_0 =  3\Omega_0^2 -\frac{G m}{r_0^2} \left(\frac{\partial \mathcal{A}}{\partial r} \right)_{0}\,+  \frac{2}{r_0}\,a^{\rm tail}_{n,0}+\left(\frac{\partial a^{\rm tail}_{n}}{\partial r} \right)_{0}\,,\\
        & \beta_0 = 2 r_0 \,\Omega_0- \frac{Gm}{r_0^2} \left(\frac{\partial \mathcal{A}}{\partial \Omega} \right)_{0} +\left(\frac{\partial a^{\rm tail}_{n}}{\partial \Omega} \right)_{0}\,,\\
        &  \gamma_0 =-\Omega_0\left[\frac{2}{r_0}+\frac{G m}{r_0^2} \, \left(\frac{\partial \mathcal{B}}{\partial u} \right)_0 \right] +\frac{1}{r_0}\left(\frac{\partial a^{\rm tail}_{\lambda}}{\partial u} \right)_{0}\,. 
    \end{align}
\end{subequations}
Now, the circular orbit is stable if the eigenmodes of the system of equations~\eqref{eq: differential system perturbation Lagrangian formalism} are purely oscillatory or damped. Let $\de^{\sigma t}$ denote the time-dependence of an eigenmode, where $\sigma \in \mathbb{C}$ is the associated eigenfrequency. We readily find that the eigenfrequencies are $\sigma=0$ and $\sigma=\pm\sqrt{\alpha_0+\beta_0\,\gamma_0}$. Clearly then, the criterion for stability of the circular orbit and hence the definition of the ISCO obtained with the equations of motion (EoM) is
\begin{equation}
\label{eq: Lagrangian criterion}
    \hat{C}_0^{\rm EoM} \equiv -\alpha_0-\beta_0\,\gamma_0 >0\,.
\end{equation}
The explicit expression for $\hat{C}_{0}^{\rm EoM}$ will be obtained in section \ref{sec tail part}, where we focus on the computation of the tail contributions in the coefficients \eqref{eq: stability coefficients}.

\subsection{Analysis in Hamiltonian formalism}\label{subsection:hamiltonian formalism}

Similarly to the equations of motion, the 4PN (reduced) Hamiltonian $\mathcal{H}$~\cite{JaraS15,BiniD13,DJS14,DJS15eob} contains both an instantaneous part $\mathcal{H}_{\rm inst}$ and a tail contribution that turns out to be simply given, at this order, by
\begin{equation}\label{eq:Htail}
    \mathcal{H}_{\rm tail}=-\mathcal{L}_{\rm tail}=-\frac{G^2 m}{5 c^8 }Q_{ij}^{(3)}\, \mathcal{T}_{ij}^{(3)}\,,
\end{equation}
which must be seen as a \textit{functional} of the phase space variables because of the time integration in~\eqref{L_tail}. Furthermore, an order reduction\footnote{By order reduction, we mean using the equations of motion to eliminate the accelerations and higher time derivatives.} of the accelerations in the tail Lagrangian has to be made in order to define the Hamiltonian.

We shall study the linear perturbations induced by the full dynamics, in order to check that the resulting criterion is consistent with that derived from the analysis based on equations of motion in Sec.~\ref{subsection:analysis EOM}. This will provide yet another confirmation that both formalisms encode the same physics. As before, we shall keep the tail part of the dynamics, now $\mathcal{H}_{\rm tail}$, unspecified, postponing its detailed analysis to
Sec.~\ref{subsec: computation tail hamiltonian}. The explicit expression of the instantaneous part $\mathcal{H}_{\rm inst}$ of the Hamiltonian is provided in Appendix~\ref{Appendix: explicit expressions of A and B}.

Following~\cite{BI03CM}, and denoting the ADM coordinates by capital letters, let us introduce the spherical ADM coordinates $(R,\Theta,\Psi)$, which can be reduced to polar ADM coordinates $(R,\Psi)$ assuming that the orbital motion takes place in the plane $\Theta=\pi/2$. Then, $\mathcal{H} = \mathcal{H}[R,\Psi, P^2]$, where $\mathcal{H} = H/\mu$ and $P^2 \equiv \boldsymbol{P}^2$ is the (reduced) linear momentum squared given by the familiar decomposition
\begin{equation}
    P^2=P_R^2+\frac{P_\Psi^2}{R^2}\,,
\end{equation}
where $P_R$ is the radial component of $\boldsymbol{P}$ and $P_\Psi$ is the angular momentum; hence we have $\mathcal{H} = \mathcal{H}[R,P_R,\Psi,P_\Psi]$. The corresponding Hamiltonian equations are given  by
\begin{subequations}\label{eq:EOM Hamiltonian}
    \begin{align}
        & \dot{R} = \frac{\partial \mathcal{H}_{\rm inst}}{\partial P_R}  +\frac{\delta \mathcal{H}_{\rm tail}}{\delta P_R} \label{eq:EOM Hamiltonian_a}\,, \\
        & \dot{P}_R = -\frac{\partial \mathcal{H}_{\rm inst}}{\partial R}  -\frac{\delta \mathcal{H}_{\rm tail}}{\delta R} \label{eq:EOM Hamiltonian_b}\,, \\
        & \dot{\Psi}= \frac{\partial \mathcal{H}_{\rm inst}}{\partial P_\Psi}  +\frac{\delta \mathcal{H}_{\rm tail}}{\delta P_\Psi} \label{eq:EOM Hamiltonian_c}\,,\\
         & \dot{P}_\Psi = -\frac{\delta \mathcal{H}_{\rm tail}}{\delta \Psi} \,, \label{eq:EOM Hamiltonian_d}
    \end{align}
\end{subequations}
where we have separated the instantaneous and tail contributions and used \textit{functional} derivatives for the latter, in contrast with ordinary partial derivatives for the former. Note that the dependence on the phase $\Psi$ appears only in the tail contribution. 

A given circular orbit is characterized by a radius $R=R_0$ and the condition $P_R=0$. The second condition allows us, \textit{via} Eq.~\eqref{eq:EOM Hamiltonian_a}, to express the \textit{non-conserved}\footnote{The angular momentum $P_\Psi$ is not conserved as a consequence of the non-local tail contribution appearing on the right hand side of~\eqref{eq:EOM Hamiltonian_d}. The conserved ``Noetherian'' angular momentum is discussed in Sec.~III B of~\cite{BBBFMb}.} angular momentum $P_\Psi \equiv P_\Psi^0$ as a function of $R_0$ and $\Psi_{0} \equiv \Omega_0\,t$ through
\begin{equation}
    \frac{\partial \mathcal{H}_{\rm inst}}{\partial R}[R_0,0,P_{\Psi}^{0}]+\frac{\delta \mathcal{H}_{\rm tail}}{\delta R}[R_0,0,\Psi_{0},P_{\Psi}^{0}]=0\,.
\end{equation}
Note that the condition 
\begin{equation}
    \frac{\partial \mathcal{H}_{\rm inst}}{\partial P_R}[R_0,0,P_{\Psi}^{0}]+\frac{\delta \mathcal{H}_{\rm tail}}{\delta P_R}[R_0,0,\Psi_{0},P_{\Psi}^{0}]=0\,,
\end{equation}
is trivially satisfied, because $\mathcal{H}$, as a conservative Hamiltonian, contains (even functionally) only even powers of $P_R$. Knowing $R_0$ and $P_\Psi^{0}(R_0)$, we can use~\eqref{eq:EOM Hamiltonian_c} to determine the orbital frequency as a function of $R_0$:
\begin{equation}
\label{eq:orbital frequency hamiltonian formalism}
   \Omega_0= \frac{\partial \mathcal{H}_{\rm inst}}{\partial P_\Psi}[R_0,0,P_{\Psi}^{0}]+\frac{\delta \mathcal{H}_{\rm tail}}{\delta P_\Psi}[R_0,0,\Psi_{0},P_{\Psi}^{0}]\,.
\end{equation}
As in Sec.~\ref{subsection:analysis EOM}, let us now consider a small perturbation of this circular orbit, which leads to a new orbit defined by $R_0+\delta R$, $\delta P_R$, $P_\Psi^0+\delta P_\Psi$ and $\Psi_0+ \delta \Psi$ (which, in turn, implies a shift of the orbital frequency to $\Omega_0+\delta \Omega$). Taking into account the fact that $\mathcal{H}$ contains only even powers of $P_R$ and the rotational invariance of circular orbits,\footnote{In principle, one could expect a $\delta \Psi$ contribution on the right side of Eqs \eqref{eq: delta R} and \eqref{eq: delta P_Psi}. To see why such terms are absent, notice that a mere redefinition of the origin of the angle coordinate, by means of a constant $\delta \Psi$, does not affect the value of $\dot{R}$ or $\dot{P}_\Psi$, because the motion remains unaltered.} the linearisation of Eqs.~\eqref{eq:EOM Hamiltonian} about the circular orbit readily gives
\begin{subequations}
\label{eq: differential system perturbation Hamiltonian formalism}
    \begin{align}
         &\label{eq: delta R}\delta \dot{R} =\sigma_0\, \delta P_R\,,  \\
          &\dot{\delta P_R} = -\pi_0\,\delta R-\rho_0\,\delta P_\Psi \,,  \\
         &\delta \dot{\Psi} = \zeta_0\, \delta R +\tau_0\, \delta P_\Psi \,,\\
          &\label{eq: delta P_Psi}\delta \dot{P}_\Psi = -\theta_0\, \delta P_R \,,
      \end{align}
\end{subequations}
where we have introduced the coefficients
\begin{subequations}
   \begin{align}
     &\sigma_0 = \frac{\partial^2 \mathcal{H}_{\rm inst}}{\partial P_R^2 }[R_0,0,P_{\Psi}^{0}]+\frac{\delta^2 \mathcal{H}_{\rm tail}}{\delta P_R^2}[R_0,0,\Psi_{0},P_{\Psi}^{0}] \,,\\
       & \pi_0 =\frac{\partial^2 \mathcal{H}_{\rm inst}}{\partial R^2 }[R_0,0,P_{\Psi}^{0}]+\frac{\delta^2 \mathcal{H}_{\rm tail}}{\delta R^2}[R_0,0,\Psi_{0},P_{\Psi}^{0}] \,, \\
    & \label{eq: def rho0} \rho_0 = \frac{\partial^2 \mathcal{H}_{\rm inst}}{\partial P_\Psi \partial R}[R_0,0,P_{\Psi}^{0}]+\frac{\delta^2 \mathcal{H}_{\rm tail}}{\delta P_{\Psi} \delta R }[R_0,0,\Psi_{0},P_{\Psi}^{0}]   \,,\\
     & \label{eq: def zeta0} \zeta_0 = \frac{\partial^2 \mathcal{H}_{\rm inst}}{\partial R \partial P_\Psi }[R_0,0,P_{\Psi}^{0}]+\frac{\delta^2 \mathcal{H}_{\rm tail}}{\delta R\, \delta P_{\Psi}}[R_0,0,\Psi_{0},P_{\Psi}^{0}]  \,,\\
   & \tau_0 = \frac{\partial^2 \mathcal{H}_{\rm inst}}{\partial P_\Psi^2 }[R_0,0,P_{\Psi}^{0}]+\frac{\delta^2 \mathcal{H}_{\rm tail}}{\delta P_\Psi^2}[R_0,0,\Psi_{0},P_{\Psi}^{0}] \,,\\
    &\theta_0 =\frac{\delta^2 \mathcal{H}_{\rm tail}}{\delta P_{R}\, \delta \Psi}[R_0,0,\Psi_{0},P_{\Psi}^{0}]\,.
   \end{align}
\end{subequations}
The eigenfrequencies of the system~\eqref{eq: differential system perturbation Hamiltonian formalism} are $\sigma=0$ and $\sigma=\pm\sqrt{-\pi_0\,\sigma_0+\rho_0\,\theta_0}$. Thus, with the Hamiltonian formalism circular orbits are stable if 
\begin{equation}
\label{eq: Hamiltonian criterion}
   \hat{C}_0^{\rm Ham} \equiv \pi_0\,\sigma_0 -\rho_0 \, \theta_0>0\,.
\end{equation}
We will find in the next section that the above Hamiltonian criterion coincides with the one derived from the equations of motion. To show this is not immediate, as the intermediate calculations in both approaches turn out to be quite different, 
and their eventual agreement inspires some confidence in our final result.

\section{Perturbation of the 4PN tail term}
\label{sec tail part}

\subsection{Variation of the tail integrals}
\label{subsection:generic variations of the tail integrals}

In both  approaches discussed in the previous section, the 4PN dynamics involves non-local tail terms which depend on integrals of the form
\begin{equation}
\label{Tijalpha}
\mathcal{T}_{ij}^{(\alpha)}(t) \equiv \mathop{\mathrm{Pf}}\limits_{2r/c}\int_{-\infty}^{+\infty} \frac{ Q_{ij}^{(\alpha)}(t')}{\lvert t-t' \rvert}\, \mathrm{d}t'
\equiv 
\int_{0}^{\infty} \mathrm{d}\tau \, \mathrm{ln}\left( \frac{c\tau}{2r}\right) \left[ Q_{ij}^{(\alpha+1)}(t-\tau)-Q_{ij}^{(\alpha+1)}(t+\tau) \right]\,,
\end{equation}
where $Q_{ij}^{(\alpha)}$ denotes the $\alpha^{\rm th}$ time derivative ($\alpha \geqslant 1$) of the Newtonian quadrupole moment of the binary system, $Q_{ij} = m \nu r^2 (n^i n^j - \frac{1}{3}\, \delta^{ij})$ in the CM frame, and where Pf is the Hadamard Partie finie. The difficulty in computing the above quantity, or its perturbation, comes from the fact that the integral depends \textit{on the entire history} of the binary system,\footnote{For comparison with the GSF calculation, our set-up must discard dissipative interactions. Thus we give up on causality and extend the domain of integration of the tail integrals to the \emph{entire} binary source's history (both past and future).} and it is indeed \textit{prima facie} not clear how the perturbation of the circular orbit will affect this integral. The perturbation studied here should be seen as an overall deformation of the full wordline of the binary. 

The perturbed Newtonian system is described by an elliptic motion and it is convenient, following previous works~\cite{ABIQ08tail,BBBFMb}, to use  a discrete Fourier decomposition of the Newtonian quadrupole moment:
\begin{equation}
\label{Fourier transform definition}
	Q_{ij}= \sum_{p \,  \in \, \mathbb{Z}}  \,\mathop{Q}_{p}{}_{\!\!ij}  \, \de^{\di p \ell} \quad \text{with} \quad \mathop{Q}_{p}{}_{\!\!ij} =\int_{0}^{2\pi} \frac{\dd \ell}{2\pi}\,Q_{ij}\,\de^{-\di p\ell }\,,
\end{equation}
where $\ell \equiv n (t-T)$ is the mean anomaly of the binary motion, involving the orbital frequency or mean motion $n=2\pi/P$ (where $P$ is  the orbital period --- there is indeed no orbital precession at Newtonian order) and $T$ is the instant of passage to the periastron. The mean motion is given by Kepler's third law as $n=\sqrt{Gm/a^3}$. The explicit expressions of the Fourier coefficients ${}_{p}{Q}_{ij}$ (which depend only on the semi-major axis $a$ and the eccentricity $e$) in terms of combinations of Bessel functions are given in Appendix~\ref{Appendix : quadrupole Fourier coefficients}. 

For a circular orbit, of radius $r_0$, the above decomposition is restricted to the modes $p=0$ and $p=\pm 2$. Substituting it into \eqref{Tijalpha} and using the identity~\cite{Gradshteyn} 
\begin{align}
\label{eq: integral Fourier log and exp}
    & \int_{0}^{\infty} \dd\tau \, \mathrm{ln}\left( \frac{c\tau}{2r}\right) \de^{-\di p n \tau} =  \frac{\di}{p n} \,\biggl[\ln\left(\frac{2|p|n r}{c}\right) + \gamma_{\rm E} + \frac{\di \pi}{2} \mathrm{sign}(p)\biggr]\,,
\end{align}
where $\gamma_\text{E}$ is the Euler constant and $\mathrm{sign}(p)$ is the sign function, we find
\begin{align}\label{Tij0}
\mathcal{T}_{ij,0}^{(\alpha)} &= -2 Q_{ij,0}^{(\alpha)}\,\Lambda_0\,,
\end{align}
where $Q_{ij,0}^{(\alpha)}$ is the $\alpha^{\rm th}$ time derivative (again, with $\alpha \geqslant 1$) of the quadrupole for a circular orbit, and where we have defined for later convenience
\begin{align}\label{Lambda0}
	\Lambda_0 \equiv \ln\left(\frac{4n_0 r_0}{c}\right) + \gamma_{\rm E} \equiv \frac{1}{2}\ln\left(\frac{16 G m}{r_0 c^2}\right) + \gamma_{\rm E} \,.
\end{align}

Let us turn now to the computation of the linear perturbation  of $\mathcal{T}_{ij}^{(\alpha)}$, assuming that we go from our reference circular trajectory, of radius $r_0$, to a new trajectory, elliptical in general, characterized by the orbital parameters $a=a_0+\delta a$ (with $a_0=r_0$), $e=\delta e$, and $T=T_0+\delta T$. Since
\begin{equation}
	\mathcal{T}_{ij}^{(\alpha)} = -2 Q_{ij}^{(\alpha)}\, \ln\left(\frac{r}{s}\right)  + \mathop{\mathrm{Pf}}\limits_{2s/c}\int_{-\infty}^{+\infty} \frac{ Q_{ij}^{(\alpha)}(t')}{\lvert t-t' \rvert}\, \dd t'\,,
\end{equation}
with $s$ an arbitrary constant length, the variation of this expression  yields
\begin{align}\label{eq: decomposition of delta Tij alpha}
     \delta  \mathcal{T}_{ij}^{(\alpha)} & = -2 \delta Q_{ij}^{(\alpha)}\, \ln\left(\frac{r_0}{s}\right)- 2 Q_{ij,0}^{(\alpha)}\frac{\delta r}{r_0} + \mathop{\mathrm{Pf}}\limits_{2s/c}\int_{-\infty}^{+\infty} \frac{ \delta Q_{ij}^{(\alpha)}(t')}{\lvert t-t' \rvert}\, \dd t'\nn \\
     & = - 2 Q_{ij,0}^{(\alpha)}\, \frac{\delta r}{r_0} + \mathop{\mathrm{Pf}}\limits_{2r/c}\int_{-\infty}^{+\infty} \frac{ \delta Q_{ij}^{(\alpha)}(t')}{\lvert t-t' \rvert}\, \dd t'\,.
\end{align}
Thus, the first term comes from the variation of the Hadamard scale $r=r(t)$. The variation of $Q_{ij}^{(\alpha)}$ itself, using the Fourier decomposition~\eqref{Fourier transform definition} and taking into account the fact that all the Fourier modes depend quadratically on $a$ (see Appendix~\ref{Appendix : quadrupole Fourier coefficients}), so  that
\begin{equation}\label{eq:variation delta Qij p}
  \delta \!  \mathop{Q}_{p}{}_{\!\!ij} = 2 \mathop{Q}_{p}{}_{\!\!ij,0}\,\frac{\delta a}{a_0}+ \mathop{Q}_{p}{}^\prime_{\!\!ij,0}\,\delta e,\quad~\text{with}~\quad \mathop{Q}_{p}{}^\prime_{\!\!ij,0}\equiv \left(\frac{\partial}{\partial e}\mathop{Q}_{p}{}_{\!\!ij}\right)\Bigr|_{a=r_0,e=0}\,, 
\end{equation}
leads to the expression
\begin{align}\label{dQalpha}
    \delta Q_{ij}^{(\alpha)} &= \sum_{p \,  \in \, \mathbb{Z}}  \mathop{Q}_{p}{}_{\!ij,0}  \bigl(\di p n_0\bigr)^{\alpha}   \left(2\, \frac{\delta a}{a_0}+  \left( \alpha + \di p \ell_0\right)\frac{\delta n}{n_0}- \di p n_0 \delta T  \right)
  \de^{\di p \ell_0}\nn\\
  & +\sum_{p \,  \in \, \mathbb{Z}} \mathop{Q}_{p}{}^\prime_{\!\!ij,0} \bigl(\di p n_0\bigr)^{\alpha} \,\delta e \,\de^{\di p \ell_0}\,,
\end{align} 
where we posed $\ell_0=n_0(t-T_0)$. We can now substitute this expression into~\eqref{eq: decomposition of delta Tij alpha} and use again  the identity~\eqref{eq: integral Fourier log and exp}, at least for most of the terms whose time dependence is confined to the complex exponential. There is however an extra linear time dependence in the term proportional to $\ell_0$, which requires the use of the supplementary identity
\begin{align}
	\label{eq: integral Fourier log and exp2}
	& \int_{0}^{\infty} \dd\tau \,\tau\, \mathrm{ln}\left( \frac{c\tau}{2r}\right) \de^{-\di p n \tau} =  \frac{1}{(p n)^2} \,\biggl[\ln\left(\frac{2|p|n r}{c}\right) + \gamma_{\rm E} + \frac{\di \pi}{2} \mathrm{sign}(p)-1\biggr]\,,
\end{align}
immediately obtained by differentiating~\eqref{eq: integral Fourier log and exp} with respect to $n$. After integration, we obtain
\begin{align}\label{eq: integral delta Qij alpha intermediary result}
&\mathop{\mathrm{Pf}}\limits_{2r/c}\int_{-\infty}^{+\infty} \frac{ \delta Q_{ij}^{(\alpha)}(t')}{\lvert t-t' \rvert}\, \mathrm{d}t' = -2 Q_{ij,0}^{(\alpha)}\frac{\delta n}{n_0} \nn\\ &\qquad\qquad -2  \sum_{p \,  \in \, \mathbb{Z}}   \bigl(\di p n_0\bigr)^{\alpha} \left\{ \mathop{Q}_{p}{}_{\!ij,0} \left(2\, \frac{\delta a}{a_0}+  \left( \alpha + \di p \ell_0\right)\frac{\delta n}{n_0}- \di p n_0 \delta T  \right)+ \mathop{Q}_{p}{}^\prime_{\!\!ij,0}\,\delta e\right\}
\biggl[\ln\left(\frac{2|p| n_0 r_0}{c}\right) + \gamma_{\rm E}\biggr]\de^{\di p \ell_0}\,,
\end{align}
which can be conveniently reorganised using~\eqref{dQalpha}, and remembering that only $p=\pm 2$ contribute for a background circular orbit, and also that (as readily checked from Appendix~\ref{Appendix : quadrupole Fourier coefficients}) only the $p=\pm 1$ and $p=\pm 3$ contribute to the derivative of the quadrupole modes with respect to the eccentricity $e$, as defined in~\eqref{eq:variation delta Qij p}. This gives
\begin{align}
 \label{eq: integral delta Qij alpha final step}
\mathop{\mathrm{Pf}}\limits_{2r/c}\int_{-\infty}^{+\infty} \frac{ \delta Q_{ij}^{(\alpha)}(t')}{\lvert t-t' \rvert}\, \mathrm{d}t' =&-2 Q_{ij,0}^{(\alpha)}\, \frac{\delta n}{n_0} -2 \Lambda_0\,  \delta Q_{ij}^{(\alpha)} + \chi_{ij,0}^{(\alpha)} \,\delta e\,,
\end{align}
where we have used the notation~\eqref{Lambda0} and posed 
\begin{align}\label{def delta chi alpha}
\chi_{ij,0}^{(\alpha)} \equiv \Re \biggl\{4\ln{(2)}\,\bigl(\di  n_0\bigr)^{\alpha}\de^{\di \ell_0} \mathop{Q}_{1}{}^\prime_{\!\!ij,0} -4\ln{\left(\frac{3}{2}\right)} \bigl(3\di  n_0\bigr)^{\alpha}\de^{3 \di \ell_0} \mathop{Q}_{3}{}^\prime_{\!\!ij,0}\biggr\}\,.
\end{align}

In practice, as will be seen in the next two subsections, we will consider several scalar quantities obtained by contracting~\eqref{Tijalpha} with time derivatives of the quadrupole moment, \textit{i.e.} of the form
\begin{equation}
	\label{eq:contraction I alpha beta def}
	\mathcal{I}^{(\alpha,\beta)} \equiv  Q_{ij}^{(\alpha)}\,\mathcal{T}_{ij}^{(\beta)},
\end{equation}
where $\alpha$ and $\beta\geqslant 1$ are some positive integers. Using the results~\eqref{Tij0},~\eqref{eq: decomposition of delta Tij alpha} and~\eqref{eq: integral delta Qij alpha final step}, we find that the linear variation of~\eqref{eq:contraction I alpha beta def} is given by
\begin{align}
\label{delta I beta alpha final expression}
   \delta  \mathcal{I}^{(\alpha,\beta)}= & -2 \Lambda_0\, \delta \!\left(Q_{ij}^{(\alpha)}\, Q_{ij}^{(\beta)} \right) -2\,  Q_{ij,0}^{(\alpha)}\, Q_{ij,0}^{(\beta)}  \left(\frac{\delta n}{n_0}+\frac{\delta r}{r_0}\right)+Q_{ij,0}^{(\alpha)}\,  \chi_{ij,0}^{(\beta)} \,\delta e\,.
\end{align}
The final step consists in rewriting these linear perturbations in terms of the variation of the coordinates, \textit{i.e.} $\delta r$, $\delta u$ and $\delta \Omega$, instead of the variation of the orbital elements $\delta n$ and $\delta e$, as required by the form~\eqref{eq: differential system perturbation Lagrangian formalism} of the perturbed equations of motion. This can easily be done by resorting to standard expressions for the elliptic motion. For example, the expression of the conserved angular momentum (per unit mass) in terms of $a$ and $e$ reads
\begin{align}
    r^2\Omega= \sqrt{Gm\, a(1-e^2)}\,,
\end{align}
from which one obtains, at linear order, combining also with $n=\sqrt{Gm/a^3}$,
\begin{subequations}\label{eq: expression delta a in terms of delta r and delta Omega}
	\begin{align}
    \frac{\delta a}{a_0} &= 4\,\frac{\delta r}{r_0}+2\,\frac{\delta \Omega}{\Omega_0}\,,\\
    \frac{\delta n}{n_0} &= -6\,\frac{\delta r}{r_0}-3\,\frac{\delta \Omega}{\Omega_0}\,.
\end{align}
\end{subequations}
For the terms proportional to $\delta e$, one can check that the only combinations that can appear in~\eqref{delta I beta alpha final expression} are $\cos\ell_0\,\delta e$ (when $\alpha + \beta$ is even) and $\sin\ell_0\,\delta e$ (when $\alpha + \beta$ is odd). Linearising the parametric representation of the elliptical motion,
\begin{equation}
     r=\frac{a(1-e^2)}{1+e\cos\psi}\,,\qquad\quad u\equiv\dot r=\frac{a(1-e^2)}{(1+e\cos\psi)^2}\,e\, \Omega \sin\psi\,,
\end{equation}
gives, at linear order and in the limit $e=0$ for the reference motion,
\begin{subequations}\label{eq: cos and sin delta e}
\begin{align}
     \cos\ell_0\,\delta e &= \frac{\delta a}{a}-\frac{\delta r}{r}=3\,\frac{\delta r}{r_0}+ 2\,\frac{\delta \Omega}{\Omega_0}\,,\\
     \sin\ell_0\,\delta e &= \frac{\delta u}{\Omega_0 r_0}\,.
\end{align}
\end{subequations}
Using~\eqref{eq: expression delta a in terms of delta r and delta Omega} and~\eqref{eq: cos and sin delta e}, one can thus express~\eqref{delta I beta alpha final expression} in terms of the variations $\delta r$, $\delta u$ and $\delta \Omega$. In the Hamiltonian formalism, these variations correspond respectively to $\delta R$, $\delta P_R$ and $\delta (P_{\Psi}/R^2)$ so it is straightforward to rewrite~\eqref{delta I beta alpha final expression} in terms of the phase space variables appropriate for the Hamiltonian.

\subsection{Computations based on the equations of motion}
\label{subsection:tail perturbation equations of motion}

Up to 3PN order, the dynamics is purely local, so that any perturbation of the equations of motion is straightforwardly carried out. It is easy to extend this calculation to the instantaneous contributions at 4PN order and  we do not need to provide detail here.  However, at 4PN order, one must also take into account the non-local (in time) tail term, derived from  the Lagrangian~\eqref{L_tail}, in the equations of motion~\eqref{eq: EOM Lagrangian first form}. 
It is convenient to decompose this term into a non-local contribution \textit{stricto-sensu} and a local contribution, 
\begin{align}\label{eq: split tail}
	a^{i}_{\text{tail}} = a^{i}_{\text{tail}}\big|_{\text{non-local}} + a^{i}_{\text{tail}}\big|_{\text{local}}\,.
\end{align}
The non-local part reads (see Sec.~IV in~\cite{BBFM17})
\begin{align}
 \label{tail_non-local}
a^{i}_{\text{tail}}\big|_{\text{non-local}} = -\frac{4G^2 m}{5c^8}
	\,x^j \,\mathcal{T}_{ij}^{(6)}  = - \frac{4G^2 m}{5c^{8}}\, x^{j}\int_{0}^{\infty} \mathrm{d}\tau \, \mathrm{ln}\left( \frac{c\tau}{2r}\right)\left[ Q_{ij}^{(7)}(t-\tau)-Q_{ij}^{(7)}(t+\tau) \right]\,.
\end{align}
The local piece of the tail term, which comes from the variation of the Hadamard regularization scale $\tau_0=2r/c$ in Eq.~\eqref{L_tail} (\textit{cf.} Footnote~\ref{footnote: partie finie def}), is given by~\cite{BBFM17}  
\begin{align}\label{tail_local}
a^{i}_{\text{tail}}\big|_{\text{local}} = \frac{8G^2 m}{5c^8}
	\,x^j\left[\left(Q_{ij}^{(3)}\ln r\right)^{(3)}-Q_{ij}^{(6)}\ln
	r\right] -\frac{2G^2}{5c^8\nu}
	\frac{n^i}{r}\,Q_{jk}^{(3)}\,Q_{jk}^{(3)}\,.
\end{align}
Note that  only the conservative part of the equations of motion is considered here: we neglect the dissipative, radiation reaction part of the equations of motion which involves also a tail term at 4PN order, given by (see Sec.~VI in~\cite{BBFM17})
\begin{align}\label{eq: dissipative part of tail acceleration}
	a^{i}_{\text{tail}}\big|_{\text{diss}} = - \frac{4G^2 m}{5c^{8}}\, x^{j}\int_{0}^{\infty} \mathrm{d}\tau \, \mathrm{ln}\left( \frac{c\tau}{2r}\right)\left[ Q_{ij}^{(7)}(t-\tau) + Q_{ij}^{(7)}(t+\tau) \right]\,.
\end{align}

We study the modification of the tail term under a linear perturbation of the circular orbit. While the change of the local part~\eqref{tail_local} is quite easy to deduce, that of the non-local piece \eqref{tail_non-local} is less trivial. 
We need to control the non-local parts of the coefficients $\alpha_0$, $\beta_0$ and $\gamma_0$ in the differential system~\eqref{eq: differential system perturbation Lagrangian formalism}, that is the radial and tangential components of Eq.~\eqref{tail_non-local} which read in terms of the notation~\eqref{eq:contraction I alpha beta def}
\begin{subequations}\label{analambda} 
	\begin{align}
		a^{n}_{\text{tail}}\big|_{\text{non-local}} &= - \frac{4G^2}{5c^{8}}\,\frac{\mathcal{I}^{(0,6)}}{\nu r} \,,\\ a^{\lambda}_{\text{tail}}\big|_{\text{non-local}} &= \frac{4G^2}{5c^{8}}\, \frac{1}{\nu \Omega r} \left( \frac{u}{ r}\,\mathcal{I}^{(0,6)} -\frac{1}{2}\,\mathcal{I}^{(1,6)} \right)\,.
	\end{align}
\end{subequations}
On a circular orbit, we find --- using~\eqref{Fourier transform definition}, \eqref{Tij0} and the circular modes~\eqref{eq: Fourier coefficients circular case} --- that
\begin{align}\label{eq: valeur I0}
	\mathcal{I}_0^{(0,6)} &= 64\, m^2 \nu^2 r_0^4\,\Omega_0^{6} \,\Lambda_0\,,
	\qquad\quad \mathcal{I}_0^{(1,6)} = 0\,,
\end{align}
where we recall that $\Lambda_0$ has been defined in~\eqref{Lambda0} and we have reinstalled $\Omega_0=n_0=\sqrt{Gm/r_0^3}$. The vanishing of this last integral only holds because we neglected the dissipative tail effect~\eqref{eq: dissipative part of tail acceleration}, an assumption we imposed in order to define the ISCO  as 
in numerical calculations that will be discussed in Section \ref{section:conclusion}.

To determine how the equations of motion~\eqref{eq: coupled system EOM Lagrangian} are affected by a small perturbation of the circular orbit, we must vary the integrals in~\eqref{analambda}. This is easily carried out by applying the general result \eqref{delta I beta alpha final expression}, which yields 
\begin{subequations}
	\label{eq:delta IJ final expression}
\begin{align}
 \delta \mathcal{I}^{(0,6)} &=  \,m^2 \nu^2 r_0^4 \,\Omega_0^6 \bigg[ \Bigl(-320 +512 \Lambda_0-2176\,\ln(2)+2187\,\ln(3) \Bigr) \frac{\delta r}{r_0} \nn \\
     &\qquad\qquad\qquad +\left(-192 +\frac{1664}{3} \Lambda_0-\frac{4352}{3}\,\ln(2)+1458\,\ln(3) \right) \frac{ \delta \Omega}{\Omega_0}\bigg] \,,\\
 \delta \mathcal{I}^{(1,6)} &=  m^2 \nu^2 r_0^4 \,\Omega_0^7 \bigg(  1080\,\Lambda_0  -1464\,\ln(2)+1458\,\ln(3) \bigg) \frac{\delta u}{\Omega_0 r_0} \,.
\end{align}
\end{subequations}
A good sanity check of Eqs.~\eqref{eq:delta IJ final expression} is to restrict them to the particular case of a transition from a circular orbit to \textit{another} circular orbit. Indeed, in this case Eqs.~\eqref{eq:delta IJ final expression} should match the expression obtained directly by varying the integrals for circular orbits given by~\eqref{eq: valeur I0}. For a transition between two circular orbits, Kepler's third law states that $\delta \Omega/\Omega_0 = - 3\delta r/(2 r_0)$ and we have $\delta u=0$, then Eqs.~\eqref{eq:delta IJ final expression} reduce to
\begin{equation}
	\delta \mathcal{I}^{(0,6)} = - 64 \,m^2 \nu^2 r_0^4 \, \Omega_0^6 \left(\frac{1}{2}+5 \Lambda_0 \right) \frac{\delta r}{r_0}\,,\qquad\quad \delta \mathcal{I}^{(1,6)} = 0\,, \quad \quad \text{(for circular perturbation)},
\end{equation}
which agrees with the direct variation of Eqs.~\eqref{eq: valeur I0} with respect to $r_0$. Finally, the results~\eqref{eq:delta IJ final expression} enable us to determine the stability coefficients $\alpha_0$, $\beta_0$ and $\gamma_0$ defined in \eqref{eq: stability coefficients}, and from there to obtain the stability criterion \eqref{eq: Lagrangian criterion}.

\subsection{Computation based on the Hamiltonian formalism}
\label{subsec: computation tail hamiltonian}

In the Hamiltonian formalism, up to 4PN order, the tail part of the Hamiltonian is simply given by Eq.~\eqref{eq:Htail}. Therefore, the tail contributions to the Hamiltonian equations are found to be (see \textit{e.g.}~\cite{BBBFMb, BL17})
\begin{subequations}
\label{functional derivatives of Htail}
    \begin{align}
      &\frac{\delta \mathcal{H}_{\rm tail}}{\delta R} =  -\frac{2 G^2}{5 c^8 \nu}\left(\frac{\partial Q^{(3)}_{ij}}{\partial R}\,\mathcal{T}^{(3)}_{ij}-\frac{1}{R}\, Q_{ij}^{(3)}\, Q_{ij}^{(3)}  \right)\,,\\
&  \frac{\delta \mathcal{H}_{\rm tail}}{\delta (P_R, \Psi,P_\Psi)} =  -\frac{2 G^2}{5 c^8 \nu}\,\frac{\partial Q^{(3)}_{ij}}{\partial (P_R, \Psi,P_\Psi)}\,\mathcal{T}^{(3)}_{ij}\,. 
    \end{align}
\end{subequations}
Plugging in~\eqref{functional derivatives of Htail} the Newtonian expression of $Q^{(3)}_{ij}$ (see, for instance, Eq.~(3.4a) of~\cite{BBBFMb}), we find
\begin{subequations}
\label{Hamiltonian equations explicit}
    \begin{align}
         & \dot{R} = \frac{\partial \mathcal{H}_{\rm inst}}{\partial P_R}  +\frac{4 G^3 m^2}{5 c^8 }\,\frac{1}{R^2}\,\mathcal{T}^{(3)}_{nn}\,,\\
         & \dot{P}_R = -\frac{\partial \mathcal{H}_{\rm inst}}{\partial R}  + \frac{8 G^3 m^2}{5 c^8}\,\frac{1}{R^3} \left[P_R\,\mathcal{T}^{(3)}_{nn}+\frac{6 P_{\Psi}}{R} \,\mathcal{T}^{(3)}_{n\lambda}-\frac{G m^2 \nu}{R^2}\left( \frac{2}{3}\,P_R^2+\frac{8 P_{\Psi}^2}{R^2}\right)\right]\,,\\
        & \dot{\Psi}= \frac{\partial \mathcal{H}_{\rm inst}}{\partial P_\Psi}   +\frac{16 G^3 m^2}{5 c^8 }\,\frac{1}{R^3}\,\mathcal{T}^{(3)}_{n\lambda }\,,\\ 
        \label{P Psi dot equation}
        & \dot{P}_\Psi =  -\frac{8 G^3 m^2}{5 c^8 } \,\frac{1}{R^2}\left[P_R\,\mathcal{T}^{(3)}_{n\lambda}+\frac{2 P_{\Psi}}{R} \left(\mathcal{T}^{(3)}_{\lambda \lambda}-\mathcal{T}^{(3)}_{nn}  \right) \right]\,,
    \end{align}
\end{subequations}
where $\mathcal{T}^{(3)}_{nn} \equiv n^{i}\,n^{j}\,  \mathcal{T}^{(3)}_{ij}$, $\mathcal{T}^{(3)}_{n\lambda} \equiv n^{i}\,\lambda^{j}\,  \mathcal{T}^{(3)}_{ij}$ and $\mathcal{T}^{(3)}_{\lambda\lambda} \equiv \lambda^{i}\,\lambda^{j}\,  \mathcal{T}^{(3)}_{ij}$. At Newtonian order, $v^{i} = P_R\, n^i + P_{\Psi} \lambda^i/R$ so that
\begin{subequations}
    \begin{align}
            &\mathcal{T}^{(3)}_{nn} = \frac{1}{m \nu}\,\frac{\mathcal{I}^{(0,3)}}{R^2}\,,\\  
            &\mathcal{T}^{(3)}_{n\lambda} =\frac{1}{m \nu}\,\frac{1}{P_{\Psi}}\left(\frac{1}{2}\,\mathcal{I}^{(1,3)}-\frac{P_R}{R}\,\mathcal{I}^{(0,3)}\right)\,,\\
&\mathcal{T}^{(3)}_{\lambda\lambda} =\frac{1}{m \nu}\,\frac{R^2}{ P_{\Psi}^2}\left(\frac{1}{2}\,\mathcal{I}^{(2,3)}-\frac{P_R}{R}\,\mathcal{I}^{(1,3)}+\left(\frac{P_R^2}{R^2}+\frac{G m}{R^3} \right)\mathcal{I}^{(0,3)}\right)\,,
    \end{align}
\end{subequations}
where we still employ the notation~\eqref{eq:contraction I alpha beta def}. 
The tail contributions in Eqs.~\eqref{Hamiltonian equations explicit} then amount to 
\begin{subequations}\label{Expressions dots}
	\begin{align}
		& \dot{R}\rvert_{\rm tail}= \frac{4 G^3 m}{5 c^8 \nu} \frac{\mathcal{I}^{(0,3)}}{R^4}\,,\\
		&\dot{P_R}\rvert_{\rm tail}= \frac{8 G^3 m}{5 c^8 \nu} \,\left( 3\,\frac{\mathcal{I}^{(1,3)}}{R^4} -\frac{G m^3 \nu^2}{R^5}\left( \frac{2}{3}\,P_R^2+\frac{8 P_{\Psi}^2}{R^2}\right)\right)\,,\\
		&\dot{\Psi}\rvert_{\rm tail} = \frac{8 G^3 m}{5 c^8 \nu}\,\frac{\mathcal{I}^{(1,3)}}{R^3 P_{\Psi}}\,,\\
		&\dot{P_\Psi}\rvert_{\rm tail}= -\frac{8 G^3 m}{5 c^8 \nu } \,\left( -\frac{3}{2}\frac{P_R}{R^2 P_\Psi } \mathcal{I}^{(1,3)} +\frac{\mathcal{I}^{(2,3)}}{R P_\Psi}\right)\,.
	\end{align}
\end{subequations}

Using~\eqref{Fourier transform definition}, \eqref{Tij0} and Eqs.~\eqref{eq: Fourier coefficients circular case}, one finds that, on a circular orbit, 
\begin{equation}
	\label{Expression L on circular orbit}
	\mathcal{I}_0^{(0,3)} = 0\,,\qquad\quad   \mathcal{I}_0^{(1,3)} = 16\, m^2 \nu^2 R_0^4\,\Omega_0^{4} \, \Lambda_0\,,\qquad\quad \mathcal{I}_0^{(2,3)} = 0\,.
\end{equation}
To specify the perturbed dynamics~\eqref{eq: differential system perturbation Hamiltonian formalism}, we need to vary the integrals in~\eqref{Expressions dots}, which is again done by applying the general result of Section~\ref{subsection:generic variations of the tail integrals}. The only point to remember is to use the identity $P_{\Psi}=\Omega R^2$ (valid at Newtonian order) to express $\delta \Omega$ as a combination of $\delta R$ and $\delta P_{\Psi}$. With that remark in mind, Eq.~\eqref{delta I beta alpha final expression} gives
\begin{subequations}\label{delta KLM final expression}
\begin{align}
	\delta \mathcal{I}^{(0,3)} &=  m^2 \nu^2 R_0^4 \,\Omega_0^3 \bigg(  \frac{8}{3}\,\Lambda_0 +\frac{88}{3}\,\ln(2)-27\,\ln(3) \bigg) \frac{\delta P_R}{\Omega_0 R_0} \,,\\
	\delta \mathcal{I}^{(1,3)} &=  m^2 \nu^2 R_0^4 \,\Omega_0^4 \bigg[ \Bigl(16 -48 \Lambda_0+48\,\ln(2)-54\,\ln(3) \Bigr) \frac{\delta R}{R_0} \nn \\
	&\qquad\qquad\qquad +\Bigl(-48 +32 \Lambda_0-96\,\ln(2)+108\,\ln(3) \Bigr) \frac{\delta P_{\Psi}}{\Omega_0 R_0^2}\bigg] \,,\\
	\delta \mathcal{I}^{(2,3)} &=  m^2 \nu^2 R_0^4 \,\Omega_0^5 \bigg(  24\Lambda_0  -120\,\ln(2)+108\,\ln(3) \bigg) \frac{\delta P_R}{\Omega_0 R_0} \,.
\end{align}
\end{subequations}
The specialization of~\eqref{delta KLM final expression} to the particular case of a transition between two circular orbits yields 
\begin{equation}
	\delta \mathcal{I}^{(0,3)} = 0\,, \quad \delta \mathcal{I}^{(1,3)} = - 16 \,m^2 \nu^2 R_0^4 \, \Omega_0^4 \left(\frac{1}{2}+2 \Lambda_0 \right) 
	\frac{\delta R}{R_0}\,,\quad \delta \mathcal{I}^{(2,3)} = 0\,,  \quad \text{(for circular perturbation)},
\end{equation}
which agrees with the direct variation of Eqs.~\eqref{Expression L on circular orbit} with respect to $R_0$ (same consistency check as before). We have thus derived all the necessary ingredients to compute the stability coefficients introduced in \eqref{eq: differential system perturbation Hamiltonian formalism} and the corresponding Hamiltonian criterion \eqref{eq: Hamiltonian criterion}.

As a side remark, it is also interesting to note that the coefficients $\rho_0$ and $\zeta_0$, defined by Eqs. \eqref{eq: def rho0} and \eqref{eq: def zeta0}, differ because of the \emph{non-commutativity} of the mixed functional derivatives of $\mathcal{H}_{\rm tail}$ with respect to $R$ and $P_\Psi$. More precisely, one can show that the ``commutator'' 
\begin{align}
 \frac{\delta^2 \mathcal{H}_{\rm tail}}{\delta P_{\Psi} \delta R } - \frac{\delta^2 \mathcal{H}_{\rm tail}}{\delta R \delta P_{\Psi}}=-\frac{2 G^2}{5 c^8 \nu} \left( \frac{\partial Q^{(3)}_{ij}}{\partial R}\,\frac{\delta \mathcal{T}^{(3)}_{ij} }{\delta P_\Psi}-\frac{\partial Q^{(3)}_{ij}}{\partial P_\Psi}\,\frac{\delta \mathcal{T}^{(3)}_{ij} }{\delta R}-\frac{2}{R}\,Q^{(3)}_{ij}\,\frac{\partial Q^{(3)}_{ij}}{\partial P_\Psi}\right)\,,
\end{align}
reduces, on a circular orbit, to 
\begin{align}
     \rho_0 - \zeta_0 &=-\frac{2 G^2}{5 c^8 \nu} \left( -\frac{3}{R_0}\,\frac{\delta \mathcal{I}^{(3,3)} }{\delta P_\Psi}-\frac{1}{\sqrt{G m R_0}}\frac{\delta \mathcal{I}^{(3,3)} }{\delta R}-\frac{2}{R_0 \sqrt{G m R_0}}\,Q^{(3)}_{ij,0}\,Q^{(3)}_{ij,0}\right)\nn\\
     &=\left(-\frac{1152}{5}+384\,\mathrm{ln}(2)-432\,\mathrm{ln}(3)\right)\,\frac{ G^5 m^5 \nu}{ R_0^5 \sqrt{G m R_0}}\,.
\end{align}

\section{Results}
\label{section:results}

Following Sec.~\ref{subsection:analysis EOM}, one can now compute explicitly the stability criterion~\eqref{eq: Lagrangian criterion} based on the 4PN equations of motion~\eqref{eq: EOM Lagrangian first form}, by inserting the  expressions of the coefficients $\mathcal{A}$ and $\mathcal{B}$ in Appendix~\ref{Appendix: explicit expressions of A and B}, and the tail contributions derived in Sec.~\ref{subsection:tail perturbation equations of motion}. Setting
\begin{equation}\label{gamma_circ}
\gamma \equiv \frac{G m}{r_0 c^2}\,,
\end{equation}
where $r_0$ is the separation between the two bodies (radius of the circular orbit) expressed in harmonic coordinates, we find that the stability criterion~\eqref{eq: Lagrangian criterion} is given by
\begin{align}\label{eq: result criterion lagrangian full}
       \hat{C}_0^{\rm EoM} &= \frac{c^6 \gamma^3}{G^2 m^2}   \Bigg[ 1+\gamma\bigl(-9+\nu\bigr)+ \gamma^2 \left( 30+\frac{65}{4}\,\nu+\nu^2\right)\nn\\
        &\quad\qquad \quad +\gamma^3 \left(-70+  \left[ -\frac{29927}{840}-\frac{451\pi^2}{64}+22\,\mathrm{ln}\left( \frac{r}{r_0'}\right)\right] \nu+\frac{19}{2}\,\nu^2+\nu^3\right)\nn\\
         &\quad\qquad\quad + \gamma^4 \bigg(135+  \bigg[ -\frac{80622469}{100800}+\frac{242971\pi^2}{3072}-466\,\mathrm{ln}\left( \frac{r}{r_0'}\right)+\frac{5792}{15}\,\gamma_{\rm E}+\frac{2896}{15}\,\mathrm{ln}(16\gamma) \nn\\
         & \quad \qquad\qquad\qquad -\frac{8864}{15}\,\mathrm{ln}(2)+\frac{2916}{5}\,\mathrm{ln}(3)\bigg]\nu 
         +\left(\frac{403}{112}-\frac{1087\pi^2}{64}+\frac{1168}{3}\,\mathrm{ln}\left( \frac{r}{r_0'}\right)\right) \nu^2  +\frac{51}{4} \nu^3+ \nu^4
         \bigg)\Bigg]\,,
\end{align}
where $r_0'$ is an arbitrary UV-type length scale introduced in the regularization of the 3PN and 4PN terms in the equations of motion in harmonic coordinates, and which drops out in any gauge-independent expression~\cite{BBFM17}. The first two lines coincide with the result obtained at 3PN order in~\cite{BI03CM}.\footnote{The value of the undetermined coefficient in the result of~\cite{BI03CM} was later obtained as $\lambda=-\frac{1987}{3080}$~\cite{BDE04}.} The last two lines are the novel contributions due to the 4PN terms.

It is very useful to rewrite the stability criterion~\eqref{eq: result criterion lagrangian full} in a gauge-invariant form by expressing it in terms of the  circular orbital frequency $\Omega_0$, or, equivalently, of the familiar PN parameter $x$ defined by Eq.~\eqref{eq: def_x}. The harmonic-coordinate-dependent parameter $\gamma$ is related to the invariant parameter $x$ at 4PN order by (see \textit{e.g.}~\cite{BBFM17}) 
\begin{align}\label{eq: relation gamma x}
	\gamma &= x \Biggl[1  + \left(1 -  \frac{\nu}{3} \right) x + \left(1 -  \frac{65}{12}\, \nu\right) x^2 \nn\\
	&\qquad\quad + \left(1 +  \left[- \frac{2203}{2520} -  \frac{41}{192} \pi^2  +  \frac{22}{3} \ln\left(\frac{x}{x'_{0}}\right)\right]\nu + \frac{229}{36}\, \nu^2 + \frac{1}{81}\, \nu^3 \right) x^3 \nn\\
	&\qquad\quad + \left(1 + \left[- \frac{2067859}{33600} -  \frac{5411}{3072} \pi^2 - 38 \ln\left(\frac{x}{x'_{0}}\right) -  \frac{256}{15}\gamma_\text{E}-\frac{128}{15} \ln(16 x)\right) \right]\nu \nn\\
	&\qquad\qquad\qquad\left. + \left[\frac{153613}{15120} + \frac{6049}{576} \pi^2 + \frac{992}{9} \ln\left(\frac{x}{x'_{0}}\right) \right]\nu^2 -  \frac{1261}{324}\, \nu^3 + \frac{1}{243}\, \nu^4 \right) x^4 \Biggr]\,,
\end{align}
where we denote $x'_0 \equiv Gm/(r'_0 c^2)$. Plugging~\eqref{eq: relation gamma x} into~\eqref{eq: result criterion lagrangian full}, and defining the dimensionless quantity
\begin{equation}
	C_\text{ISCO} \equiv \frac{(Gm)^2}{c^6x^3} \hat{C}_0^\text{EoM}\,,
\end{equation}
we obtain the invariant criterion (now independent of the scale $r'_0$) already presented in Eq.~\eqref{eq: criterion result in intro}.

To confirm this result by a fully independent calculation,  we used the Hamiltonian approach detailed in Sec.~\ref{subsection:hamiltonian formalism}. In this case,  the stability criterion is given by~\eqref{eq: Hamiltonian criterion}, which we evaluated using the explicit expression of the 4PN Hamiltonian, recalled in Appendix~\ref{Appendix: explicit expressions of A and B} and the  results of Sec.~\ref{subsec: computation tail hamiltonian} for the perturbation of the tail terms. Defining
\begin{equation}\label{Gamma_circ}
	\Gamma \equiv \frac{G m}{R_0 c^2}\,,
\end{equation}
where $R_0$ is the radius of the circular orbit in ADM coordinates, we obtained for 
the criterion~\eqref{eq: Hamiltonian criterion} the expression
\begin{align}\label{eq: result criterion hamiltonian full}
	\hat{C}_0^\text{Ham} &= \frac{c^6 \Gamma^3}{G^2 m^2} \Bigg[ 1+\Gamma(-9+\nu)+ \Gamma^2 \left( \frac{117}{4}+\frac{43}{8}\,\nu+\nu^2\right)+\Gamma^3 \left(-61+  \left( \frac{4777}{48}-\frac{325\pi^2}{64}\right)\,\nu-\frac{31}{8}\,\nu^2+\nu^3\right)\nn\\
	&+ \Gamma^4 \bigg(\frac{783}{8}+  \left(- \frac{82386103}{57600}+\frac{780221\pi^2}{24576} +\frac{928}{3}\,\gamma_{\rm E}+\frac{464}{3}\,\mathrm{ln}(16\Gamma)-\frac{8864}{15}\,\mathrm{ln}(2)+\frac{2916}{5}\,\mathrm{ln}(3)\right)\,\nu \nn\\
    & \qquad \quad+\left(-\frac{2454599}{9600}+\frac{241281\pi^2}{16384}\right)\,\nu^2 -\frac{977}{256}\,\nu^3+ \nu^4 \bigg)\Bigg]\,.
\end{align}
Substituting the relation between the parameter $\Gamma$ and the gauge-invariant variable $x$, which we derived up to 4PN order,
\begin{align}\label{eq: relation Gamma x}
	\Gamma &= x  \Bigg[ 1+x\left(1-\frac{\nu}{3}\right)+ x^2 \left( \frac{5}{4}-\frac{43}{24}\,\nu\right)+x^3 \left(\frac{7}{4}+  \left( \frac{1133}{144}-\frac{167\pi^2}{192}\right)\,\nu+\frac{85}{36}\,\nu^2+\frac{1}{81}\,\nu^3\right)\nn\\
	&\qquad + x^4 \bigg(\frac{21}{8}+  \left( -\frac{3019459}{57600}+\frac{86513\pi^2}{24576} +\frac{128}{15}\,\gamma_{\rm E} +\frac{64}{15}\,\mathrm{ln}\left(16 x \right)\right)\,\nu \nn\\
		& \qquad \qquad \quad+\left(-\frac{1214003}{86400}+\frac{247933\pi^2}{147456}\right)\,\nu^2  -\frac{45685}{20736}\,\nu^3+ \frac{1}{243}\,\nu^4
		\bigg)\Bigg]\,,
\end{align}
one recovers the same expression \eqref{eq: criterion result in intro} for 
\begin{equation}
	C_\text{ISCO} \equiv \frac{(Gm)^2}{c^6x^3} \hat{C}_0^\text{Ham}\,.
\end{equation}

\section{Transformation from harmonic to ADM coordinates}\label{section:transformation}

We have shown in the previous sections that using either the equations of motion in harmonic coordinates or the Hamiltonian formalism in ADM coordinates leads to the same stability criterion, when the latter is expressed in terms of the observable binary's orbital frequency. Still it is instructive to show how the two approaches can be related by providing the explicit ``contact'' transformation linking the harmonic coordinates associated with the equations of motion (or Lagrangian) to the ADM coordinates associated with the Hamiltonian. In particular, in the case of circular orbits, the contact transformation should reduce to a relation of the form $\Gamma(\gamma)$, and therefore permit to cross-check the overall consistency of both approaches.

We now derive the contact transformation from harmonic to ADM coordinates at 4PN order [see, infra, Eq.~\eqref{eq Gamma gamma}]. More precisely, most of the discussion below will first focus on the instantaneous part of the 4PN Lagrangian in harmonic coordinates, which means that the tail contribution of the Lagrangian will be ignored. Correspondingly, we shall consider in a first stage the 4PN Hamiltonian {\it without} the tail part, knowing that the tail part of the Hamiltonian is related to the tail part in the Lagrangian by Eq.~\eqref{eq:Htail}. The contact transformation between the instantaneous parts of the Lagrangian and Hamiltonian at 4PN order has already been implemented (\textit{e.g.} in Ref.~\cite{BBFM17}) and we revisit it below, providing some explicit expressions for the contact transformation and the final Lagrangian. As we will see at the end of this section, when one takes into account the tail parts, a further shift of the coordinates is required to reach the final ADM coordinates, due to the replacement of accelerations in the tail part of the Lagrangian in Eq.~\eqref{eq:Htail}, which should be ordinary, \textit{i.e.} devoid of accelerations.

In harmonic coordinates, denoted below by a hat, it is known that the Lagrangian is a generalized one~\cite{DD81b_e}, thus the instantaneous part of the 4PN Lagrangian is of the type
\begin{equation}\label{Lharm}
	L_\text{4PN, harm}^{\rm inst} \equiv L\bigl[\hat{\boldsymbol{y}}_{\da}(t),\hat{\boldsymbol{v}}_{\da}(t),\hat{\boldsymbol{a}}_{\da}(t)\bigr]\,,
\end{equation}
depending not only on the positions $\hat{\boldsymbol{y}}_{\da}(t)$ and velocities $\hat{\boldsymbol{v}}_{\da}(t)$ of the particles (labelled by $\da, \db$, \textit{etc.}) but also, starting at  2PN order, on their accelerations $\hat{\boldsymbol{a}}_{\da}(t)=\dd\hat{\boldsymbol{v}}_{\da}/\dd t$. Most of our considerations below are valid for $N$-body systems ($\da, \db = 1, \dots, N$); for two-body systems the (generalized) variation of the Lagrangian~\eqref{Lharm} gives the instantaneous part of the equations of motion~\eqref{eq: EOM Lagrangian first form} after order reduction of the accelerations and restriction to the center of mass frame. Furthermone one can always assume (at any PN order) that the Lagrangian~\eqref{Lharm} is linear in accelerations and does not contain derivatives of accelerations (this can be proved recursively by combining total time derivatives and suitable ``double-zero'' terms~\cite{DS85,DS91}).

Since the Hamiltonian in ADM coordinates  is associated, \textit{via} a Legendre transform, with an ordinary Lagrangian, \textit{i.e.} without accelerations, our first task is to identify contact transformations that 
enable us to remove all accelerations from the harmonic Lagrangian, by following general principles developed in~\cite{DS85,DS91} and, more specifically, by extending to 4PN order the calculation of~\cite{ABF01} where the contact transformation was given up to 3PN order. In an arbitrary new coordinate system, the motion of the particles is described by their new positions  $\boldsymbol{y}_{\da}(t)$ and velocities $\boldsymbol{v}_{\da}(t)$. Their dynamics is physically equivalent to that of \eqref{Lharm} if and only if 
the new Lagrangian $L_{\rm new}$, which depends on $\boldsymbol{y}_{\da}(t)$, $\boldsymbol{v}_{\da}(t)$
and, in general, accelerations and derivatives of accelerations, is equal to \eqref{Lharm}  (up to total derivatives), \textit{i.e.}
\begin{equation}
	\label{Lag_transf}
	L_{\rm new}\bigl[\boldsymbol{y}_{\da}(t),\boldsymbol{v}_{\da}(t), \,\dots\bigr] = L\bigl[\hat{\boldsymbol{y}}_{\da}(t),\hat{\boldsymbol{v}}_{\da}(t),\hat{\boldsymbol{a}}_{\da}(t)\bigr]\,.
\end{equation}
Our goal is then to find the general transformation of variables such that the corresponding new Lagrangian is ordinary, \textit{i.e.} depends only on positions and velocities, and, finally, to identify among this subfamily of transformations the one for which the variables coincide with the ADM ones.  

In practice, we consider a generic infinitesimal contact transformation of the form
\begin{equation}\label{coord_transf}
	\hat{y}_\da^i(t) = y_\da^i(t) + \delta Y_\da^i(t)\,,
\end{equation}
where, since the accelerations  appear in the harmonic Lagrangian at 2PN order, 
$\delta Y_\da^i$ starts at 2PN and must be controlled up to 4PN order. 
Assuming the contact transformation at 4PN to depend on accelerations, \textit{i.e.}
\begin{equation}\label{contact struct}
	\delta Y_\da^i(t) = \delta Y_\da^i\bigl[\boldsymbol{y}_\db(t),\boldsymbol{v}_\db(t),\boldsymbol{a}_\db(t)\bigr]\,,
\end{equation}
the new Lagrangian~\eqref{Lag_transf} will 
in general depend on positions, velocities, accelerations and first and second time derivatives of accelerations, $\boldsymbol{b}_{\da}=\dd\boldsymbol{a}_{\da}/\dd t$ and $\boldsymbol{c}_{\da}=\dd\boldsymbol{b}_{\da}/\dd t$ respectively. Substituting~\eqref{coord_transf} into the right-hand side of~\eqref{Lag_transf} and expanding up to quadratic order in terms of the particle positions and their time derivatives, in order to obtain an  expression that is  valid up to 4PN order, we obtain
\begin{align}\label{eq: lagdummy}
	L_{\rm new}\bigl[\boldsymbol{y}_{\da},\boldsymbol{v}_{\da},\boldsymbol{a}_{\da},\boldsymbol{b}_{\da},\boldsymbol{c}_{\da}\bigr] &= L\bigl[\boldsymbol{y}_{\da},\boldsymbol{v}_{\da},\boldsymbol{a}_{\da}\bigr] + \sum_{\da} \frac{\partial L}{\partial y_\da^{i}}\,\delta Y_\da^{i}  +\sum_{\da} \frac{\partial L}{\partial v_\da^{i}}\,\delta V_\da^{i} +\sum_{\da} \frac{\partial L}{\partial a_\da^{i}}\,\delta A_\da^{i}\nn\\
	& + \frac{1}{2} \sum_{\da,\db} \frac{\partial^2 L_N}{\partial y_\da^{i}\, \partial y_\db^{j}}\,\delta Y_\da^{i}\,\delta Y_\db^{j} +  \frac{1}{2} \sum_{\da,\db} \frac{\partial^2 L_N}{\partial v_\da^{i}\, \partial v_\db^{j}}\,\delta V_\da^{i}\,\delta V_\db^{j} + \calO\left(\frac{1}{c^{10}}\right)\,.
\end{align}
Since the variations $\delta Y_\da^i$ start at 2PN order, the quadratic terms simply involve the second order derivatives of the Newtonian Lagrangian given by (with $r_{\da\db}=\vert\boldsymbol{y}_{\da}-\boldsymbol{y}_{\db}\vert$)
\begin{equation}\label{eq Newtonian LN}
	L_N\bigl[\boldsymbol{y}_{\da},\boldsymbol{v}_{\da}\bigr] = \sum_\da\biggl[\frac{1}{2}m_\da \boldsymbol{v}_\da^2 + \sum_{\db>\da}\frac{G m_\da m_\db}{r_{\da\db}}\biggr]\,.
\end{equation}
For later use we  also  define  the Newtonian acceleration
\begin{equation}\label{eq: N acc}
	N_{\da}^{i} \equiv \frac{1}{m_\da}\frac{\partial L_N}{\partial y_\da^i} = - \sum_{\db \not=  \da} \frac{G m_\db}{r_{\da\db}^2} \,n_{\da\db}^{i}\,.
\end{equation}

Next we introduce the notation for the functional derivative of the Lagrangian together with the definitions of the momenta conjugate to the position and velocity,
\begin{subequations}
\begin{align}
	\frac{\delta L}{\delta y_\da^{i}} &\equiv  \frac{\partial L}{\partial y_\da^{i}}-\frac{\mathrm{d}}{\mathrm{d}t} \left( \frac{\partial L}{\partial v_\da^{i}}\right) +\frac{\mathrm{d}^2}{\mathrm{d}t^2} \left( \frac{\partial L}{\partial a_\da^{i}}\right)\,, \label{dLdy}\\ p_\da^{i} &\equiv \frac{\delta L}{\delta v_\da^{i}} = \frac{\partial L}{\partial v_\da^{i}} - \frac{\mathrm{d}}{\mathrm{d}t} \left( \frac{\partial L}{\partial a_\da^{i}}\right)\,, \\ q_\da^{i} &\equiv \frac{\delta L}{\delta a_\da^{i}} = \frac{\partial L}{\partial a_\da^{i}}\,.\label{conjugate}
\end{align}
\end{subequations}
Performing some integrations by parts we rewrite the linear terms in~\eqref{eq: lagdummy} with the help of the functional derivative~\eqref{dLdy} and modulo the total time derivative of
\begin{align}\label{Lag1}
	R \equiv  \sum_{\da} \biggl[ p_\da^{i}\,\delta Y_\da^{i} + q_\da^{i}\,\delta V_\da^{i} \biggr]\,.
\end{align}
This total time derivative can be absorbed into the equivalent Lagrangian $L' = L_{\rm new} - \dd R/\dd t$ 
given by
\begin{align}\label{Lag1}
	L'\bigl[\boldsymbol{y}_{\da},\boldsymbol{v}_{\da},\boldsymbol{a}_{\da},\boldsymbol{b}_{\da}\bigr] 
	&=
	L\bigl[\boldsymbol{y}_{\da},\boldsymbol{v}_{\da},\boldsymbol{a}_{\da}\bigr] + \sum_{\da}  \frac{\delta L}{\delta y_\da^{i}}\,\delta Y_\da^{i} \nn\\
	& + \frac{1}{2} \sum_{\da,\db} \frac{\partial^2 L_N}{\partial y_\da^{i}\, \partial y_\db^{j}}\,\delta Y_\da^{i}\,\delta Y_\db^{j} 
	+   \frac{1}{2} \sum_{\da,\db} \frac{\partial^2 L_N}{\partial v_\da^{i}\, \partial v_\db^{j}}\,\delta V_\da^{i}\,\delta V_\db^{j} + \calO\left(\frac{1}{c^{10}}\right) \,.
\end{align}
Since the Lagrangian up to 4PN order is linear in the accelerations, the functional derivative $\delta L/\delta y_\da^{i}$ contains terms linear and quadratic in the accelerations $\boldsymbol{a}_{\da}$, as well as terms linear in the derivatives of accelerations $\boldsymbol{b}_{\da}=\dd\boldsymbol{a}_{\da}/\dd t$, while the $\boldsymbol{c}_\da$-dependent terms have been absorbed into the total derivative. From now on we shall work modulo a total time derivative and no longer give the expression of that total time derivative (which is irrelevant to the dynamics).

The contact transformation we shall apply takes the following structure
\begin{equation}\label{delta_Y}
	\delta Y_{\da}^{i} = \frac{1}{m_\da} \biggl( Q_{\da}^{i}  + \calX_{\da}^{i} + \sum_\db a_\db^j \calY_{\db\da}^{ji} \biggr) \,, \qquad Q_{\da}^{i} \equiv q_{\da}^{i} + \frac{\partial F}{\partial v_{\da}^{i}}\,,
\end{equation}
where $q_{\da}^{i}$ is the conjugate momentum of the velocity as defined by~\eqref{conjugate} and depends on positions $\boldsymbol{y}_{\db}$ and velocities $\boldsymbol{v}_{\db}$, where $F=F(\boldsymbol{y}_{\da},\boldsymbol{v}_{\da})$ is a generic function of the positions and velocities --- left unspecified at this stage --- depending on many free parameters (even so for the two-body case considered here)  which will be adjusted so as to recover the final ADM  Hamiltonian. Both $q_{\da}^{i}$ and the freely specifiable $F$ involve 2PN, 3PN and 4PN contributions. The extra term $\calX_{\da}^{i}$ in~\eqref{delta_Y} has 3PN and 4PN contributions, while the coefficient $\calY_{\db\da}^{ji}$ of the explicit acceleration in~\eqref{delta_Y} is purely 4PN. The terms at 2PN and 3PN orders have already been calculated in~\cite{ABF01} and we shall determine the 4PN terms. For any choice of the function $F$ there will be a unique determination of the ``unknown'' $\calX_{\da}^{i}$ and $\calY_{\db\da}^{ji}$ making the final Lagrangian acceleration-free.

Before obtaining the expression of $\calX_{\da}^{i}$ and $\calY_{\db\da}^{ji}$, it is useful to rewrite the terms quadratic in $\delta V_{\da}^{i}$ in~\eqref{Lag1} in the following manner: we substitute the Newtonian Lagrangian~\eqref{eq Newtonian LN} and replace $\delta V_{\da}^{i}$ by its expression derived from~\eqref{delta_Y} (up to 2PN only, \textit{i.e.} involving only $Q_{\da}^{i}$). This yields
\begin{align}
	\label{term_quad_v}
	\frac{1}{2} \sum_{\da,\db} \frac{\partial^2 L_N}{\partial v_\da^{i}\, \partial v_\db^{j}}\,\delta V_\da^{i}\,\delta V_\db^{j} &=
	\frac{1}{2} \sum_{\da} {m_\da}\,\left(\frac{\dd  \delta Y_{\da}^{i} }{\dd t}\right)^2 =
	-\frac{1}{2} \sum_{\da} \frac{\mathrm{d}^2 Q_{\da}^{i}}{\mathrm{d}t^2} \,\delta Y_{\da}^{i} + \calO\left(\frac{1}{c^{10}}\right)\,,
\end{align}
where the second equality is valid up to a total time derivative. This leads us to consider the equivalent Lagrangian 
\begin{align}\label{Lag2}
	L'' = L + \sum_{\da} T^i_\da\,\delta Y_\da^{i} + \frac{1}{2} \sum_{\da,\db} \frac{\partial^2 L_N}{\partial y_\da^{i}\, \partial y_\db^{j}}\,\delta Y_\da^{i}\,\delta Y_\db^{j} + \calO\left(\frac{1}{c^{10}}\right) \,,
\end{align}
where the coefficient of the linear term in $\delta Y_\da^{i}$ is now
\begin{align}\label{def_T}
	T^i_\da \equiv \frac{\delta L}{\delta y_\da^{i}} -\frac{1}{2} \frac{\mathrm{d}^2 Q_{\da}^{i}}{\mathrm{d}t^2} \,.
\end{align}
The last term in~\eqref{Lag2} does not depend on accelerations and therefore will just play a spectator role in what follows. The crucial element is the  structure of $T^i_\da$ in terms of $\boldsymbol{a}_{\da}$ and $\boldsymbol{b}_{\da}$ up to 2PN order, of the form
\begin{equation}\label{decomp_hat_T}
	T^i_\da = m_\da \biggl[ - a_{\da}^{i} + N_{\da}^{i} +  \calC_{\da}^{i} +  \sum_\db \Bigl(a_\db^j \,\calC_{\db\da}^{ji} + b_\db^j \,\calD_{\db\da}^{ji}\Bigr) + \sum_{\db,\dc} a_\db^j a_\dc^k \,\calD_{\dc\db\da}^{kji}	\biggr] + \calO\left(\frac{1}{c^{6}}\right)\,.
\end{equation}
Here $N_{\da}^{i}$ is the Newtonian acceleration~\eqref{eq: N acc}, and both $\calC_{\da}^{i}$ and $\calC_{\db\da}^{ji}$ are composed of 1PN and 2PN terms, while both $\calD_{\db\da}^{ji}$ and $\calD_{\dc\db\da}^{kji}$ are made of 2PN terms only. Thus accelerations appear linearly at 1PN order, while derivatives of accelerations and terms quadratic in accelerations appear at 2PN order.

Plugging Eqs.~\eqref{delta_Y} and~\eqref{decomp_hat_T} into~\eqref{Lag2}, it is easy to identify the acceleration-dependent terms. Then the terms linear in $\boldsymbol{b}_{\da}$ can be
transformed, modulo a total time derivative, in terms linear and quadratic in accelerations, which involve the combination
\begin{equation}
	\frac{\dd}{\dd t}\Bigl[ \calD_{\da\db}^{ij} Q_\db^j\Bigr]=\Lambda_\da^i+\sum_\db a_\db^j\, \Lambda_{\db\da}^{ji}\,,
\end{equation}
where both $\Lambda_{\da}^{i}$ and $\Lambda_{\db\da}^{ji}$ are purely 4PN quantities. Finally, the terms quadratic in accelerations are reduced to terms linear in accelerations by means of the already-mentioned ``double-zero'' technique~\cite{DS85}.\footnote{In practice, we use the identity $$\sum_{\da,\db} q_{\da\db}^{ij}a_\da^i a_\db^j=\sum_{\da,\db}q_{\da\db}^{ij} (a_\da^i-N_\da^i)(a_\db^j- N_\db^j)+\sum_{\da,\db}q_{\da\db}^{ij} (a_\da^iN_\db^j+ N_\da^ia_\db^j) -\sum_{\da,\db} q_{\da\db}^{ij}N_\da^i N_\db^j,$$
where $q_{\da\db}^{ij}$ is a purely 4PN quantity. The first term on the right hand side is the double-zero term which does not contribute to the equations of motion at 4PN order.} As a result, we obtain the following equivalent Lagrangian, which depends linearly on the accelerations:
\begin{align}\label{Lag3}
	L''' &= L - \sum_\da a_\da^i q_\da^i + \sum_\da v_\da^i \frac{\partial F}{\partial y_{\da}^{i}} \\ 
	&+ \sum_\da \bigl(N_\da^i+\calC_\da^i\bigr)\Bigl[ Q_{\da}^{i} + \calX_{\da}^{i}\Bigr] + \sum_{\da,\db} N_\da^iN_\db^j\Bigl[\calY_{\db\da}^{ji} +  \Lambda_{\db\da}^{ji} - \sum_\dc \calD_{\db\da\dc}^{jik} Q_\dc^k\Bigr] + \frac{1}{2} \sum_{\da,\db} \frac{\partial^2 L_N}{\partial y_\da^{i}\, \partial y_\db^{j}}\,\frac{Q_{\da}^{i} Q_{\db}^{j}}{m_\da m_\db} \nn\\ & + \sum_\da a_\da^i\Biggl\{ - \calX_\da^i - \Lambda_\da^i + \sum_\db\biggl(\calC_{\da\db}^{ij}\Bigl(Q_\db^j + \calX_\db^j\Bigr) + N_\db^j\biggl[- \calY_{\db\da}^{ji} - \Lambda_{\da\db}^{ij} - \Lambda_{\db\da}^{ji}
	+ \sum_\dc\Bigl(\calD_{\da\db\dc}^{ijk} + \calD_{\db\da\dc}^{jik}\Bigr)Q_{\dc}^{k}\biggr]\biggr)\Biggr\} + \calO\left(\frac{1}{c^{10}}\right)\,.\nn
\end{align}
The sum of the first two terms, $\tilde{L} = L - \sum_\da a_\da^i q_\da^i$, is nothing but the original harmonic-coordinates Lagrangian in which the accelerations are set to zero (since $L$ is linear in the accelerations). All remaining accelerations (occurring at 3PN and 4PN orders) are therefore explicitly displayed in the third line of~\eqref{Lag3}, and it is now trivial to cancel them out by setting
\begin{subequations}
	\begin{align}
		\calX_{\da}^{i} &= -\Lambda_{\da}^{i}+\sum_\db  \calC_{\da\db}^{ij}\biggl[ Q_{\db}^{j} + \sum_{\dc} \calC_{\db\dc}^{jk} Q_{\dc}^{k} \biggr] + \calO\left(\frac{1}{c^{10}}\right) \,,\\
		\calY_{\db\da}^{ji} &= -\Lambda_{\da\db}^{ij} - \Lambda_{\db\da}^{ji} + \sum_\dc \Bigl(\calD_{\da\db\dc}^{ijk} + \calD_{\db\da\dc}^{jik}\Bigr) Q_{\dc}^{k} + \calO\left(\frac{1}{c^{10}}\right) \,.
	\end{align}
\end{subequations}
With this choice, which completely specifies the contact transformation~\eqref{delta_Y} for any choice of the function $F$, we finally obtain the following ordinary Lagrangian, with all accelerations removed, up to 4PN order: 
\begin{align}\label{Lagord}
	L_\text{ord}\bigl[\boldsymbol{y}_{\da},\boldsymbol{v}_{\da}\bigr]  &= L\bigl[\boldsymbol{y}_{\da},\boldsymbol{v}_{\da},\boldsymbol{a}_{\da}\bigr] - \sum_\da a_\da^i q_\da^i + \sum_\da v_\da^i \frac{\partial F}{\partial y_{\da}^{i}} \nn\\ 
	&+ \sum_\da \bigl(N_\da^i+\calC_\da^i\bigr)\biggl[ Q_{\da}^{i} - \Lambda_{\da}^{i} + \sum_\db \calC_{\da\db}^{ij} Q_{\db}^{j} + \sum_{\db,\dc} \calC_{\da\db}^{ij} \calC_{\db\dc}^{jk} Q_{\dc}^{k} \biggr]\nn\\ & + \sum_{\da,\db} N_\da^i N_\db^j \biggl[ - \Lambda_{\da\db}^{ij} + \sum_\dc \calD_{\da\db\dc}^{ijk} Q_{\dc}^{k} \biggr] + \frac{1}{2} \sum_{\da,\db} \frac{\partial^2 L_N}{\partial y_\da^{i}\, \partial y_\db^{j}}\,\frac{Q_{\da}^{i} Q_{\db}^{j}}{m_\da m_\db} + \calO\left(\frac{1}{c^{10}}\right)\,.
\end{align}

Restricting the analysis back to the two-body case, we parametrize the function $F$ by a set of dimensionless arbitrary parameters (each associated with a term whose dimension is compatible with $F$) --- this defines a large class of ordinary 4PN Lagrangians~\eqref{Lagord}. We then obtain the associated class of Hamiltonians by a standard Legendre transformation, and finally select the parameters so as to recover the (instantaneous) 4PN Hamiltonian in ADM coordinates (in a general frame and for any orbits). We find the solution to be unique; this of course confirms that the 4PN Hamiltonian~\cite{JaraS15,BiniD13,DJS14,DJS15eob} and the 4PN equations of motion (or 4PN Fokker Lagrangian)~\cite{BBBFMa,BBBFMb,BBBFMc,MBBF17,BBFM17} carry the same physical content. The details of the resulting contact transformation, \textit{i.e.} Eq.~\eqref{delta_Y} in which all terms are now explicitly and uniquely determined, are reported in Appendix~\ref{app:contact-transf}.

We can then find the relation between the 2-body relative distances expressed in harmonic coordinates and in ADM coordinates, respectively, using~\eqref{coord_transf} with the explicit form of~\eqref{contact struct} given in App.~\ref{app:contact-transf}. Going to the CM frame and specializing the result to circular orbits, one thus obtains the relation between the distance $r_0$ in harmonic coordinates and the distance $R_0$ in ADM coordinates, as $r_0 = R_0 + \delta R_0^\text{inst}$ where $\delta R_0^\text{inst}$ is obtained explicitly from the contact transformation $\delta Y_\da^i(t)$ in~\eqref{coord_transf}.\footnote{We remind that in Eq.~\eqref{coord_transf} the hat variable $\hat{y}_\da^i(t)$ in the left-hand side is the harmonic-coordinates variable, while $y_\da^i(t)$ is to be identified in the end of the construction with the ADM variable. Hence, for instance, $r_0=\vert\hat{y}_1^i-\hat{y}_2^i\vert$ for circular orbits.} Here the mention ``inst'' is a reminder that for the moment we have obtained the contact transformation only between the instantaneous parts of the Lagrangian and Hamiltonian, as mentioned earlier. For the final step, one must indeed take into account the tail contributions. As discussed in Section V.C of~\cite{BBBFMa}, the tail part of the Hamiltonian~\eqref{eq:Htail} is obtained after Eq.~\eqref{L_tail} has been reduced to an ordinary Lagrangian. This order reduction of the higher order terms in the tail part of the Lagrangian is equivalent to a coordinate shift, explicitly given by Eq.~(5.15) of~\cite{BBBFMa}. When restricted to a circular orbit (in the center of mass frame), this shift leads to the following tail-induced modification of the relation between the true distances $r_0$ and $R_0$:
\begin{align}
\label{shift_ADM}
	r_0 = R_0 \left[ 1 + \frac{128}{5} \frac{G^4 m^4\nu}{c^8 R_0^4} \Lambda_0 \right] + \delta R_0^\text{inst}\,,
\end{align}
where we employ the notation~\eqref{Lambda0}. Rewriting this relation between $r_0$ and $R_0$ in terms of the parameters $\gamma$ and $\Gamma$, one finally gets
\begin{align}\label{eq Gamma gamma}
	\gamma &= \Gamma \Bigg[ 1+ \Gamma^2 \left( -\frac{1}{4}-\frac{29}{8}\,\nu\right)+\Gamma^3 \left(  \left(\frac{3163}{1680}+\frac{21\pi^2}{32}-\frac{22}{3}\,\mathrm{ln}\left( \frac{R}{r_0'}\right)\right)\,\nu+\frac{3}{8}\,\nu^2\right)\nn\\
	& \quad \qquad + \Gamma^4 \bigg(\frac{1}{16}+  \left( -\frac{3137}{2304}-\frac{64771\pi^2}{8192}+\frac{202}{3}\,\mathrm{ln}\left( \frac{R}{r_0'}\right)-\frac{128}{5}\,\gamma_{\rm E} -\frac{64}{5}\,\mathrm{ln}\left(16 \Gamma \right)\right)\,\nu \nn\\
	& \qquad \qquad \qquad
	+\left(\frac{9737}{9600}+\frac{476545\pi^2}{49152}-120\,\mathrm{ln}\left( \frac{R}{r_0'}\right)\right)\,\nu^2
	+\frac{5}{256}\,\nu^3
	\bigg)\Bigg]\,.
\end{align}
One can check that this relation is consistent with the expressions of $\gamma(x)$ and $\Gamma(x)$, respectively given by~\eqref{eq: relation gamma x} and~\eqref{eq: relation Gamma x}, which were obtained independently. 

\section{Comparison with numerical and EOB methods}\label{section:comparison}

\begin{figure}[t!]
	\begin{subfigure}[b]{0.49\linewidth}
		\centering
		\includegraphics[width=\linewidth]{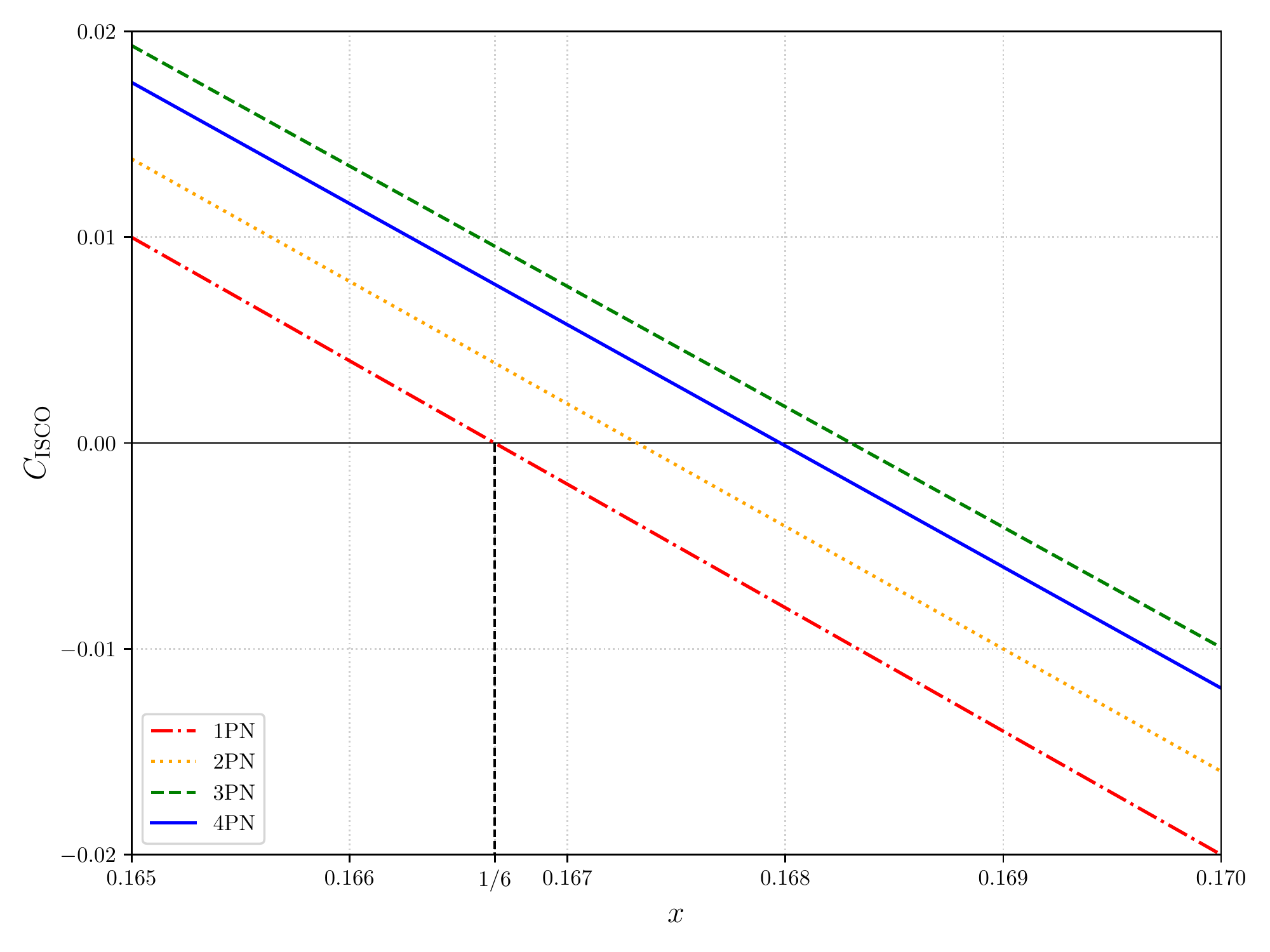}
		\caption{$\nu=0.01$}
		\label{Figure critere nu small}
	\end{subfigure}\hfill
	\begin{subfigure}[b]{0.49\linewidth}
		\centering
		\includegraphics[width=\linewidth]{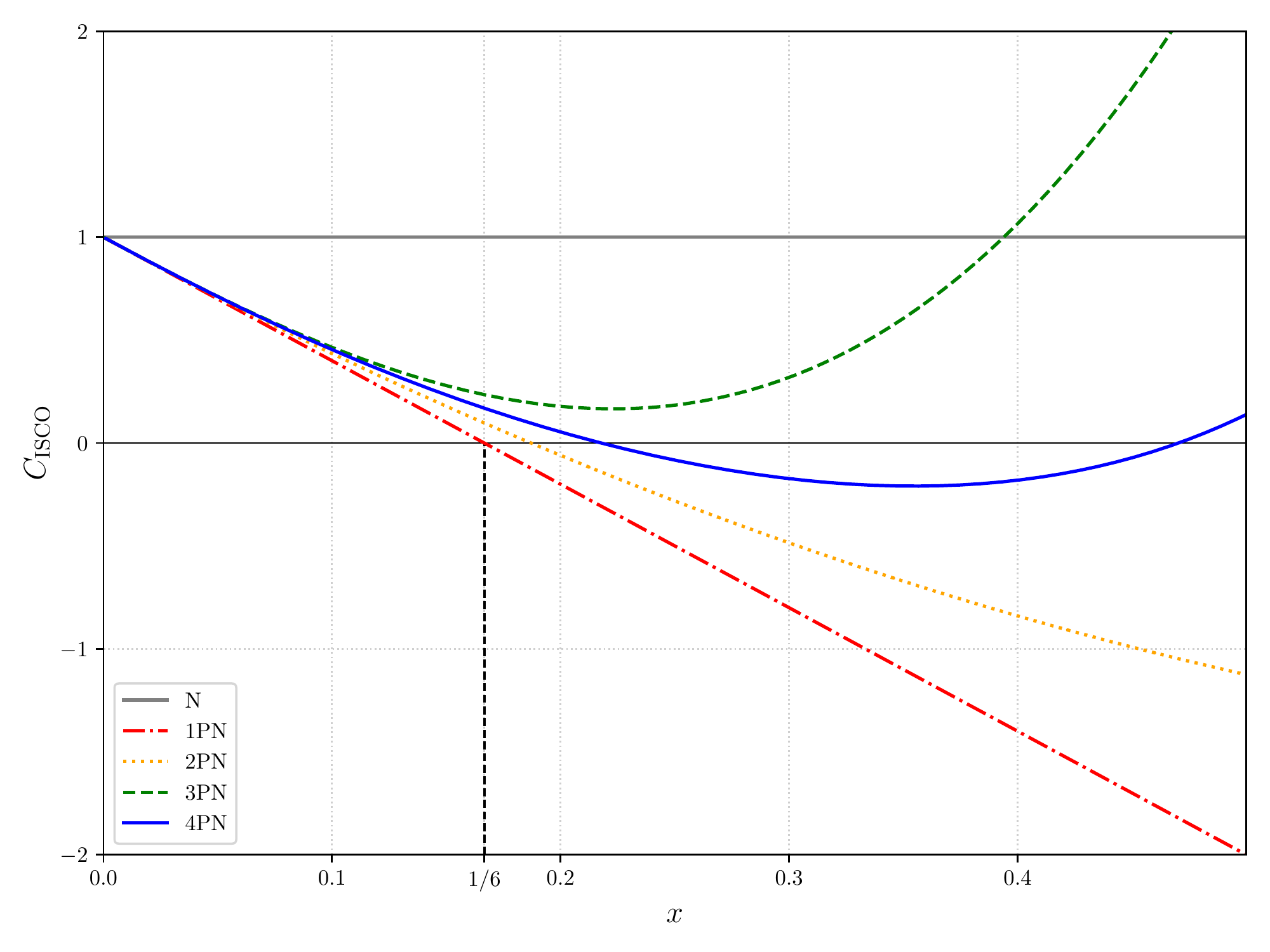}
		\caption{$\nu=1/4$}
		\label{Figure critere nu large}
	\end{subfigure}\hfill
	\caption{Instability threshold function $C_{\rm ISCO}(x)$, from 1PN to 4PN orders, for (a) $\nu=0.01$ and (b) $\nu=1/4$.}
	\label{Figure critere}
\end{figure}

\begin{figure}[h!]
	\centering
	\includegraphics[scale=0.6]{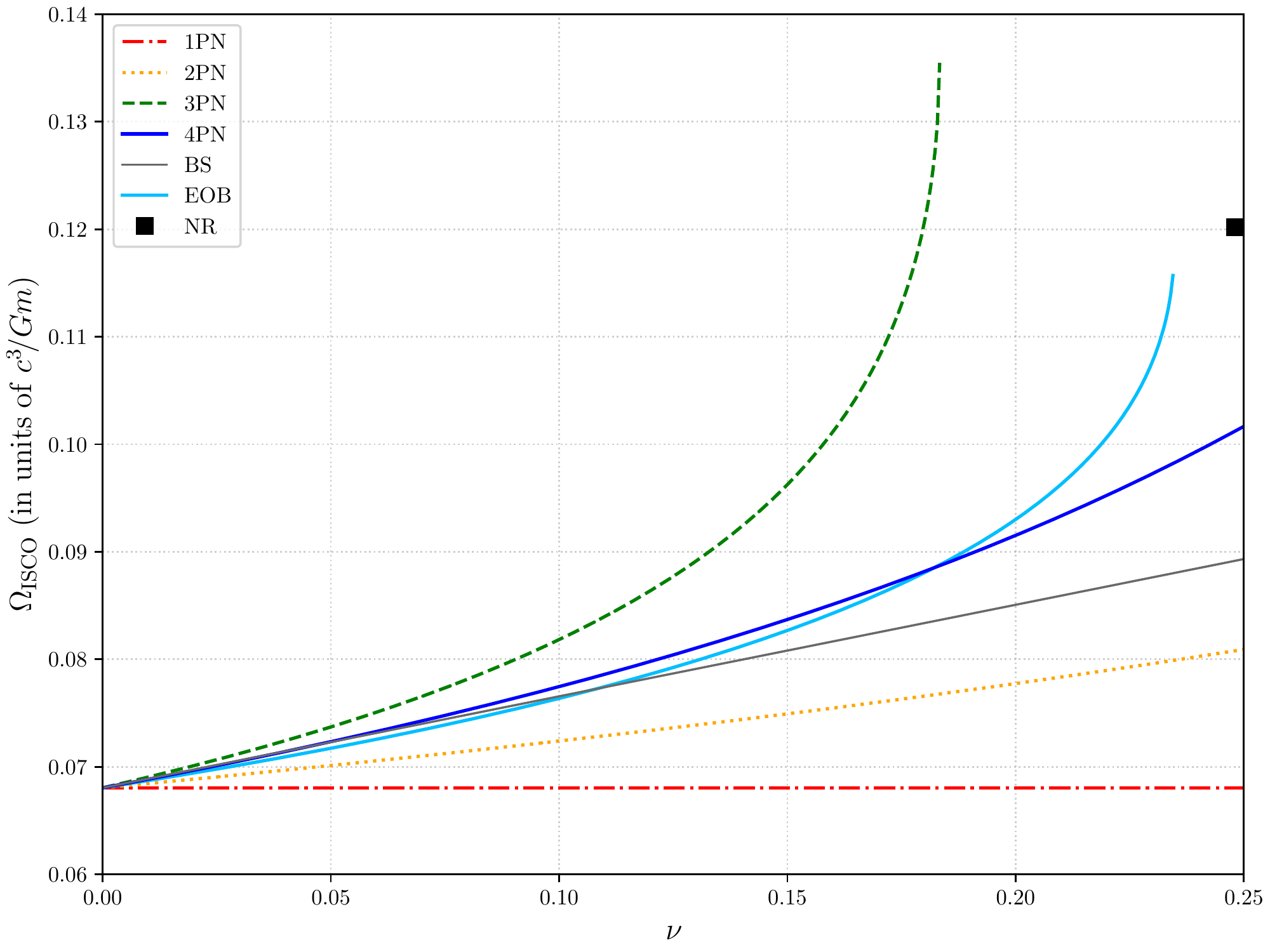}
	\caption{The ISCO frequency $\Omega_\text{ISCO}$ [in units of $c^3/(G m)$] in terms of the mass ratio $\nu$, from 1PN to 4PN orders. The Barack-Sago (BS)~\cite{BarackS09} first-order GSF result given by~\eqref{BSresult} is presented as a solid grey line. The ISCO estimate from numerical relativity for equal masses~\cite{CCGPf06} is shown as a black square. The EOB prediction at 4PN using the criterion in~\eqref{eq: criterion 4PN EOB} below is plotted as a solid skyblue  curve. In contrast with our 4PN result, one finds no ISCO above some critical mass ratio both for the  criterion at 3PN ($\nu_\text{c}\simeq 0.183$) and for the EOB criterion at 4PN  ($\nu'_\text{c} \simeq 0.234$).\
    }\label{figure ISCO}
\end{figure}

Let us return to our main result for the 4PN invariant stability criterion, Eq.~\eqref{eq: criterion result in intro}, which we reproduce here for better readability: $C_\text{ISCO}>0$ where
\begin{align}\label{eq: criterion result discussion section}
	C_\text{ISCO} &= 1 -6 x+14\nu\, x^2 + \nu \left[ \frac{397}{2}-\frac{123 \pi^2}{16} -14\nu  \right] x^3\nn\cr
	& \quad + \nu \Bigg[ -\frac{215729}{180} + \frac{58265 \pi^2}{1536} +\frac{5024}{15}\,\gamma_{\rm E} +\frac{2512}{15}\,\mathrm{ln}\left(x \right) +\frac{1184}{15}\,\mathrm{ln}(2)+\frac{2916}{5}\,\mathrm{ln}(3) \\
	& \qquad \quad + \left( -\frac{4223}{6}+\frac{451\pi^2}{16}\right) \nu  +\frac{196}{27}\,\nu^2 \Bigg]x^4 + \calO(x^5)\,.
\end{align}
Each successive power of $x = (G m \Omega_0/c^3)^{2/3}$ in this expression corresponds to the associated PN order  and encapsulates all interactions at said order, determined either in their Lagrangian or Hamiltonian formulations, and without resorting to \textit{ad hoc} resummation techniques. Accordingly, the ISCO at a given PN order is obtained by retaining the powers in $x$ up to that order. The onset of instability is by construction the lowest root of the equation $C_\text{ISCO}(x) = 0$, which we denote $x_\text{ISCO}$: the compact binary is dynamically stable for $x<x_\text{ISCO}$, and unstable otherwise. The criterion~\eqref{eq: criterion result discussion section} is plotted in Figure~\ref{Figure critere} at successive
orders up to 4PN, for an extreme mass ratio $\nu=0.01$ and for equal masses $\nu=1/4$. The dependence of $x_\text{ISCO}$, or equivalently $\Omega_\text{ISCO}$, on the mass ratio $\nu$ is plotted in Figure~\ref{figure ISCO}, from 1PN to 4PN orders.
A remarkable property of \eqref{eq: criterion result discussion section} is that it yields the \textit{exact} Schwarzschild value $x^\text{Schw}_\text{ISCO}=1/6$ when $\nu\to 0$, for all known PN orders. The finite mass-ratio corrections appear only from the 2PN order on. At 2PN order the ISCO is given by
\begin{align}
\label{x 2PN}
	x_\text{ISCO}^{\rm 2PN}=\frac{3}{14\nu}\biggl(1-\sqrt{1-\frac{14\nu}{9}}\biggr)\,,
\end{align}
and we find that $x_{\rm ISCO}^{\rm 2PN}$ grows monotonically with $\nu$, reaching $x_{\rm ISCO}^{\rm 2PN}\simeq 0.187$ for the equal mass case $\nu=1/4$. As can be seen in Fig.~\ref{figure ISCO}, the 3PN contributions increase the value of the ISCO, making it more ``inward'', \textit{i.e.} $x_{\rm ISCO}^{\rm 3PN}>x_{\rm ISCO}^{\rm 2PN}$ for each given $\nu$. Furthermore, as stressed in \cite{BI03CM}, the 3PN criterion has no root when $\nu$ is above a critical value $\nu_\text{c}\simeq 0.183$, thus any circular binary orbit is stable for  $\nu \geqslant \nu_\text{c}$, which contradicts the prediction from NR that there should be an ISCO at $\nu=1/4$ given by the black square in Fig.~\ref{figure ISCO}.

In addition to being more involved than the 3PN one, the new 4PN contribution in~\eqref{eq: criterion result discussion section} includes a $\ln(x)$ term which renders the function $C_{\rm ISCO}(x)$ non-polynomial. At 4PN order, we find that, in contrast to the 3PN result, an ISCO exists for any mass ratio $\nu$. Moreover, its value is located between the 2PN prediction and the 3PN one: $ x_\text{ISCO}^\text{2PN}<x_\text{ISCO}^\text{4PN}<x_\text{ISCO}^\text{3PN}$, where the second inequality obviously holds only when the 3PN value is defined (\textit{i.e.} for $\nu<\nu_c$). The maximum of $x_\text{ISCO}^\text{4PN}$ is again reached for $\nu=1/4$, and yields $x_\text{ISCO}^\text{4PN}(1/4)\simeq 0.218$, which is significantly larger than the corresponding maximum of $x_{\rm ISCO}^{\rm 2PN}$.

An important assessment of our stability criterion at 4PN order is how it fares with the numerical calculation of the ISCO shift by the gravitational self-force (GSF) perturbation approach. This fully relativistic numerical calculation was performed at first order in the mass ratio by Barack and Sago~\cite{BarackS09}. Later it was also done by Le~Tiec, Barausse and Buonanno~\cite{LBB12} and Akcay \textit{et al.}~\cite{Akcay12} using the first law of black hole mechanics~\cite{LBW12}. The generalization to the conservative ISCO shift of a Kerr black hole has been obtained in~\cite{IBDLNSTW14}.

The comparison of the stability criterion has already been done at 3PN order by Favata~\cite{F11a} who concluded that the agreement is better than any other 3PN order estimate based on uncalibrated methods, including all hybrid, resummed or EOB methods. Here we  find that the addition of the new 4PN  contribution to
the criterion~\eqref{eq: criterion result discussion section} makes the comparison even better, nicely improving the agreement to the ISCO shift~\cite{BarackS09, LBB12, Akcay12}. Adopting the notation in~\cite{F11a} we define at any given $n$PN order the ISCO shift by
\begin{subequations}
\begin{align}
	x_\text{ISCO}^\text{$n$PN} &= \frac{1}{6}\Bigl[1+c_x^\text{$n$PN}\nu+\calO\left(\nu^2\right)\Bigr]\,,\\
	\Omega_\text{ISCO}^\text{$n$PN} &= \frac{c^3}{6^{3/2}G m}\Bigl[1+c_\Omega^\text{$n$PN}\nu+\calO\left(\nu^2\right)\Bigr]\,,
\end{align}
\end{subequations}
where $c_x^\text{$n$PN}$ and $c_\Omega^\text{$n$PN} = (3/2) c_x^\text{$n$PN}$ are dimensionless coefficients. Our analytical results for $c_\Omega^\text{$n$PN}$  up to 4PN order are presented in Table~\ref{tab:ISCO}.
\begin{table}[h] \centering \renewcommand{\arraystretch}{1.5} \begin{tabular}{|>{\centering\arraybackslash}m{2cm}||>{\centering\arraybackslash}m{5cm}|>{\centering\arraybackslash}m{2cm}||>{\centering\arraybackslash}m{2cm}|} \hline \parbox[c][1cm][c]{2cm}{\centering Order $n$} & \parbox[c][1cm][c]{5cm}{\centering $c_\Omega^{n\text{PN}}$} & \parbox[c][1cm][c]{2cm}{\centering Num. value} & \parbox[c][1cm][c]{2cm}{\centering $\Delta c_\Omega^{n\text{PN}}$} \\ 
\hline
\hline \parbox[c][0.5cm][c]{1cm}{1PN} & 0 & 0 & $-100\%$ \\ 
\hline \parbox[c][1cm][c]{1cm}{2PN} & $\vcenter{$\begin{aligned} \frac{7}{12} \end{aligned}$}$ & 0.583 & $-53\%$  \\ 
\hline \parbox[c][1cm][c]{1cm}{3PN} & $\vcenter{$\begin{aligned} \frac{565}{288}-\frac{41}{768}\pi^2 \end{aligned}$}$ & 1.435 & $+15\%$  \\ 
\hline  \parbox[c][1.8cm][c]{1cm}{4PN} & $\vcenter{$\begin{aligned} &\frac{89371}{155520} + \frac{157}{405}\gamma_\text{E} - \frac{12583}{1327104}\pi^2 \\ &- \frac{83}{810}\ln(2) + \frac{1559}{3240}\ln(3)  \end{aligned}$}$ & 1.162  & $-7\%$ \\
\hline  \parbox[c][1.8cm][c]{1cm}{4PN$^\text{EOB}$} & $\vcenter{$\begin{aligned} &\frac{81419}{103680} + \frac{2}{9}\gamma_\text{E} - \frac{3923}{884736}\pi^2 \\ &+ \frac{1}{3}\ln(2) - \frac{1}{9}\ln(3)  \end{aligned}$}$ & 0.979  & $-22\%$ \\ \hline
\end{tabular} \caption{Analytic predictions for the ISCO shift up to 4PN order, with corresponding numerical values and deviation $\Delta c_\Omega^{n\text{PN}}$ from the GSF result~\cite{BarackS09,LBB12,Akcay12} as defined by Eq.~\eqref{DeltacOmega}. In the last line is shown the prediction for the ISCO shift at 4PN order given by the EOB criterion~\eqref{eq: criterion 4PN EOB}.
}\label{tab:ISCO} \end{table}

On the other hand, the GSF result of Barack and Sago~\cite{BarackS09}, working in Lorenz gauge, reads in brute form
\begin{align}
	\hat{\Omega}_\text{ISCO}
	= \frac{c^3}{6^{3/2}G m_2}\Bigl[1+\hat{c}_\Omega^\text{BS}q+\calO\left(q^2\right)\Bigr]\,,
\end{align}
where the mass ratio is $q=m_1/m_2$, with $m_2$ the black hole mass ($m_1\ll m_2$), and where the coefficient is calculated numerically as $\hat{c}_\Omega^\text{BS}\simeq 0.4869$. However, as pointed out by Damour~\cite{D10sf}, the Lorenz gauge used in the GSF calculation is not asymptotically flat, and one must therefore implement a correction given by $\Omega = \hat{\Omega}[1-q/\sqrt{18}+\calO(q^2)]$ in order to match our definition of the gauge invariant $\Omega$ (see~\cite{D10sf,F11a}) --- recall that $\Omega$ is only invariant with respect to the class of coordinate systems that are asymptotically flat. Such correction, together with the one coming from the change of parameters $(m_2,q)\longrightarrow(m,\nu)$, yields
\begin{subequations}\label{BSresult}
	\begin{align}
	\Omega_\text{ISCO}
	&= \frac{c^3}{6^{3/2}G m}\Bigl[1+c_\Omega^\text{BS}\nu+\calO\left(\nu^2\right)\Bigr]\,,\\
	\text{where}\quad c_\Omega^\text{BS} &= \hat{c}_\Omega^\text{BS}+1-\frac{1}{\sqrt{18}}\simeq 1.2512\,.
\end{align}
\end{subequations}
The numerical value of $c_\Omega^\text{BS}$ (which is denoted $c_\Omega^\text{ren}$ in~\cite{F11a}) is to be directly compared to the PN results, namely the successive PN coefficients $c_\Omega^\text{$n$PN}$ in Table~\ref{tab:ISCO}.
Let
\begin{align}\label{DeltacOmega}
\Delta c_\Omega^\text{$n$PN} \equiv \frac{c_\Omega^\text{$n$PN}-c_\Omega^\text{BS}}{c_\Omega^\text{BS}}\,,
\end{align}
be the error of the $n$PN prediction compared to the exact GSF result. We find that while $\Delta c_\Omega^\text{$3$PN}=+15\%$~\cite{F11a}, we have $\Delta c_\Omega^\text{$4$PN}=-7\%$. There is clearly a substantial improvement (the error decreases overall with the PN order, see Table~\ref{tab:ISCO}), which is satisfying bearing in mind that the PN approximation is used here in the strong field regime (in principle outside its domain of validity) and in standard non-resummed Taylor-expanded form.

We have also presented in Fig.~\ref{figure ISCO}  (similarly to the lower left plot of Fig.~2 in~\cite{F11a}) the rough estimate $\Omega_{\rm ISCO}^{\rm NR}\simeq 0.12$ of the ISCO angular velocity [in units of $c^3/(G m)$] for $\nu=1/4$, deduced from numerical relativity (NR) computations~\cite{CCGPf06}, \textit{cf.} the black square in Fig.~\ref{figure ISCO}. Our 4PN ISCO estimate $\Omega_{\rm ISCO}^{\rm 4PN}\simeq 0.102$ agrees to within $15\%$ with the numerical value for equal masses, and confirms the tendency that the ISCO gets more ``inward'' when the mass ratio is increased. Again, the 4PN result represents a significant improvement compared with the 2PN ISCO estimate for $\nu=1/4$,  $\Omega_{\rm ISCO}^{\rm 2PN}\simeq 0.081$, which corresponds to a $33\%$ relative error from the numerical relativity result, whereas the 3PN calculation predicts no ISCO for such a large value of $\nu$.

Note that Favata~\cite{F11a} proposed (before the 4PN dynamics was calculated) a heuristic pseudo-4PN criterion by assuming that the 4PN contribution to $C_{\rm ISCO}$, which corresponds to the last two lines of~\eqref{eq: criterion result discussion section} is of the form $c_{\rm 4PN}\nu x^4$, where $c_{\rm 4PN}$ is a constant coefficient that is tuned to ensure a perfect agreement with the GSF result in the limit $\nu\to 0$. It turns out that the associated pseudo-curve for $\Omega_{\rm ISCO}$ as a function of $\nu$ (see the lower left plot of Fig.~2 in~\cite{F11a}), away from small values of $\nu$, is above the exact 4PN curve computed in the present work, leading to an estimate of $\Omega_{\rm ISCO}$ for $\nu=1/4$ which is $31\%$ above the NR result, while the exact 4PN gives an estimate which is $15\%$ below.

A different 4PN criterion can be calculated using the EOB formalism~\cite{BuonD99,BuonD00,DNorleans,DJS15eob}. By combining Eq. (4.14) of~\cite{D10sf} with the 4PN EOB metric coefficient given by Eq.~(8.1a) in~\cite{DJS15eob}, we readily find
\begin{align}\label{eq: criterion 4PN EOB}
	C_\text{ISCO}^{\rm EOB} &= 1 -6 x+5\nu\, x^2 + \nu \left[ \frac{827}{6}-\frac{41 \pi^2}{8} -2\nu  \right] x^3\nn\\
	& \quad + \nu \Bigg[ -\frac{39421}{120} + \frac{27565 \pi^2}{1024} +192\,\gamma_{\rm E} +96\,\mathrm{ln}\left(x \right) +384\,\mathrm{ln}(2) \nn\\
	& \qquad \quad 
    + \left( -\frac{4843}{12}+\frac{943\pi^2}{64}\right) \nu  +\frac{7}{27}\,\nu^2 \Bigg]x^4 + \calO(x^5)\,.
\end{align}
This differs from our standard PN criterion~\eqref{eq: criterion result discussion section} already from the 2PN order. The orbital frequencies $\Omega_\text{ISCO}^{\rm EOB}$ associated with this EOB criterion are plotted in Fig.~\ref{figure ISCO}. Similarly to the 3PN  criterion discussed before,  the EOB criterion at 4PN order~\eqref{eq: criterion 4PN EOB} predicts no ISCO about a critical value of $\nu$, here for $\nu \geqslant \nu'_\text{c}$ with $\nu'_\text{c} \simeq 0.234$ (which is higher than the critical value $\nu_\text{c}\simeq 0.183$ for the 3PN criterion and close to the upper value $\nu=1/4$). We have also added in Table~\ref{tab:ISCO} the EOB prediction for the ISCO shift at 4PN order, which is not as good as the one from the standard 4PN expansion as compared to the GSF result ($-22\%$ versus $-7\%$ relative difference).

\section{Conclusion}\label{section:conclusion}

We have obtained an improved stability criterion for compact binary circular orbits (without spins) by expanding to 4PN order the gauge-invariant criterion that was derived in~\cite{BI03CM} at 3PN order. Such criterion yields a notion of the ISCO which is valid for arbitrary-mass compact binaries. Going from 3PN to 4PN requires significant work as the 4PN correction involves the tail term, which is non-local in time, and whose stability analysis is somewhat delicate. To reach our final result, we have performed two completely independent calculations, the first based
on the perturbation of the 4PN equations of motion in harmonic coordinates, the second based on the perturbation of the 4PN Hamiltonian in ADM coordinates. The 4PN tail term in both approaches is treated using the discrete Fourier series of the Keplerian orbit (without orbital precession). No time averaging is applied in our procedure. However the perturbation of the orbit is restricted to conservative effects, neglecting the dissipative radiation reaction contributions. As we have shown, the intermediate steps in the two approaches are quite different and the agreement of the final results is very satisfying and confirms the gauge invariance of the stability criterion. 

As an important check, we have compared our 4PN criterion with numerical relativity results, following the analysis of Favata~\cite{F11a} at 3PN order. In particular, we find a 7\% relative agreement with the ISCO shift (due to the finite mass-ratio correction with respect to the Schwarzschild value) as computed by the first-order gravitational self-force program~\cite{BarackS09, LBB12, Akcay12}. Such good agreement between our 4PN result and the numerical ISCO shift may appear surprising since the PN expansion is not expected to be adequate in the regime where the binary system reaches the ISCO, however it confirms the efficiency of the PN expansion when considered in non-resummed Taylor expanded form, and used to predict directly observable (gauge invariant) quantities. For comparable masses we also find a rather good agreement with numerical relativity, of the order of 15\% relative. 

Finally, we derived a distinct effective-one-body (EOB) stability criterion by combining results from Refs.~\cite{D10sf} and~\cite{DJS15eob}, and found that it does not perform as well as our standard 4PN criterion when compared with the GSF result in the small mass ratio limit. Furthermore, in contrast with the NR results~\cite{CCGPf06}, the EOB  criterion at 4PN order provides no ISCO value in the strictly equal mass case.

\acknowledgments

We thank Thibault Damour and Nathalie Deruelle for the suggestion to look at the comparison with the 4PN EOB prediction which can be deduced from Refs.~\cite{D10sf,DJS15eob}.

\appendix

\section{Equations of motion and Hamiltonian at 4PN order}\label{Appendix: explicit expressions of A and B}

The functions $\mathcal{A}$ and $\mathcal{B}$ parametrize the instantaneous (non-tail) part of the equations of motion~\eqref{eq: EOM Lagrangian first form} in harmonic coordinates and in the center-of-mass frame. They are presented as a sequence of PN coefficients, together with a split in powers of $G$ at 4PN order:
\begin{subequations}\label{notationA}
	\begin{align}
	\mathcal{A} &= \mathcal{A}_\text{N} +
	\frac{1}{c^2}\mathcal{A}_\text{1PN} +
	\frac{1}{c^4}\mathcal{A}_\text{2PN} +
	\frac{1}{c^6}\mathcal{A}_\text{3PN} +
	\frac{1}{c^8}\mathcal{A}_\text{4PN} + \calO\left(\frac{1}{c^{10}}\right)\,,\\
	\text{with}\quad\mathcal{A}_\text{4PN} &= \mathcal{A}^{(0)}_\text{4PN} + G\,\mathcal{A}^{(1)}_\text{4PN} + G^2 \mathcal{A}^{(2)}_\text{4PN} + G^3 \mathcal{A}^{(3)}_\text{4PN}+ G^4 \mathcal{A}^{(4)}_\text{4PN}+ G^5 \mathcal{A}^{(5)}_\text{4PN} \,,
	\end{align}
\end{subequations}
and similarly for $\mathcal{B}$. Neglecting the dissipative 2.5PN and 3.5PN contributions we have
\begin{subequations}\label{calABcoeff}
	\begin{align}
		\mathcal{A}_\text{1PN} &= -\frac{3\,\dot{r}^2\,\nu}{2} + v^2 +
		3\,\nu\,v^2-\frac{G m}{r}\left(4 +2\,\nu \right) \,,\\
		\mathcal{A}_\text{2PN} &= \frac{15\,\dot{r}^4\,\nu}{8} -
		\frac{45\,\dot{r}^4\,\nu^2}{8} - \frac{9\,\dot{r}^2\,\nu\,v^2}{2} +
		6\,\dot{r}^2\,\nu^2\,v^2 + 3\,\nu\,v^4 -
		4\,\nu^2\,v^4 \nn\\ & + \frac{G m}{r}\left(
		-2\,\dot{r}^2 - 25\,\dot{r}^2\,\nu - 2\,\dot{r}^2\,\nu^2 -
		\frac{13\,\nu\,v^2}{2} + 2\,\nu^2\,v^2 \right) + \frac{G^2m^2}{r^2}\,\left( 9 + \frac{87\,\nu}{4}
		\right)\,,\\
		\mathcal{A}_\text{3PN} &= -\frac{35\,\dot{r}^6\,\nu}{16} +
		\frac{175\,\dot{r}^6\,\nu^2}{16} -
		\frac{175\,\dot{r}^6\,\nu^3}{16}+\frac{15\,\dot{r}^4\,\nu\,v^2}{2}
		- \frac{135\,\dot{r}^4\,\nu^2\,v^2}{4} +
		\frac{255\,\dot{r}^4\,\nu^3\,v^2}{8} \nn\\& -
		\frac{15\,\dot{r}^2\,\nu\,v^4}{2} +
		\frac{237\,\dot{r}^2\,\nu^2\,v^4}{8} -\frac{45\,\dot{r}^2\,\nu^3\,v^4}{2} + \frac{11\,\nu\,v^6}{4} -
		\frac{49\,\nu^2\,v^6}{4} + 13\,\nu^3\,v^6 \nn\\ & +
		\frac{G m}{r}\left( 79\,\dot{r}^4\,\nu -
		\frac{69\,\dot{r}^4\,\nu^2}{2} - 30\,\dot{r}^4\,\nu^3 -
		121\,\dot{r}^2\,\nu\,v^2 + 16\,\dot{r}^2\,\nu^2\,v^2 +
		20\,\dot{r}^2\,\nu^3\,v^2+\frac{75\,\nu\,v^4}{4} + 8\,\nu^2\,v^4 -
		10\,\nu^3\,v^4 \right) \nn\\ & +
		\frac{G^2m^2}{r^2}\,\left( \dot{r}^2 +
		\frac{32573\,\dot{r}^2\,\nu}{168} + \frac{11\,\dot{r}^2\,\nu^2}{8} -
		7\,\dot{r}^2\,\nu^3 + \frac{615\,\dot{r}^2\,\nu\,\pi^2}{64} -
		\frac{26987\,\nu\,v^2}{840} + \nu^3\,v^2 - \frac{123\,\nu\,\pi^2\,v^2}{64} \right.\nn\\&\quad\quad\quad
		\left. -
		110\,\dot{r}^2\,\nu\,\ln \Big(\frac{r}{r'_0}\Big) + 22\,\nu\,v^2\,\ln
		\Big(\frac{r}{r'_0}\Big) \right) +\frac{G^3m^3}{r^3}\left( -16 - \frac{437\,\nu}{4} -
		\frac{71\,\nu^2}{2} + \frac{41\,\nu\,{\pi }^2}{16} \right)\,,\\
		\mathcal{A}^{(0)}_\text{4PN} &= \left(\frac{315}{128} \nu -  \frac{2205}{128} \nu^2 +
		\frac{2205}{64} \nu^3 -  \frac{2205}{128} \nu^4\right)
		\dot{r}^8 + \left(- \frac{175}{16} \nu + \frac{595}{8}
		\nu^2 -  \frac{2415}{16} \nu^3  + \frac{735}{8} \nu^4\right) \dot{r}^6 v^{2} \nn\\& +
		\left(\frac{135}{8} \nu -  \frac{1875}{16} \nu^2 + \frac{4035}{16} \nu^3 -
		\frac{1335}{8} \nu^4\right) \dot{r}^4 v^{4} + \left(- \frac{21}{2} \nu +
		\frac{1191}{16} \nu^2 -  \frac{327}{2} \nu^3 + 99 \nu^4\right) \dot{r}^2 v^{6} \nn\\&
		+
		\left(\frac{21}{8} \nu -  \frac{175}{8} \nu^2 + 61 \nu^3  - 54 \nu^4\right)
		v^{8}
		\,,\\
		\mathcal{A}^{(1)}_\text{4PN} &= \frac{m}{r} \biggl(\frac{2973}{40} \nu \dot{r}^6
		+ 407 \nu^2 \dot{r}^6+ \frac{181}{2} \nu^3 \dot{r}^6 - 86 \nu^4 \dot{r}^6 
		+ \frac{1497}{32} \nu \dot{r}^4 v^{2}
		-  \frac{1627}{2} \nu^2 \dot{r}^4 v^{2} - 81 \nu^3 \dot{r}^4 v^{2} + 228 \nu^4 \dot{r}^4 v^{2} 
		\nn\\&\quad\quad\quad -  \frac{2583}{16} \nu \dot{r}^2 v^{4} + \frac{1009}{2} \nu^2 \dot{r}^2 v^{4} 
		+ 47 \nu^3 \dot{r}^2 v^{4} - 104 \nu^4 \dot{r}^2 v^{4} + \frac{1067}{32} \nu v^{6} - 58 \nu^2 v^{6} - 44 \nu^3 v^{6} 
		+ 58 \nu^4 v^{6}\biggr)
		\,,\\
		\mathcal{A}^{(2)}_\text{4PN} &= \frac{m^2}{r^2} \left(\frac{2094751}{960} \nu \dot{r}^4
		+ \frac{45255}{1024} \pi^2 \nu \dot{r}^4
		+ \frac{326101}{96} \nu^2 \dot{r}^4
		-  \frac{4305}{128} \pi^2 \nu^2 \dot{r}^4
		-  \frac{1959}{32} \nu^3 \dot{r}^4 - 126 \nu^4 \dot{r}^4
		\right.\nn\\&\quad\quad\quad\left. - 1155 \nu^2 \ln\Big(\frac{r}{r'_{0}}\Big) \dot{r}^4
		+ 385 \nu \ln\Big(\frac{r}{r'_{0}}\Big) \dot{r}^4
		-  \frac{1636681}{1120} \nu \dot{r}^2 v^{2}  -  \frac{12585}{512} \pi^2 \nu \dot{r}^2 v^{2} -  \frac{255461}{112} \nu^2 \dot{r}^2 v^{2}
		\right.\nn\\&\quad\quad\quad\left.+ \frac{3075}{128} \pi^2 \nu^2 \dot{r}^2 v^{2}
		-  \frac{309}{4} \nu^3 \dot{r}^2 v^{2}
		+ 63 \nu^4 \dot{r}^2 v^{2} - 605 \nu \ln\Big(\frac{r}{r'_{0}}\Big) \dot{r}^2 v^{2}
		+ 825 \nu^2 \ln\Big(\frac{r}{r'_{0}}\Big) \dot{r}^2 v^{2}
		+ \frac{1096941}{11200} \nu v^{4}
		\right.\nn\\&\quad\quad\quad\left.+ \frac{1155}{1024} \pi^2 \nu v^{4}
		+ \frac{7263}{70} \nu^2 v^{4}
		-  \frac{123}{64} \pi^2 \nu^2 v^{4}
		+ \frac{145}{2} \nu^3 v^{4}
		- 16 \nu^4 v^{4} + 88 \nu \ln\Big(\frac{r}{r'_{0}}\Big) v^{4} - 66 \nu^2 \ln\Big(\frac{r}{r'_{0}}\Big) v^{4}
		\right) \,,\\
		\mathcal{A}^{(3)}_\text{4PN} &= \frac{m^3}{r^3} \left(-2 \dot{r}^2
		+ \frac{1297943}{8400} \nu \dot{r}^2
		-  \frac{2969}{16} \pi^2 \nu \dot{r}^2
		+ \frac{1255151}{840} \nu^2 \dot{r}^2
		+ \frac{7095}{32} \pi^2 \nu^2 \dot{r}^2
		- 17 \nu^3 \dot{r}^2 - 24 \nu^4 \dot{r}^2
		\right.\nn\\
		&\quad\quad\quad\left. +2232 \nu \ln\Big(\frac{r}{r'_{0}}\Big) \dot{r}^2
		- 3364 \nu^2 \ln\Big(\frac{r}{r'_{0}}\Big) \dot{r}^2
		+ \frac{1237279}{25200} \nu v^{2}
		+ \frac{3835}{96} \pi^2 \nu v^{2}
		-  \frac{693947}{2520} \nu^2 v^{2} -  \frac{229}{8} \pi^2 \nu^2 v^{2}
		\right.\nn\\
		&\quad\quad\quad\left. + \frac{19}{2} \nu^3 v^{2}
		- \frac{1336}{3} \nu \ln\Big(\frac{r}{r'_{0}}\Big) v^{2}
		+  \frac{1616}{3} \nu^2 \ln\Big(\frac{r}{r'_{0}}\Big) v^{2}\right) \,,\\
		\mathcal{A}^{(4)}_\text{4PN} &= \frac{m^4}{r^4} \left(25
		+ \frac{6625537}{12600} \nu
		-  \frac{4543}{96} \pi^2 \nu
		+ \frac{477763}{720} \nu^2
		+ \frac{3}{4} \pi^2 \nu^2 + \frac{334}{3} \nu \ln\Big(\frac{r}{r'_{0}}\Big)  - \frac{514}{3} \nu^2 \ln\Big(\frac{r}{r'_{0}}\Big)
		\right)\,,\\
		\mathcal{B}_\text{1PN} &= -4\,\dot{r} + 2\,\dot{r}\,\nu\,,\\
		\mathcal{B}_\text{2PN} &= \frac{9\,\dot{r}^3\,\nu}{2} +
		3\,\dot{r}^3\,\nu^2 -\frac{15\,\dot{r}\,\nu\,v^2}{2} -
		2\,\dot{r}\,\nu^2\,v^2 +
		\frac{G m}{r}\left( 2\,\dot{r} + \frac{41\,\dot{r}\,\nu}{2} +
		4\,\dot{r}\,\nu^2 \right)\,,\\
		\mathcal{B}_\text{3PN} &= -\frac{45\,\dot{r}^5\,\nu}{8} +
		15\,\dot{r}^5\,\nu^2 + \frac{15\,\dot{r}^5\,\nu^3}{4} +
		12\,\dot{r}^3\,\nu\,v^2 - \frac{111\,\dot{r}^3\,\nu^2\,v^2}{4}
		-12\,\dot{r}^3\,\nu^3\,v^2 -\frac{65\,\dot{r}\,\nu\,v^4}{8} +
		19\,\dot{r}\,\nu^2\,v^4 + 6\,\dot{r}\,\nu^3\,v^4
		\nn\\& + \frac{G m}{r}\left(
		\frac{329\,\dot{r}^3\,\nu}{6} + \frac{59\,\dot{r}^3\,\nu^2}{2} +
		18\,\dot{r}^3\,\nu^3 - 15\,\dot{r}\,\nu\,v^2 - 27\,\dot{r}\,\nu^2\,v^2
		- 10\,\dot{r}\,\nu^3\,v^2 \right) \nn\\&
		+ \frac{G^2m^2}{r^2}\,\left( -4\,\dot{r} -
		\frac{18169\,\dot{r}\,\nu}{840} + 25\,\dot{r}\,\nu^2 +
		8\,\dot{r}\,\nu^3 - \frac{123\,\dot{r}\,\nu\,\pi^2}{32} + 44\,\dot{r}\,\nu\,\ln \Big(\frac{r}{r'_0}\Big)
		\right)\,,\\
		\mathcal{B}^{(0)}_\text{4PN} &= \left(\frac{105}{16} \nu
		-  \frac{245}{8} \nu^2
		+ \frac{385}{16} \nu^3
		+ \frac{35}{8} \nu^4\right) \dot{r}^7
		+ \left(- \frac{165}{8} \nu
		+ \frac{1665}{16} \nu^2
		-  \frac{1725}{16} \nu^3
		-  \frac{105}{4} \nu^4\right) \dot{r}^5 v^{2}\nn\\
		& + \left(\frac{45}{2} \nu
		-  \frac{1869}{16} \nu^2
		+ 129 \nu^3
		+ 54 \nu^4\right) \dot{r}^3 v^{4}
		+ \left(- \frac{157}{16} \nu
		+ 54 \nu^2
		- 69 \nu^3
		- 24 \nu^4\right) \dot{r} v^{6}\,,\\
		\mathcal{B}^{(1)}_\text{4PN} &= \frac{m}{r} \left(- \frac{54319}{160} \nu \dot{r}^5
		-  \frac{901}{8} \nu^2 \dot{r}^5
		+ 60 \nu^3 \dot{r}^5
		+ 30 \nu^4 \dot{r}^5
		+ \frac{25943}{48} \nu \dot{r}^3 v^{2}
		\right.\nn\\&\quad\quad\quad\left.+ \frac{1199}{12} \nu^2 \dot{r}^3 v^{2}  -  \frac{349}{2} \nu^3 \dot{r}^3 v^{2}
		- 98 \nu^4 \dot{r}^3 v^{2}
		-  \frac{5725}{32} \nu \dot{r} v^{4} -  \frac{389}{8} \nu^2 \dot{r} v^{4}
		+ 118 \nu^3 \dot{r} v^{4}
		+ 44 \nu^4 \dot{r} v^{4}\right)
		\,,\\
		\mathcal{B}^{(2)}_\text{4PN} &= \frac{m^2}{r^2} \left(- \frac{9130111}{3360} \nu \dot{r}^3
		-  \frac{4695}{256} \pi^2 \nu \dot{r}^3
		-  \frac{184613}{112} \nu^2 \dot{r}^3
		+ \frac{1845}{64} \pi^2 \nu^2 \dot{r}^3
		+ \frac{209}{2} \nu^3 \dot{r}^3  + 74 \nu^4 \dot{r}^3
		\right.\nn\\&\quad\quad\quad\left. + 440 \nu \ln\Big(\frac{r}{r'_{0}}\Big) \dot{r}^3
		+ 550 \nu^2 \ln\Big(\frac{r}{r'_{0}}\Big) \dot{r}^3 + \frac{8692601}{5600} \nu \dot{r} v^{2}
		+ \frac{1455}{256} \pi^2 \nu \dot{r} v^{2}
		+ \frac{58557}{70} \nu^2 \dot{r} v^{2}
		-  \frac{123}{8} \pi^2 \nu^2 \dot{r} v^{2}
		\right.\nn\\
		& \quad\quad\quad\left.- 70 \nu^3 \dot{r} v^{2}
		- 34 \nu^4 \dot{r} v^{2} - 154 \nu \ln\Big(\frac{r}{r'_{0}}\Big) \dot{r} v^{2}
		- 264 \nu^2 \ln\Big(\frac{r}{r'_{0}}\Big) \dot{r} v^{2}
		\right) \,,\\
		\mathcal{B}^{(3)}_\text{4PN} &= \frac{m^3}{r^3} \left(2
		-  \frac{619267}{525} \nu
		+ \frac{791}{16} \pi^2 \nu
		-  \frac{28406}{45} \nu^2
		-  \frac{2201}{32} \pi^2 \nu^2 + 66 \nu^3
		+ 16 \nu^4 - \frac{1484}{3} \nu \ln\Big(\frac{r}{r'_{0}}\Big)
		+ 1268 \nu^2 \ln\Big(\frac{r}{r'_{0}}\Big)
		\right) \dot{r}\,.
\end{align}\end{subequations}

The instantaneous (non-tail) part of the reduced 4PN Hamiltonian $\mathcal{H}=H/\mu$, where we denote $P^2\equiv\bm{P}^2$ and $P_R\equiv\bm{N}\cdot\bm{P}$ with capital letters referring to ADM coordinates, reads, following the same convention as in~\eqref{notationA},
\begin{subequations}\label{HADMcm}
	\begin{align}
		\mathcal{H}_\text{N} &= \frac{P^2}{2} -\frac{G m}{R}\,,\\ 
		\mathcal{H}_\text{1PN} &=
		\left(- \frac{1}{8} +
		\frac{3\,\nu}{8}\right)P^4 + \frac{G m}{R}\left( - \frac{{P_R}^2\,\nu}{2} -
		\frac{3\,P^2}{2} - \frac{\nu\,P^2}{2} \right)+\frac{G^2m^2}{2R^2} \,,\\ 
		\mathcal{H}_\text{2PN} &=
		\left(\frac{1}{16} -
		\frac{5\nu}{16} + \frac{5\nu^2}{16}\right)P^6
		\nn\\& + \frac{G m}{R}\left( - \frac{3\,{P_R}^4\,\nu^2}{8}
		- \frac{{P_R}^2\,P^2\,\nu^2}{4} + \frac{5\,P^4}{8} -
		\frac{5\,\nu\,P^4}{2} - \frac{3\,\nu^2\,P^4}{8}
		\right)\nn\\& + \frac{G^2m^2}{R^2}\,\left(
		\frac{3\,{P_R}^2\,\nu}{2} + \frac{5\,P^2}{2} + 4\,\nu\,P^2 \right)
		+\frac{G^3m^3}{R^3}\left( -\frac{1}{4} -
		\frac{3\,\nu}{4} \right)\,, \\ 
		\mathcal{H}_\text{3PN} &=
		\left(-\frac{5}{128} + \frac{35\nu}{128} -
		\frac{35\nu^2}{64} + \frac{35\nu^3}{128}\right)P^8
		\nn\\& + \frac{G m}{R}\left(
		-\frac{5\,{P_R}^6\,\nu^3}{16} + \frac{3\,{P_R}^4\,P^2\,\nu^2}{16} -
		\frac{3\,{P_R}^4\,P^2\,\nu^3}{16} + \frac{{P_R}^2\,P^4\,\nu^2}{8}
		\right.\nn\\&\quad\qquad\quad~ \left. -
		\frac{3\,{P_R}^2\,P^4\,\nu^3}{16}-\frac{7\,P^6}{16} +
		\frac{21\,\nu\,P^6}{8} - \frac{53\,\nu^2\,P^6}{16} -
		\frac{5\,\nu^3\,P^6}{16} \right) \nn\\& +
		\frac{G^2m^2}{R^2}\,\left( \frac{5\,{P_R}^4\,\nu}{12} +
		\frac{43\,{P_R}^4\,\nu^2}{12} + \frac{17\,{P_R}^2\,P^2\,\nu}{16}
		+
		\frac{15\,{P_R}^2\,P^2\,\nu^2}{8} - \frac{27\,P^4}{16} +
		\frac{17\,\nu\,P^4}{2} + \frac{109\,\nu^2\,P^4}{16} \right)
		\nn\\& + \frac{G^3m^3}{R^3}\,\left(
		-\frac{85\,{P_R}^2\,\nu}{16} - \frac{7\,{P_R}^2\,\nu^2}{4} -
		\frac{25\,P^2}{8} - \frac{335\,\nu\,P^2}{48}
		- \frac{23\,\nu^2\,P^2}{8}
		- \frac{3\,{P_R}^2\,\nu\,\pi^2}{64} + \frac{\nu\,P^2\,\pi^2}{64}
		\right)\nn\\& + \frac{G^4m^4}{R^4}\, \left( \frac{1}{8} +
		\frac{109\,\nu}{12} - \frac{21\,\nu\,\pi^2}{32} \right) \,,\\
		\mathcal{H}^{(0)}_\text{4PN} &= \left(
		\frac{7}{256}
		-\frac{63}{256}\nu
		+\frac{189}{256}\nu^2
		-\frac{105}{128}\nu^3
		+\frac{63}{256}\nu^4
		\right)P^{10} \,,\\ 
		\mathcal{H}^{(1)}_\text{4PN} &=
		\frac{m}{R}\Biggl\{
		\frac{45}{128} P^8
		-\frac{45}{16} P^8\nu
		+\left(
		\frac{423}{64} P^8
		-\frac{3}{32} P_R^2 P^6
		-\frac{9}{64} P_R^4 P^4
		\right)\nu^2
		\nn\\ &
		\quad + \left(
		-\frac{1013}{256} P^8
		+\frac{23}{64} P_R^2 P^6
		+\frac{69}{128} P_R^4 P^4
		-\frac{5}{64} P_R^6 P^2
		+\frac{35}{256} P_R^8
		\right)\nu^3
		\nn\\&
		\quad + \left(
		-\frac{35}{128} P^8
		-\frac{5}{32} P_R^2 P^6
		-\frac{9}{64} P_R^4 P^4
		-\frac{5}{32} P_R^6 P^2
		-\frac{35}{128} P_R^8
		\right)\nu^4
		\Biggr\}\,,\\
		\mathcal{H}^{(2)}_\text{4PN} &=
		\frac{m^2}{R^2}\Biggl\{
		\frac{13}{8} P^6
		+ \left(
		-\frac{791}{64}P^6
		+\frac{49}{16} P_R^2 P^4
		-\frac{889}{192} P_R^4 P^2
		+\frac{369}{160} P_R^6
		\right)\nu
		\nn\\&
		\quad + \left(
		\frac{4857}{256} P^6
		-\frac{545}{64} P_R^2 P^4
		+\frac{9475}{768} P_R^4 P^2
		-\frac{1151}{128} P_R^6
		\right)\nu^2
		\nn\\&
		\quad + \left(
		\frac{2335}{256} P^6
		+\frac{1135}{256} P_R^2 P^4
		-\frac{1649}{768} P_R^4 P^2
		+\frac{10353}{1280} P_R^6
		\right)\nu^3
		\Biggr\}\,,\\ 
		\mathcal{H}^{(3)}_\text{4PN} &=
		\frac{m^3}{R^3}\Biggl\{ \frac{105}{32} P^4
		\nn\\&\quad+ \left[ \left(\frac{2749}{8192}\pi^2-\frac{589189}{19200}\right) P^4
		+ \left(\frac{63347}{1600} - \frac{1059}{1024}\pi^2\right) P_R^2 P^2 
		+ \left(\frac{375}{8192}\pi^2-\frac{23533}{1280}\right) P_R^4 \right]\nu
		\nn\\&\quad + \bigg[ \left(\frac{18491}{16384}\pi^2 - \frac{1189789}{28800}\right) P^4
		- \left(\frac{127}{3} + \frac{4035}{2048}\pi^2\right) P_R^2 P^2
		+ \left(\frac{57563}{1920} - \frac{38655}{16384}\pi^2 \right) P_R^4
		\bigg]\nu^2
		\nn\\&\quad
		+ \bigg(
		-\frac{553}{128} P^4
		-\frac{225}{64} P_R^2 P^2
		-\frac{381}{128} P_R^4
		\bigg)\nu^3
		\Biggr\}\,,\\
		\mathcal{H}^{(4)}_\text{4PN} &=
		\frac{m^4}{R^4}\Biggl\{
		\frac{105}{32}P^2
		+ \left[  \left(\frac{185761}{19200} - \frac{21837}{8192}\pi^2\right) P^2
		+ \left(\frac{3401779}{57600} - \frac{28691}{24576}\pi^2\right) P_R^2 \right]\nu
		\nn\\[1ex]&\quad
		+ \left[ \left(\frac{672811}{19200} - \frac{158177}{49152}\pi^2\right) P^2
		+ \left(-\frac{21827}{3840} + \frac{110099}{49152}\pi^2\right) P_R^2 \right]\nu^2
		\Biggr\}\,,\nn\\
		\mathcal{H}^{(5)}_\text{4PN} &=
		\frac{m^5}{R^5}\Biggl\{
		-\frac{1}{16}
		+ \left({-\frac{169199}{2400} + \frac{6237}{1024}\pi^2}\right) \, \nu
		+ \left(-\frac{1256}{45} + \frac{7403}{3072}\pi^2\right)\,\nu^2
		\Biggr\}\,,
	\end{align}
\end{subequations}

\section{Fourier decomposition of the quadrupole moment}\label{Appendix : quadrupole Fourier coefficients}

We provide the expressions of the discrete Fourier coefficients of the trace-free Newtonian quadrupole moment in terms of combinations of Bessel functions. They are defined by
\begin{equation}\label{Fourier}
	Q_{ij}(t) =
	\sum_{p=-\infty}^{+\infty}\,\mathop{Q}_{p}{}_{\!\!ij}\,\de^{\di
		\,p\,\ell} \quad\text{with}\quad \mathop{Q}_{p}{}_{\!\!ij}=
	\int_0^{2\pi}\frac{\dd\ell}{2\pi}\,Q_{ij}\,\de^{-\di\, p\,\ell}\,,
\end{equation}
where $\ell=n(t-T)$ is the mean anomaly of the binary motion, with $n=2\pi/P$ the orbital frequency (or mean motion) corresponding to the orbital period $P$, and $T$ is the instant of passage at periastron. The Fourier coefficients ${}_{p}{Q}_{ij}$ are functions of the orbit's eccentricity $e$ and semi-major axis $a$ to Newtonian order (we have $n=\sqrt{G m/a^3}$) and satisfy ${}_{-p}{Q}_{ij}={}_{p}{\overline{Q}}_{ij}$, with the overbar denoting the complex conjugation. Following the Appendix A of~\cite{ABIQ08tail} they read in a Cartesian basis $(x,y,z)$:
\begin{subequations}\label{IabFourier1}
\begin{align}
\mathop{Q}_{p}{}_{\!\!xx} &= m \nu a^2 \biggl\{ {\left(\frac{1}{6}+\frac{3}{2}
	e^2\right) J_p\left(p e\right)} {+\left(-\frac{7}{8}
	e-\frac{3}{8} e^3\right) \left(J_{p-1}\left(p
	e\right)+J_{p+1}\left(p e\right)\right)}\nn\\&\quad +\left(\frac{1}{4}+\frac{1}{4} e^2\right) \left(J_{p-2}\left(p
	e\right)+J_{p+2}\left(p e\right)\right) +\left(-\frac{1}{8} e+\frac{1}{24} e^3\right)
	\left(J_{p-3}\left(p e\right)+J_{p+3}\left(p e\right)\right) \biggr\}\,,\\
\mathop{Q}_{p}{}_{\!\!xy} &= m \nu a^2 \di
	\sqrt{1-e^2}\left\{\frac{5}{8} e \left(-J_{p-1}\left(p
	e\right)+J_{p+1}\left(p e\right)\right)\right.\nn\\&\quad
	 +\left(-\frac{1}{4}-\frac{1}{4} e^2\right) \left(J_{p+2}\left(p
	e\right)-J_{p-2}\left(p e\right)\right) 
 \left.+\frac{1}{8} e \left(J_{p+3}\left(p
	e\right)-J_{p-3}\left(p e\right)\right)\right\}\,,\\
\mathop{Q}_{p}{}_{\!\!yy} &= m \nu a^2 \biggl\{{\left(\frac{1}{6}-e^2\right)
	J_p\left(p e\right)} +\left(\frac{3}{8}
	e+\frac{1}{4} e^3\right) \left(J_{p-1}\left(p
	e\right)+J_{p+1}\left(p e\right)\right)\nn\\&\quad 
    -\frac{1}{4} \left(J_{p-2}\left(p e\right)+J_{p+2}\left(p
	e\right)\right) +\left(\frac{1}{8} e-\frac{1}{12}
	e^3\right)
	\left(J_{p-3}\left(p e\right)+J_{p+3}\left(p e\right)\right) \biggr\}\,,\\
\mathop{Q}_{p}{}_{\!\!zz} &= m \nu a^2 \biggl\{\left(-\frac{1}{3}-\frac{1}{2}
	e^2\right) J_p\left(p e\right) +\left(\frac{1}{2}
	e+\frac{1}{8} e^3\right) \left(J_{p-1}\left(p
	e\right)+J_{p+1}\left(p e\right)\right)\nn\\&\quad
    -\frac{1}{4} e^2 \left(J_{p-2}\left(p e\right)+J_{p+2}\left(p
	e\right)\right) +\frac{1}{24} e^3
	\left(J_{p-3}\left(p e\right)+J_{p+3}\left(p e\right)\right)\biggr\}\,.
\end{align}
\end{subequations}
We present also some alternative expressions (following the Appendix B of~\cite{BBBFMb}), based on the decomposition
\begin{equation}\label{Fourierdecompose}
	\mathop{Q}_{p}{}_{\!\!ij} = \mathop{\mathcal{A}}_{p}
	m_0^{i}m_0^{j} + \mathop{\mathcal{B}}_{p}
	\,\overline{m}_0^{i}\overline{m}_0^{j} + \mathop{\mathcal{C}}_{p}
	\ell^{\langle i}\ell^{j\rangle}\,,
\end{equation}
where we denote $\bm{m}_0=\frac{1}{\sqrt{2}}(\bm{i}+\di\,\bm{j})$ and $\overline{\bm{m}}_0=\frac{1}{\sqrt{2}}(\bm{i}-\di\,\bm{j})$, with $\bm{i}$, $\bm{j}$ being two fixed orthonormal basis vectors in the orbital plane, $\bm{i}$ pointing towards the orbit's periastron, and $\bm{\ell}$ perpendicular to the orbit. With this decomposition the coefficients are, for $p\not= 0$:
\begin{subequations}\label{IabFourier2}
	\begin{align}
		\mathop{\mathcal{A}}_{p} &= \frac{m\nu\,a^2}{p^2 e^2}\biggl\{ 2 e
		J_{p-1}\left(p e\right)\left( p
		(1-e^2)+\sqrt{1-e^2}\right)\nn\\& \qquad\qquad + J_{p}\left(p
		e\right)\left(-2(p+1)\left(1+\sqrt{1-e^2}\right) +
		e^2\left[1+2p\left(1+\sqrt{1-e^2}\right)\right]\right)\biggr\}\,,\\
		\mathop{\mathcal{B}}_{p} &= \frac{m\nu\,a^2}{p^2 e^2}\biggl\{ 2 e
		J_{p-1}\left(p e\right)\left( p
		(1-e^2)-\sqrt{1-e^2}\right)\nn\\& \qquad\qquad + J_{p}\left(p
		e\right)\left(-2(p+1)\left(1-\sqrt{1-e^2}\right) +
		e^2\left[1+2p\left(1-\sqrt{1-e^2}\right)\right]\right)\biggr\}\,,\\
		\mathop{\mathcal{C}}_{p} &= \frac{m\nu\,a^2}{p^2} J_{p}\left(p e\right)\,,
	\end{align}
\end{subequations}
and, when $p=0$:
\begin{subequations}\label{casp0}
	\begin{align}
		\mathop{\mathcal{A}}_{0} = \mathop{\mathcal{B}}_{0} =
		\frac{5}{4}m\nu\,a^2\,e^2\,,\qquad\mathop{\mathcal{C}}_{0} =
		-\frac{1}{2}m\nu\,a^2\left(1+\frac{3}{2}e^2\right)\,.
	\end{align}
\end{subequations}
The divergence of the coefficients~\eqref{IabFourier2} when $e\to 0$ is only apparent and the two decompositions~\eqref{IabFourier1} and~\eqref{IabFourier2} are equivalent. Note that for circular orbits we have (with other modes being zero):
\begin{subequations}
\begin{align}
	&\mathop{Q}_{2}{}_{\!\!xx} =
	\mathop{Q}_{-2}{}_{\!\!xx} = \frac{1}{4}m\,\nu \,a^2\,,
	&\mathop{Q}_{2}{}_{\!\!yy} =
	\mathop{Q}_{-2}{}_{\!\!yy} = -\frac{1}{4}m\,\nu
	\,a^2\,,\\&\mathop{Q}_{2}{}_{\!\!xy} =
	-\mathop{Q}_{-2}{}_{\!\!xy} = -\frac{\di}{4}m\,\nu a^2\,,
	&\mathop{Q}_{0}{}_{\!\!xx} =
	\mathop{Q}_{0}{}_{\!\!yy} =
	-\frac{1}{2}\mathop{Q}_{0}{}_{\!\!zz} = \frac{1}{6}m\,\nu
	\,a^2\,.
\end{align}
\end{subequations}
or, equivalently,
\begin{align}
\label{eq: Fourier coefficients circular case}
	& \mathop{Q}_{2}{}_{\!\!ij}  = \frac{m \nu a^2}{2}
	\overline{m}_0^{i}\overline{m}_0^{j}\,, & &
	\mathop{Q}_{-2}{}_{\!\!ij}  = \frac{m \nu a^2}{2}
	m_0^{i}m_0^{j}\,, & & \mathop{Q}_{0}{}_{\!\!ij} 
	=-\frac{m \nu a^2}{2} \ell^{\langle i}\ell^{j\rangle}.
\end{align}

\section{Results for the transformation from harmonic to ADM coordinates}
\label{app:contact-transf}

The contact transformation between the 4PN equations of motion in harmonic coordinates and the 4PN Hamiltonian in ADM coordinates (both without the tail term) is defined by~\eqref{delta_Y}, \textit{i.e.}, for two bodies
\begin{equation}\label{delta_Y_1}
	\delta Y_{1}^{i} = \frac{1}{m_1} \left( q_{1}^{i} + \frac{\partial F}{\partial v_{1}^{i}} + \calX_{1}^{i} + a_1^j \calY_{11}^{ji} + a_2^j \calY_{21}^{ji}\right) \,,
\end{equation}
where $q_{1}^{i}$ is the conjugate momentum of the velocity 1, $F$ is a function of positions and velocities fully specified by the match to ADM coordinates, and the quantities $\calX_{1}^{i}$, $\calY_{11}^{ji}$ and $\calY_{21}^{ji}$ (and $1\leftrightarrow 2$) are uniquely determined by the choice of $F$. We have (extending~\cite{ABF01} to 4PN order) 
\begin{subequations}
\begin{align}
     F &=\frac{1}{c^4}\, F_{\rm 2PN}+\frac{1}{c^6}\,F_{\rm 3PN} + \frac{1}{c^8}\,F_{\rm 4PN} + \mathcal{O}\biggl(\frac{1}{c^{10}}\biggr)\,,\\
      F_{\rm 2PN} &= \frac{G m_1 m_2}{4} (n_{12} v_2) v_1^2 +  \frac{G^2 m_1^2 m_2}{r_{12}} \bigg(\frac{7}{4} (n_{12} v_1) - \frac{1}{4} (n_{12} v_2) \bigg) + (1 \leftrightarrow 2)\,,\\
F_{\rm 3PN} &= G m_1 m_2 \bigg(
    -\frac{1}{16} (n_{12} v_1) (n_{12} v_2)^2 v_1^2
    - \frac{5}{24} (n_{12} v_2)^3 v_1^2
    - \frac{1}{2} (n_{12} v_2) v_1^4
    + \frac{1}{8} (n_{12} v_2) v_1^2 (v_1 v_2)
    + \frac{5}{16} (n_{12} v_1) v_1^2 v_2^2
    \bigg)
    \nonumber \\
& + \frac{G^2 m_1^2 m_2}{r_{12}} \bigg(
    -\frac{91}{144} (n_{12} v_1)^3
    + \frac{21}{16} (n_{12} v_1)^2 (n_{12} v_2)
    - \frac{113}{24} (n_{12} v_1) v_1^2
    + \frac{35}{8} (n_{12} v_2) v_1^2  + \frac{195}{16} (n_{12} v_1) (v_1 v_2)
    \nonumber \\
&- \frac{3}{4} (n_{12} v_1) v_2^2
    - \frac{1}{8} (n_{12} v_2) v_2^2
    \bigg) + \frac{G^3 m_1^2 m_2^2}{r_{12}^2} \bigg(
    \frac{245}{18} (n_{12} v_1)
    - \frac{21}{32} \pi^2 (n_{12} v_1)
    \bigg)
    \nonumber \\
& + \frac{G^3 m_1^3 m_2}{r_{12}^2} \bigg(
    -\frac{25867}{2520} (n_{12} v_1)
    - \frac{3}{4} (n_{12} v_2)
    + \frac{22}{3} (n_{12} v_1) \ln\bigg(\frac{r_{12}}{r_{0}'}\bigg)
    \bigg)
    + (1 \leftrightarrow 2)\,,\\
F_{\rm 4PN} &= \frac{G^3 m_1^2 m_2^2}{r_{12}^2} \bigg(
    - \frac{305033}{11520} (n_{12}v_1)^3
    + \frac{12691}{240} (n_{12}v_1)^2 (n_{12}v_2)
    - \frac{3373 \pi^2}{8192} (n_{12}v_1)^3 + \frac{27969 \pi^2}{16384} (n_{12}v_1)^2 (n_{12}v_2)
    \nonumber \\
&  + \frac{282589}{19200} (n_{12}v_1) v_1^2
    - \frac{846689}{11520} (n_{12}v_2) v_1^2
    - \frac{3941 \pi^2}{8192} (n_{12}v_1) v_1^2
    + \frac{21355 \pi^2}{16384} (n_{12}v_2) v_1^2
    - \frac{1172149}{57600} (n_{12}v_1) (v_1 v_2)
    \nonumber \\
&  + \frac{7511}{8192} (n_{12}v_1) (v_1 v_2)
    \bigg)
    \nonumber \\
& + \frac{G^2 m_1^2 m_2}{r_{12}} \bigg(
    \frac{23}{300} (n_{12} v_1)^5
    + \frac{577}{1280} (n_{12} v_1)^4 (n_{12} v_2)
    - \frac{27229}{5760} (n_{12} v_1)^3 (n_{12} v_2)^2 + \frac{7}{24} (n_{12} v_1)^2 (n_{12} v_2)^3
    \nonumber \\
& - \frac{2051}{1440} (n_{12} v_1)^3 v_1^2
    - \frac{11587}{1280} (n_{12} v_1)^2 (n_{12} v_2) v_1^2
    - \frac{5041}{960} (n_{12} v_2)^3 v_1^2  - \frac{8629}{1920} (n_{12} v_1) (n_{12} v_2)^2 v_1^2
    \nonumber \\
& - \frac{11479}{1920} (n_{12} v_1) v_1^4
    - \frac{699}{640} (n_{12} v_2) v_1^4
    - \frac{25069}{3840} (n_{12} v_1)^3 (v_1 v_2)  - \frac{401}{128} (n_{12} v_1)^2 (n_{12} v_2) (v_1 v_2) \nonumber \\
&  + \frac{49}{48} (n_{12} v_1) (n_{12} v_2)^2 (v_1 v_2)- \frac{86999}{3840} (n_{12} v_1) v_1^2 (v_1 v_2) - \frac{29801}{1920} (n_{12} v_2) v_1^2 (v_1 v_2) - \frac{8189}{480} (n_{12} v_1) (v_1 v_2)^2
    \nonumber \\
&    - \frac{7}{48} (n_{12} v_2) (v_1 v_2)^2
    + \frac{863}{2880} (n_{12} v_1)^3 v_2^2 - \frac{5}{32} (n_{12} v_1)^2 (n_{12} v_2) v_2^2
    - \frac{3587}{480} (n_{12} v_1) v_1^2 v_2^2
    \nonumber \\
& - \frac{39}{32} (n_{12} v_2) v_1^2 v_2^2
    + \frac{617}{96} (n_{12} v_1) (v_1 v_2) v_2^2- \frac{21}{16} (n_{12} v_1) v_2^4
    - \frac{3}{32} (n_{12} v_2) v_2^4
    \bigg)
    \nonumber \\
& + G m_1 m_2 \bigg(
    \frac{5}{256} (n_{12} v_1)^4 (n_{12} v_2)^3
    + \frac{15}{128} (n_{12} v_1)^2 (n_{12} v_2)^3 v_1^2
    + \frac{19}{256} (n_{12} v_1) (n_{12} v_2)^4 v_1^2
    \nonumber \\
& - \frac{3}{64} (n_{12} v_1) (n_{12} v_2)^2 v_1^4
    - \frac{15}{256} (n_{12} v_2)^3 v_1^4
    + \frac{5}{32} (n_{12} v_2) v_1^6
    - \frac{5}{64} (n_{12} v_1)^3 (n_{12} v_2)^2 (v_1 v_2)
    \nonumber \\
&  + \frac{25}{64} (n_{12} v_1) (n_{12} v_2)^2 v_1^2 (v_1 v_2)
    + \frac{37}{192} (n_{12} v_2)^3 v_1^2 (v_1 v_2)
    - \frac{21}{32} (n_{12} v_2) v_1^4 (v_1 v_2)
    \nonumber \\
&  - \frac{5}{64} (n_{12} v_1)^2 (n_{12} v_2) (v_1 v_2)^2
    + \frac{5}{64} (n_{12} v_2) v_1^2 (v_1 v_2)^2
    + \frac{1}{32} (n_{12} v_1) (v_1 v_2)^3
    + \frac{15}{128} (n_{12} v_1)^2 (n_{12} v_2) v_1^2 v_2^2
    \nonumber \\
& 
    + \frac{15}{64} (n_{12} v_1) v_1^4 v_2^2
    + \frac{59}{256} (n_{12} v_1) v_1^2 v_2^4
    - \frac{9}{64} (n_{12} v_1) v_1^2 (v_1 v_2) v_2^2
    \bigg)
    \nonumber \\
& + \frac{G^4 m_1^4 m_2}{r_{12}^3} \bigg(
    \frac{436183}{7200} (n_{12} v_1)
    - \frac{35}{32} (n_{12} v_2)
    - \frac{220}{3} (n_{12} v_1) \ln\bigg(\frac{r_{12}}{r_0'}\bigg)
    \bigg)
    \nonumber \\
& + \frac{G^4 m_1^3 m_2^2}{r_{12}^3} \bigg(
    - \frac{137287}{10080} (n_{12} v_1)
    + \frac{1287389}{12600} (n_{12} v_2)
    + \frac{11033 \pi^2}{6144} (n_{12} v_1)
    - \frac{6909 \pi^2}{1024} (n_{12} v_2)
    \nonumber \\
& 
    - \frac{22}{3} (n_{12} v_1) \ln\bigg(\frac{r_{12}}{r_0'}\bigg)
    - \frac{40}{3} (n_{12} v_2) \ln\bigg(\frac{r_{12}}{r_0'}\bigg)
    \bigg)
    \nonumber \\
& + \frac{G^3 m_1^3 m_2}{r_{12}^2} \bigg(
    - \frac{159299}{14400} (n_{12} v_1)^3
    + \frac{2089}{1200} (n_{12} v_1)^2 (n_{12} v_2)
    - \frac{99311}{4200} (n_{12} v_1) (n_{12} v_2)^2 - \frac{63803}{6720} (n_{12} v_1) v_1^2
    \nonumber \\
& 
  - \frac{456811}{16800} (n_{12} v_2) v_1^2
    - \frac{424937}{7200} (n_{12} v_1) (v_1 v_2)
    + \frac{15}{16} (n_{12} v_2) (v_1 v_2) - \frac{1}{2} (n_{12} v_1) v_2^2
    - \frac{7}{8} (n_{12} v_2) v_2^2
    \nonumber \\
&   - 11 (n_{12} v_1) (n_{12} v_2)^2 \ln\bigg(\frac{r_{12}}{r_0'}\bigg)- \frac{55}{3} (n_{12} v_1) v_1^2 \ln\bigg(\frac{r_{12}}{r_0'}\bigg)
    + 22 (n_{12} v_1) v_2^2 \ln\bigg(\frac{r_{12}}{r_0'}\bigg)
    \bigg)
\end{align}
\end{subequations}

\begin{subequations}
\begin{align}
         q_1^{i} &=\frac{1}{c^4}\,  q_{1,\rm 2PN}^{i}+\frac{1}{c^6}\,q_{1,\rm 3PN}^{i} + \frac{1}{c^8}\,q_{1,\rm 4PN}^{i} + \mathcal{O}\biggl(\frac{1}{c^{10}}\biggr)\,,\\
q_{1,\rm 2PN}^{i} &= G m_1 m_2 \bigg\{ \bigg(-\frac{1}{8} (n_{12} v_2)^2 + \frac{7}{8}  v_2^2 \bigg) n_{12}^{i}
    - \frac{7}{4} (n_{12} v_2) v_{2}^{i} \bigg\} \,,\\
q_{1,\rm 3PN}^{i} &= G m_1 m_2 \bigg\{ 
    \bigg(
        \frac{1}{8} (n_{12} v_1) (n_{12} v_2)^3
        + \frac{1}{16} (n_{12} v_2)^4
        + \frac{3}{8} (n_{12} v_2)^2 (v_1 v_2)
        - \frac{5}{8} (n_{12} v_1)^2 v_2^2  - \frac{1}{2} (n_{12} v_1) (n_{12} v_2) v_2^2
        \nonumber \\
& - \frac{5}{16} (n_{12} v_2)^2 v_2^2
    \bigg) n_{12}^{i} 
    + \bigg(
        \frac{11}{4} (n_{12} v_2) v_1^2
        - 2 (n_{12} v_2) (v_1 v_2)
        + \frac{15}{8} (n_{12} v_2) v_2^2
    \bigg) v_{1}^{i}
    \nonumber \\
&  + \bigg(
        \frac{3}{8} (n_{12} v_1) (n_{12} v_2)^2
        + \frac{5}{12} (n_{12} v_2)^3
        - (n_{12} v_2) v_1^2
        + \frac{1}{4} (n_{12} v_2) (v_1 v_2)
        - \frac{15}{8} (n_{12} v_2) v_2^2
    \bigg) v_{2}^{i} 
    \bigg\}
    \nonumber \\
& + \frac{G^2 m_1 m_2^2}{r_{12}} \bigg\{
    \bigg(
        \frac{29}{24} (n_{12} v_2)^2
        + \frac{235}{48} v_2^2
    \bigg) n_{12}^{i}
    + \frac{235}{24} (n_{12} v_2) v_{2}^{i}
    \bigg\}
    \nonumber \\
& + \frac{G^2 m_1^2 m_2}{r_{12}} \bigg\{
    \bigg(
        -\frac{17}{6} (n_{12} v_2)^2
        + \frac{185}{16} v_1^2
        - \frac{185}{8} (v_1 v_2)
        + \frac{20}{3} v_2^2
    \bigg) n_{12}^{i}
    - \frac{235}{24} (n_{12} v_2) v_{2}^{i}
    \bigg\} \,,\\
    q_{1,\rm 4PN}^{i} &= \frac{G^3 m_1^2 m_2^2}{r_{12}^2} \bigg\{ 
    \bigg(
        -\frac{2005}{96} (n_{12} v_2)^2 
        + \frac{123 \pi^2}{128} (n_{12} v_2)^2 
        + \frac{477941}{7200} v_1^2 
        - \frac{21 \pi^2}{32} v_1^2 
        - \frac{477941}{3600} (v_1 v_2) 
        + \frac{21 \pi^2}{16} (v_1 v_2)
    \bigg) n_{12}^{i}
    \nonumber \\
& + \bigg(
        \frac{225233}{1800} - \frac{43 \pi^2}{64}
    \bigg) (n_{12} v_2) v_{1}^{i}
    + \bigg(
        \frac{1099}{144} - \frac{41 \pi^2}{64}
    \bigg) (n_{12} v_2) v_{2}^{i}
    \bigg\}
    \nonumber \\
& + \frac{G^3 m_1 m_2^3}{r_{12}^2} \bigg\{
    \bigg(
        -\frac{198097}{4200} (n_{12} v_1) (n_{12} v_2) 
        - \frac{14377}{560} v_2^2 
        - 22 (n_{12} v_1) (n_{12} v_2) \ln\bigg(\frac{r_{12}}{r_{0}'}\bigg)
    \bigg) n_{12}^{i}
    \nonumber \\
& + 44 (n_{12} v_2) \ln\bigg(\frac{r_{12}}{r_{0}'}\bigg) v_{1}^{i}
    - \frac{562}{9} (n_{12} v_2) v_{2}^{i}
    \bigg\}
    \nonumber \\
& + \frac{G^3 m_1^3 m_2}{r_{12}^2} \bigg\{ 
    \bigg(
        \frac{6397}{75} (n_{12} v_1) (n_{12} v_2) 
        + \frac{44023}{720} (v_1 v_2) 
        + \frac{937}{1440} v_2^2 
        + 44 v_1^2 \ln\bigg(\frac{r_{12}}{r_{0}'}\bigg) - 44 (v_1 v_2) \ln\bigg(\frac{r_{12}}{r_{0}'}\bigg)
    \bigg) n_{12}^{i}
        \nonumber \\
& + \bigg(
        \frac{44023}{720} (n_{12} v_1)
        + \frac{110}{3} (n_{12} v_2) \ln\bigg(\frac{r_{12}}{r_{0}'}\bigg)
    \bigg) v_{1}^{i}
    \nonumber \\
&  + \bigg(
        \frac{14377}{280} (n_{12} v_2) 
        + 44 (n_{12} v_1) \ln\bigg(\frac{r_{12}}{r_{0}'}\bigg)
        - \frac{110}{3} (n_{12} v_2) \ln\bigg(\frac{r_{12}}{r_{0}'}\bigg)
    \bigg) v_{2}^{i}
    \bigg\}
    \nonumber \\
& + \frac{G^2 m_1 m_2^2}{r_{12}} \bigg\{
    \bigg(
        -\frac{6661}{720} (n_{12} v_1) (n_{12} v_2)^3
        - \frac{4621}{320} (n_{12} v_1)^2 v_2^2
        - \frac{166}{15} (n_{12} v_1) (n_{12} v_2) v_2^2
        \nonumber \\
&- \frac{3021}{320} (n_{12} v_2)^2 v_2^2
        - \frac{8849}{480} (v_1 v_2) v_2^2
    \bigg) n_{12}^{i}
    + \bigg(
        -\frac{59}{180} (n_{12} v_2)^3
        - \frac{3679}{480} (n_{12} v_2) v_2^2
    \bigg) v_{1}^{i}
    \nonumber \\
& + \bigg(
        -\frac{3341}{480} (n_{12} v_2)^3
        - \frac{4529}{240} (n_{12} v_2) (v_1 v_2)
        - \frac{8849}{480} (n_{12} v_1) v_2^2
        - \frac{7193}{240} (n_{12} v_2) v_2^2
    \bigg) v_{2}^{i}
    \bigg\}
    \nonumber \\
& + \frac{G^2 m_1^2 m_2}{r_{12}} \bigg\{
    \bigg(
        \frac{2197}{240} (n_{12} v_1)^3 (n_{12} v_2)
        + \frac{1337}{240} (n_{12} v_1) (n_{12} v_2)^3 
        + \frac{1133}{960} (n_{12} v_2)^4
        + \frac{2063}{96} (n_{12} v_1) (n_{12} v_2) v_1^2
        \nonumber \\
&
        + \frac{6943}{384} v_1^4
        + \frac{2099}{96} (n_{12} v_1)^2 (v_1 v_2)
        - \frac{23}{60} (n_{12} v_1) (n_{12} v_2) (v_1 v_2)
        + \frac{247}{30} (n_{12} v_2)^2 (v_1 v_2)
        + \frac{2503}{96} (v_1 v_2)^2
        \nonumber \\
&  + \frac{7}{5} (n_{12} v_1) (n_{12} v_2) v_2^2
        - \frac{139}{20} (n_{12} v_2)^2 v_2^2
        + \frac{3613}{192} v_1^2 v_2^2
        - \frac{5593}{240} (v_1 v_2) v_2^2
        + \frac{2931}{320} v_2^4 
    \bigg) n_{12}^{i}
    \nonumber \\
& + \bigg(
        \frac{2099}{288} (n_{12} v_1)^3
        + \frac{3059}{96} (n_{12} v_1) (n_{12} v_2)^2
        + \frac{13549}{480} (n_{12} v_2) v_1^2
        + \frac{3613}{48} (n_{12} v_1) (v_1 v_2)
        + \frac{3733}{160} (n_{12} v_2) v_2^2
    \bigg) v_{1}^{i}
    \nonumber \\
& + \bigg(
        \frac{10223}{480} (n_{12} v_1)^2 (n_{12} v_2) 
        - \frac{3781}{160} (n_{12} v_1) (n_{12} v_2)^2
        + \frac{4621}{480} (n_{12} v_2)^3
        + \frac{4723}{96} (n_{12} v_1) v_1^2
        + \frac{6499}{240} (n_{12} v_2) (v_1 v_2)
        \nonumber \\
&+ \frac{2293}{160} (n_{12} v_1) v_2^2
        - \frac{3733}{160} (n_{12} v_2) v_2^2
    \bigg) v_{2}^{i}
    \bigg\}
    \nonumber \\
& + G m_1 m_2 \bigg\{
    \bigg(
        -\frac{5}{32} (n_{12} v_1)^3 (n_{12} v_2)^3
        - \frac{5}{64} (n_{12} v_1) (n_{12} v_2)^5
        - \frac{5}{128} (n_{12} v_2)^6
        - \frac{5}{16} (n_{12} v_1) (n_{12} v_2)^3 (v_1 v_2)
        \nonumber \\
&  - \frac{11}{64} (n_{12} v_2)^4 (v_1 v_2)
        + \frac{5}{16} (n_{12} v_1) (n_{12} v_2) (v_1 v_2)^2
        + \frac{15}{32} (n_{12} v_1)^3 (n_{12} v_2) v_2^2
        + \frac{27}{64} (n_{12} v_1)^2 (n_{12} v_2)^2 v_2^2
        \nonumber \\
& + \frac{11}{32} (n_{12} v_1) (n_{12} v_2)^3 v_2^2
        + \frac{27}{128} (n_{12} v_2)^4 v_2^2
        - \frac{23}{32} (n_{12} v_1) (n_{12} v_2) v_1^2 v_2^2
        + \frac{17}{32} (n_{12} v_1)^2 (v_1 v_2) v_2^2
        \nonumber \\
& + \frac{7}{16} (n_{12} v_1) (n_{12} v_2) (v_1 v_2) v_2^2
        + \frac{13}{32} (n_{12} v_2)^2 (v_1 v_2) v_2^2
        - \frac{31}{64} (n_{12} v_1) (n_{12} v_2) v_2^4
        - \frac{53}{128} (n_{12} v_2)^2 v_2^4
        - \frac{49}{64} (v_1 v_2) v_2^4
    \nonumber \\
& + \frac{75}{128} v_2^6
    \bigg) n_{12}^{i}
    + \bigg(
        \frac{13}{64} (n_{12} v_2)^5
        - \frac{19}{32} (n_{12} v_2)^3 v_1^2
        + \frac{1}{16} (n_{12} v_2) (v_1 v_2)^2
        - \frac{23}{32} (n_{12} v_1)^2 (n_{12} v_2) v_2^2  - \frac{77}{96} (n_{12} v_2)^3 v_2^2
        \nonumber \\
&       + \frac{93}{32} (n_{12} v_2) v_1^2 v_2^2
        - \frac{33}{16} (n_{12} v_2) (v_1 v_2) v_2^2
        + \frac{123}{64} (n_{12} v_2) v_2^4
    \bigg) v_{1}^{i} + \bigg(
        -\frac{5}{32} (n_{12} v_1)^2 (n_{12} v_2)^3
        - \frac{11}{64} (n_{12} v_1) (n_{12} v_2)^4
    \nonumber \\
&   - \frac{13}{64} (n_{12} v_2)^5
        + \frac{5}{16} (n_{12} v_1)^2 (n_{12} v_2) (v_1 v_2) + \frac{1}{16} (n_{12} v_2)^3 (v_1 v_2)
        + \frac{1}{16} (n_{12} v_2) v_1^2 (v_1 v_2)
        + \frac{3}{16} (n_{12} v_2) (v_1 v_2)^2
        \nonumber \\
&       + \frac{17}{96} (n_{12} v_1)^3 v_2^2
        + \frac{7}{32} (n_{12} v_1)^2 (n_{12} v_2) v_2^2 + \frac{13}{32} (n_{12} v_1) (n_{12} v_2)^2 v_2^2
        + \frac{77}{96} (n_{12} v_2)^3 v_2^2
        - \frac{33}{32} (n_{12} v_2) v_1^2 v_2^2
        \nonumber \\
&       + \frac{3}{16} (n_{12} v_2) (v_1 v_2) v_2^2
        - \frac{49}{64} (n_{12} v_1) v_2^4
        - \frac{123}{64} (n_{12} v_2) v_2^4
    \bigg) v_{2}^{i}
    \bigg\}  \,.
\end{align}
\end{subequations}

\begin{subequations}
\begin{align}
        \calX_{1}^{i} &=\frac{1}{c^6}\,\calX_{1,\rm 3PN}^{i} + \frac{1}{c^8}\,\calX_{1,\rm 4PN}^{i} + \mathcal{O}\biggl(\frac{1}{c^{10}}\biggr)\,,\\
       \calX_{1,\rm 3PN}^{i} &= -
\frac{G^3 m_1^3 m_2}{r_{12}^2} n_{12}^{i} - \frac{49}{4}\frac{G^3 m_1^2 m_2^2}{r_{12}^2} n_{12}^{i} - \frac{3}{4}\frac{ G^3 m_1 m_2^3}{r_{12}^2} n_{12}^{i} \nn \\
&  + \frac{G^2 m_1^2 m_2}{r_{12}} \biggl\{ \biggl( \frac{11}{8} (n_{12} v_1)^2 - \frac{1}{4} (n_{12} v_1) (n_{12} v_2) - \frac{27}{8} v_1^2 \biggr) n_{12}^{i}+ \frac{35}{8} (n_{12} v_1) v_{1}^{i} - \frac{7}{4} (n_{12} v_1) v_{2}^{i} \biggr\} \nn \\
& + \frac{G^2 m_1 m_2^2}{r_{12}} \biggl\{ \biggl( \frac{3}{8} (n_{12} v_2)^2 - \frac{1}{8} v_1^2 - \frac{15}{8} v_2^2 \biggr) n_{12}^{i}  - \biggl( \frac{1}{4} (n_{12} v_1) + \frac{3}{2} (n_{12} v_2) \biggr) v_{1}^{i} + \frac{21}{4} (n_{12} v_2) v_{2}^{i} \biggr\} \nn \\
&+ G m_1 m_2 \biggl\{ \biggl( \frac{1}{16} (n_{12} v_2)^2 v_1^2 - \frac{5}{16} v_1^2 v_2^2 \biggr) n_{12}^{i}  + \biggl( \frac{1}{8} (n_{12} v_1) (n_{12} v_2)^2 - \frac{3}{4} (n_{12} v_2) v_1^2 + \frac{7}{4} (n_{12} v_2) (v_1 v_2)\nn  \\
& - \frac{5}{8} (n_{12} v_1) v_2^2 \biggr) v_{1}^{i}  + \frac{7}{8} (n_{12} v_2) v_1^2 v_{2}^{i} \biggr\}\,,\\
\calX_{1,\rm 4PN}^{i} &= 
    -\frac{41}{32} \frac{G^4 m_1^4 m_2}{r_{12}^3} n_{12}^{i} - \frac{35}{32} \frac{G^4 m_1 m_2^4}{r_{12}^3}  n_{12}^{i} + \frac{G^4 m_1^2 m_2^3}{r_{12}^3} \biggl( \frac{241987}{10080} + \frac{63 \pi^2}{32} - \frac{88}{3} \ln\bigg(\frac{r_{12}}{r_{0}'}\bigg) \bigg) n_{12}^{i} \nn \\
    &+ \frac{G^4 m_1^3 m_2^2}{r_{12}^3} \biggl( \frac{67469}{10080} + \frac{21 \pi^2}{8} - 22 \ln\bigg(\frac{r_{12}}{r_{0}'}\bigg) \bigg) n_{12}^{i} \nn \\
    &+ \frac{G^3 m_1 m_2^3}{r_{12}^2} \biggl\{ \biggl( -\frac{3}{8} (n_{12} v_1) (n_{12} v_2) + \frac{79}{64} (n_{12} v_2)^2 - \frac{7}{8} v_1^2 + \frac{15}{16} (v_1 v_2) + \frac{57}{64} v_2^2 \biggr) n_{12}^{i} \nn \\
    &+ \biggl( -\frac{7}{4} (n_{12} v_1) - \frac{3}{4} (n_{12} v_2) \biggr) v_{1}^{i} + \biggl( \frac{15}{16} (n_{12} v_1) - \frac{97}{32} (n_{12} v_2) \biggr) v_{2}^{i} \biggr\} \nn \\
    &+ \frac{G^3 m_1^3 m_2}{r_{12}^2}  \biggl\{ \biggl( \frac{703}{192} (n_{12} v_1)^2 - \frac{33}{32} (n_{12} v_1) (n_{12} v_2) + \frac{1}{8} (n_{12} v_2)^2 + \frac{147463}{20160} v_1^2 \nn \\
    &+ \frac{5}{32} (v_1 v_2) - \frac{1}{8} v_2^2 - \frac{11}{3} v_1^2 \ln\bigg(\frac{r_{12}}{r_{0}'}\bigg) \bigg) n_{12}^{i} \nn \\
    &+ \biggl( \frac{4481}{1260} (n_{12} v_1) + \frac{33}{32} (n_{12} v_2) - \frac{22}{3} (n_{12} v_1) \ln\bigg(\frac{r_{12}}{r_{0}'}\bigg) \bigg) v_{1}^{i} \nn \\
    &+ \biggl( -\frac{27}{32} (n_{12} v_1) - \frac{15}{16} (n_{12} v_2) \biggr) v_{2}^{i} \biggr\} \nn \\
    &+ \frac{G^3 m_1^2 m_2^2}{r_{12}^2}  \biggl\{ \biggl( \frac{3899}{192} (n_{12} v_1)^2 - \frac{409}{16} (n_{12} v_1) (n_{12} v_2) + \frac{4859}{192} (n_{12} v_2)^2 \nn \\
    &- \frac{27281}{576} v_1^2 + \frac{21 \pi^2}{64} v_1^2 + \frac{2695}{32} (v_1 v_2) - \frac{7631}{192} v_2^2 \biggr) n_{12}^{i} \nn \\
    &+ \biggl( \frac{9613}{288} (n_{12} v_1) - \frac{781}{16} (n_{12} v_2) + \frac{21 \pi^2}{32} (n_{12} v_1)  \biggr) v_{1}^{i} \nn \\
    &+ \biggl( -\frac{1793}{32} (n_{12} v_1) + \frac{3569}{96} (n_{12} v_2) \biggr) v_{2}^{i} \biggr\} \nn \\
    &+ \frac{G^2 m_1^2 m_2}{r_{12}}  \biggl\{ \biggl( -\frac{175}{192} (n_{12} v_1)^4 - \frac{119}{48} (n_{12} v_1)^3 (n_{12} v_2) - \frac{23}{32} (n_{12} v_1)^2 (n_{12} v_2)^2 \nn \\
    &+ \frac{5}{16} (n_{12} v_1) (n_{12} v_2)^3 + \frac{259}{96} (n_{12} v_1)^2 v_1^2 - \frac{99}{32} (n_{12} v_1) (n_{12} v_2) v_1^2 + \frac{37}{24} (n_{12} v_2)^2 v_1^2 \nn \\
    &- \frac{1393}{192} v_1^4 - \frac{69}{64} (n_{12} v_1)^2 (v_1 v_2) + \frac{89}{16} (n_{12} v_1) (n_{12} v_2) (v_1 v_2) - \frac{1}{8} (n_{12} v_2)^2 (v_1 v_2) \nn \\
    &+ \frac{701}{64} v_1^2 (v_1 v_2) - \frac{49}{16} (v_1 v_2)^2 + \frac{29}{32} (n_{12} v_1)^2 v_2^2 - \frac{23}{16} (n_{12} v_1) (n_{12} v_2) v_2^2 \nn \\
    &- \frac{193}{48} v_1^2 v_2^2 + \frac{3}{4} (v_1 v_2) (v_2^2) \biggr) n_{12}^{i} \nn \\
    &+ \biggl( -\frac{1}{32} (n_{12} v_1)^3 + \frac{19}{8} (n_{12} v_1)^2 (n_{12} v_2) - \frac{169}{96} (n_{12} v_1) (n_{12} v_2)^2 + \frac{1697}{96} (n_{12} v_1) v_1^2 \nn \\
    &- \frac{417}{32} (n_{12} v_2) v_1^2 - \frac{515}{32} (n_{12} v_1) (v_1 v_2) + \frac{161}{12} (n_{12} v_2) (v_1 v_2) - \frac{97}{96} (n_{12} v_1) v_2^2 \biggr) v_{1}^{i} \nn \\
    &+ \biggl( \frac{349}{192} (n_{12} v_1)^3 - \frac{97}{16} (n_{12} v_1)^2 (n_{12} v_2) + \frac{41}{16} (n_{12} v_1) (n_{12} v_2)^2 - \frac{953}{64} (n_{12} v_1) v_1^2 \nn \\
    &+ \frac{77}{24} (n_{12} v_2) v_1^2 + \frac{225}{16} (n_{12} v_1) (v_1 v_2) - \frac{7}{4} (n_{12} v_2) (v_1 v_2) - \frac{53}{16} (n_{12} v_1) v_2^2 \biggr) v_{2}^{i} \biggr\} \nn \\
    &+ \frac{G^2 m_1 m_2^2 }{r_{12}} \biggl\{ \biggl( -\frac{7}{16} (n_{12} v_1) (n_{12} v_2)^3 + \frac{1}{64} (n_{12} v_2)^4 + \frac{41}{96} (n_{12} v_2)^2 v_1^2 \nn \\
    &- \frac{3}{32} v_1^4 - \frac{81}{32} (n_{12} v_2)^2 (v_1 v_2) + \frac{23}{16} (n_{12} v_1) (n_{12} v_2) v_2^2 + \frac{161}{64} (n_{12} v_2)^2 v_2^2 \nn \\
    &- \frac{145}{96} v_1^2 v_2^2 + \frac{87}{32} (v_1 v_2) v_2^2 - \frac{83}{32} v_2^4 \biggr) n_{12}^{i} \nn \\
    &+ \biggl( \frac{29}{48} (n_{12} v_1) (n_{12} v_2)^2 + \frac{3}{2} (n_{12} v_2)^3 - \frac{3}{8} (n_{12} v_1) v_1^2 - \frac{21}{4} (n_{12} v_2) v_1^2 \nn \\
    &+ \frac{643}{48} (n_{12} v_2) (v_1 v_2) - \frac{145}{48} (n_{12} v_1) v_2^2 - \frac{37}{4} (n_{12} v_2) v_2^2 \biggr) v_{1}^{i} \nn \\
    &+ \biggl( -\frac{49}{32} (n_{12} v_1) (n_{12} v_2)^2 - \frac{243}{64} (n_{12} v_2)^3 + \frac{631}{96} (n_{12} v_2) v_1^2 - \frac{127}{8} (n_{12} v_2) (v_1 v_2) \nn \\
    &+ \frac{87}{32} (n_{12} v_1) v_2^2 + \frac{839}{64} (n_{12} v_2) v_2^2 \biggr) v_{2}^{i} \biggr\} \nn \\
    &+ G m_1 m_2 \biggl\{ \biggl( -\frac{1}{16} (n_{12} v_1) (n_{12} v_2)^3 v_1^2 - \frac{1}{32} (n_{12} v_2)^4 v_1^2 + \frac{3}{64} (n_{12} v_2)^2 v_1^4 \nn \\
    &- \frac{3}{16} (n_{12} v_2)^2 v_1^2 (v_1 v_2) + \frac{3}{16} (n_{12} v_1) (n_{12} v_2) v_1^2 v_2^2 + \frac{5}{32} (n_{12} v_2)^2 v_1^2 v_2^2 \nn \\
    &- \frac{15}{64} v_1^4 v_2^2 + \frac{1}{16} v_1^2 (v_1 v_2) v_2^2 - \frac{1}{4} v_1^2 v_2^4 \biggr) n_{12}^{i} \nn \\
    &+ \biggl( -\frac{1}{8} (n_{12} v_1)^2 (n_{12} v_2)^3 - \frac{1}{16} (n_{12} v_1) (n_{12} v_2)^4 + \frac{3}{16} (n_{12} v_1) (n_{12} v_2)^2 v_1^2 \nn \\
    &+ \frac{5}{8} (n_{12} v_2)^3 v_1^2 - \frac{15}{16} (n_{12} v_2) v_1^4 - \frac{3}{4} (n_{12} v_1) (n_{12} v_2)^2 (v_1 v_2) - \frac{5}{12} (n_{12} v_2)^3 (v_1 v_2) \nn \\
    &+ \frac{21}{8} (n_{12} v_2) v_1^2 (v_1 v_2) - \frac{1}{4} (n_{12} v_2) (v_1 v_2)^2 + \frac{3}{8} (n_{12} v_1)^2 (n_{12} v_2) v_2^2 + \frac{5}{16} (n_{12} v_1) (n_{12} v_2)^2 v_2^2 \nn \\
    &- \frac{15}{16} (n_{12} v_1) v_1^2 v_2^2 - \frac{15}{8} (n_{12} v_2) v_1^2 v_2^2 + \frac{1}{4} (n_{12} v_1) (v_1 v_2) v_2^2 + \frac{15}{8} (n_{12} v_2) (v_1 v_2) v_2^2 \nn \\
    &- \frac{1}{2} (n_{12} v_1) v_2^4 \biggr) v_{1}^{i} \nn \\
    &+ \biggl( -\frac{3}{16} (n_{12} v_1) (n_{12} v_2)^2 v_1^2 - \frac{5}{24} (n_{12} v_2)^3 v_1^2 + \frac{21}{32} (n_{12} v_2) v_1^4 \nn \\
    &- \frac{1}{8} (n_{12} v_2) v_1^2 (v_1 v_2) + \frac{1}{16} (n_{12} v_1) v_1^2 v_2^2 + \frac{15}{16} (n_{12} v_2) v_1^2 v_2^2 \biggr) v_{2}^{i} \biggr\}\,.
\end{align}
\end{subequations}

\begin{subequations}
\begin{align}
   \calY_{11}^{ji} &= \frac{1}{c^8}\,\calY_{11,\rm 4PN}^{ji} + \mathcal{O}\biggl(\frac{1}{c^{10}}\biggr)\,,\qquad \calY_{21}^{ji} = \frac{1}{c^8}\,\calY_{21,\rm 4PN}^{ji} + \mathcal{O}\biggl(\frac{1}{c^{10}}\biggr)\,,\\
   \calY_{11,\rm 4PN}^{ji} &= 
     \frac{1}{4} G^2 m_1 m_2^2\, (n_{12} v_2)^2\, \delta^{ij}  + G^2 m_1^2 m_2\, \biggl\{
   \biggl(
     -\frac{49}{16} (n_{12} v_1)^2
     - \frac{49}{8} (n_{12} v_1)(n_{12} v_2)
   \biggr)\,\delta^{ij} \nonumber\\
   & + 
   \biggl(
     -\frac{29}{16} (n_{12} v_1)^2
     -\frac{15}{8} (n_{12} v_1)(n_{12} v_2)
     +\frac{1}{4} (n_{12} v_2)^2
     -\frac{49}{16} v_1^2
     +\frac{49}{8} (v_1 v_2)
     -\frac{3}{2} v_2^2
   \biggr)
     n_{12}^i n_{12}^j \nonumber\\
   &+
     \biggl(
       \frac{7}{2} (n_{12} v_1)
       -3 (n_{12} v_2)
     \biggr)
     \biggl(
       n_{12}^j v_1^i + n_{12}^i v_1^j
     \biggr)-
     \frac{45}{16} v_1^i v_1^j +
     \biggl(
       -\frac{7}{2} (n_{12} v_1)
       +\frac{7}{4} (n_{12} v_2)
     \biggr)
     \biggl(
       n_{12}^j v_2^i + n_{12}^i v_2^j
     \biggr) \nonumber\\
   & +
     \frac{35}{16} \biggl( v_1^j v_2^i + v_1^i v_2^j \biggr)
   \biggr\}\,,\\
   \calY_{21,\rm 4PN}^{ji} &= G^2 m_1^2 m_2 \biggl\{ \frac{7}{8} (n_{12} v_1) (n_{12} v_2) \delta^{ij} + \frac{1}{8} (n_{12} v_1) (n_{12} v_2) n_{12}^i n_{12}^j - \frac{1}{4} (n_{12} v_1) n_{12}^j v_1^i + \frac{7}{8} (n_{12} v_1) n_{12}^j v_2^i \nn \\
    &- \frac{5}{8} (n_{12} v_1) n_{12}^i v_2^j \biggr\} \nn \\
    & + G^2 m_1 m_2^2 \biggl\{ \frac{7}{8} (n_{12} v_1) (n_{12} v_2) \delta^{ij} + \frac{1}{8} (n_{12} v_1) (n_{12} v_2) n_{12}^i n_{12}^j - \frac{5}{8} (n_{12} v_2) n_{12}^j v_1^i+ \frac{7}{8} (n_{12} v_2) n_{12}^i v_1^j   \nn \\
    &- \frac{1}{4} (n_{12} v_2) n_{12}^i v_2^j \biggr\} \,.
\end{align}
\end{subequations}

\bibliography{ListeRef_BLL24.bib}

\end{document}